\documentclass[oupdraft]{bio}
\usepackage{natbib}
\usepackage{url} 
\usepackage{amsmath}
\usepackage[margin=1in]{geometry}
\usepackage{url}
\usepackage{graphicx}
\usepackage{amsmath, amssymb}
\usepackage{bbold}
\usepackage{mathbbol}
\usepackage{float}
\usepackage{xcolor}
\usepackage{bm} 
\usepackage{setspace}
\usepackage{multirow}
\usepackage{lscape}
\usepackage{rotating}
\usepackage{adjustbox}
\usepackage{booktabs}
\usepackage{ragged2e}
\usepackage{algorithm}
\usepackage{algorithmic}
\usepackage{caption}
\usepackage{float}
\def\*#1{\mathbf{#1}}
\def\+#1{\amsmathbb{#1}}
\def\^#1{\mathbb{#1}}
\DeclareSymbolFontAlphabet{\amsmathbb}{AMSb}%

\begin{document}
\title{Distributional data analysis via quantile functions and its application to modelling digital biomarkers of gait in Alzheimer’s Disease}

\author{RAHUL GHOSAL$^{1,\ast}$, VIJAY R. VARMA$^{2}$, DMITRI VOLFSON$^{3}$, INBAR HILLEL$^{4}$,\\ JACEK URBANEK$^{5}$,
JEFFREY M. HAUSDORFF$^{4,6,7}$, AMBER WATTS$^{8}$,\\ 
VADIM ZIPUNNIKOV$^{1}$ \\[4pt]
\textit{$^{1}$ Department of Biostatistics, Johns Hopkins Bloomberg School of Public Health, Baltimore, Maryland USA,
$^{2}$ National Institute on Aging (NIA), National Institutes of Health (NIH), Baltimore, Maryland, USA, $^{3}$Neuroscience Analytics, Computational Biology, Takeda, Cambridge, MA, USA, $^{4}$Center for the Study of Movement, Cognition and Mobility, Neurological Institute, Tel Aviv Sourasky Medical Center, Tel Aviv, Israel, $^{5}$ Department of Medicine, Johns Hopkins University School of Medicine, Baltimore Maryland, USA, $^{6}$ Department of Physical Therapy, Sackler Faculty of Medicine, and Sagol School of Neuroscience, Tel Aviv University, Tel Aviv, Israel, $^{7}$Rush Alzheimer’s Disease Center and Department of Orthopedic Surgery, Rush University Medical Center, Chicago, USA, $^{8}$ Department of Psychology, University of Kansas, Lawrence, KS, USA }
\\[2pt]
{rghosal@ncsu.edu}}

\markboth%
{R. Ghosal and others}
{Distributional data analysis via quantile functions}

\maketitle

\footnotetext{To whom correspondence should be addressed.}

\begin{abstract}
{With the advent of continuous health monitoring with wearable devices, users now generate their unique streams of continuous data such as minute-level step counts or heartbeats. Summarizing these streams via scalar summaries often ignores the distributional nature of wearable data and almost unavoidably leads to the loss of critical information. We propose to capture the distributional nature of wearable data via user-specific quantile functions (QF) and use these QFs as predictors in scalar-on-quantile-function-regression (SOQFR). As an alternative approach, we also propose to represent QFs via user-specific L-moments, robust rank-based analogs of traditional moments, and use L-moments as predictors in SOQFR (SOQFR-L). These two approaches provide two mutually consistent interpretations: in terms of quantile levels by SOQFR and in terms of L-moments by SOQFR-L. We also demonstrate how to deal with multi-modal distributional data via Joint and Individual Variation Explained (JIVE) using L-moments. The proposed methods are illustrated in a study of association of digital gait biomarkers with cognitive function in Alzheimer’s disease (AD). Our analysis shows that the proposed methods demonstrate higher predictive performance and attain much stronger associations with clinical cognitive scales compared to simple distributional summaries.}
{Wearable data; Quantile functions;  L-Moments; Scalar-on-quantile-function regression; JIVE; Alzheimer’s disease; Gait.}
\end{abstract}

\section{Introduction}
\label{sec:intro4}
Wearables now generate continuous streams of data that capture minute-level physical activity, heart rhythms and other physiological signals. These streams are a rich source of information and can be used for a deeper understanding of human behaviours and their influence on human health and disease. Common analytical practice in many epidemiological and clinical studies employing wearables is to use simple scalar summaries such as total activity count (TAC), minutes of moderate-to-vigorous-intensity physical activity (MVPA) \citep{varma2017re,bakrania2017associations} or total step count (TSC)  \citep{reider2020methods, mc2020differentiating}. However, collapsing the entire stream of data into a single metric completely ignores temporal (diurnal or time of the day) and distributional aspects of wearable data.
 
Temporal aspect of wearable data can be accounted for with functional data analysis (FDA) that treats wearable data streams as functional observations recorded over 24 hours \citep{morris2006using, xiao2015quantifying,goldsmith2016new,goldsmith2015generalized}. Association between accelerometry-estimated temporal functional profiles and variables of interest such as health outcomes, age, BMI, and others can be studied within FDA using scalar-on-function (SOFR) or function-on-scalar (FOSR) regression \citep{morris2015functional} models. In addition to diurnal (over 24-hour) modelling, FDA approaches help to model weekly and seasonal variation in accelerometric signal \citep{huang2019multilevel, xiao2015quantifying}. 
Co-registration (or warping) is often a desirable pre-processing step to make sure the amplitude and phase variations are properly separated during diurnal modelling \citep{dryden2016statistical, wrobel2019registration}. 

Distributional aspect of wearable data can be captured via modelling user-specific distributions. \cite{augustin2017modelling} proposed to summarize accelerometry activity counts recorded over 24-hours with user-specific histograms. They proposed to use these histograms as predictors in scalar-on-function regression. The main limitations of this approach include i) possibly unequal (effective) support across user-specific histograms, ii) inability to model specific quantiles of the distribution, which, as it will be demonstrated below, could be of a great practical interest, iii) scale-dependence. 
\cite{mck} developed an empirical likelihood based functional ANOVA test for comparing groups of subjects based on their ordered activity profiles to model the amounts of time spent by subjects doing activity above a certain threshold. Compositional data analysis (CoDA)  \citep{aitchison1982statistical,dumuid2019compositional,dumuid2020compositional} is another group of methods to model continuously measured wearable data. \cite{petersen2016functional} and \cite{hron2016simplicial}  developed functional compositional methods to analyse samples of densities. \cite{petersen2021modeling} provides a nice accessible tutorial style review of recent developments in that area. Distribution-on-distribution regression models were suggested by \cite{chen2021wasserstein} and  \cite{ghodrati2021distribution} who used Wasserstein-distances and optimal transport ideas. Additionally, \cite{talska2021compositional} developed a compositional scalar-on-function regression method using a centred logratio transformation \citep{aitchison1982statistical} of subject-specific densities. This has been done by mapping densities from the Bayes space of density functions \citep{van2014bayes} to a Hilbert $L^2$ space and then performing traditional functional data modelling. \cite{matabuena2020glucodensities} proposed to use subject-specific densities of glucose measures collected with continuous glucose monitoring (CGM) as predictors (as well as response) and demonstrated advantages of this approach over the use of summary measures typically employed in CGM studies. Specifically, a non-parametric kernel functional regression model was developed that employed a 2-Wasserstein distance to model scalar outcomes and glucodensity predictors. 

In this article, we put forward an alternative to the above compositional functional approaches and propose to use subject-specific quantile functions to capture the distributional nature of wearable data. Our approach overcomes the limitations i) and ii) as in \citep{augustin2017modelling} and provides a more flexible way to summarize distributional properties of wearable data. \cite{matabuena2021distributional} used a quantile-function representation of NHANES 2003-2006 accelerometry data to predict health outcomes using survey weighted nonparametric regression model that employed a 2-Wasserstein distance. Note that compared to analyzing density functions in a nonlinear space \citep{talska2021compositional} or using nonparametric methods based on distances or kernels \citep{matabuena2021distributional}, our method is semiparametric and offers direct interpretability in terms of the quantile-levels of subject-specific distribution. There is rich literature on quantile functions \citep{gilchrist2000statistical,powley2013quantile} in statistical modelling and decision theoretic analysis. Quantile functions enjoy many mathematical properties, which make them particularly useful for distributional modelling. Quantile representations have been used in symbolic data literature for regression modelling of both quantile functions responses and predictors. The approach is based on the use of $\ell_{2}$ Wasserstein distance \citep{irpino2013metric,verde2010ordinary}. \cite{zhang2011functional} proposed a method for density function estimation using a quantile synchronization approach, instead of using the cross-sectional average density which does not incorporate time-warping. Recently, \cite{yang2020quantile} proposed  quantile-function-on-scalar regression approaches for modelling quantile functions as outcomes. Having quantile functions as outcomes imposes constraints on the regression that requires a regression approximant to be a valid quantile function \cite{yang2020random}. In our regression applications, we do not have these constraints because user-specific quantile functions are used as functional predictors. 

 As a motivating study, we will focus on continuous accelerometry data collected over one week in a sample of $86$ community-dwelling older adults with mild Alzheimer's disease (AD) and cognitively normal controls (CNC)\citep{varma2017daily,gait2020vr}. AD is the most common form of dementia and cases are projected to more than double in the next 40 years \citep{addoi:10.1002/alz.12068,hebert2013alzheimer} in the United States. The absence of any permanent treatment to cure AD makes early detection of cognitive impairment paramount. Digital biomarkers have recently been considered for early detection of AD as an alternative to more invasive and expensive fluid and imaging markers \citep{kourtis2019digital}.  Specifically, digital biomarkers that reflect alterations in gait may help to predict AD due to the close relationship between complex cognitive functions and gait \citep{yogev2008role,mc2019alzheimer, gait2020vr}. In our study, subjects wore an accelerometer on their hip in order to measure continuous, community activity over 7 days. Using a validated processing pipeline \citep{weiss2014objective}, we measured 52, domain-specific gait parameters during identified episodes of sustained walking (defined as walking longer than 60 seconds) over the course of the 7-day data collection period.

\begin{figure}[H]
\centering
\includegraphics[width=1\linewidth , height=.5\linewidth]{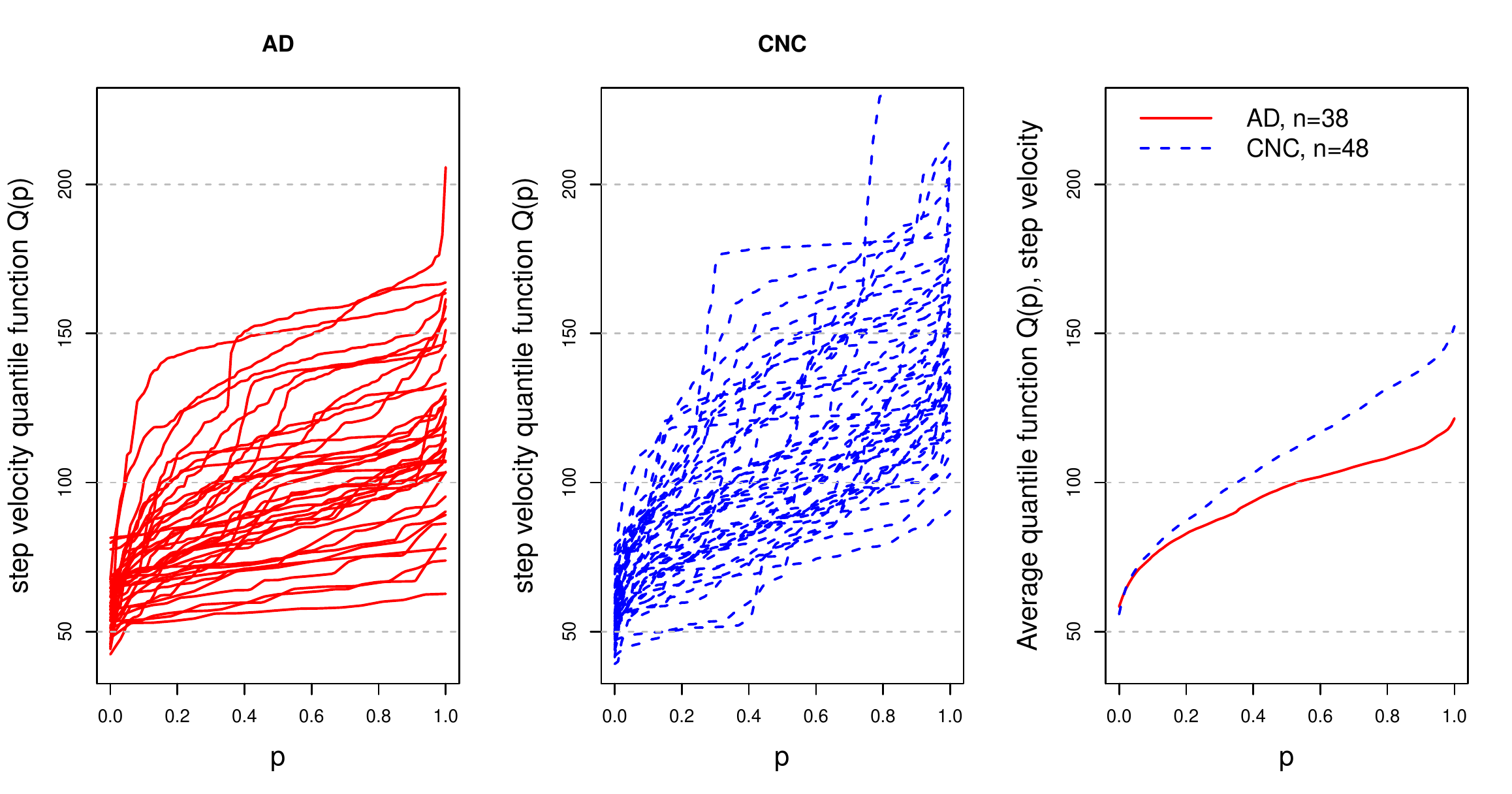}
\caption{Displayed are the individual (left two panel) and average (right panel) quantile functions of step velocity for AD (solid) and CNC (dashed).}
\label{fig:fig1}
\end{figure}

 Figure \ref{fig:fig1} shows the user-specific quantile functions of accelerometry-estimated step velocity for mild-AD and CNC group.
 Interestingly, for each group, the average quantile function is directly related to the Wasserstein barycenter of their respective distribution \citep{bigot2018upper}. The average quantile functions, therefore, can be highly informative and can identify parts of the distribution that are most discriminative between groups of interest. The largest difference between the two groups can be seen in the upper quantiles of step velocity. This supports the point that quantile functions can be useful for distributional representations.

 Three main methodological contributions of this article are as follows. First, we propose to  capture  the  distributional aspect of wearable data via user-specific quantile functions and use them as predictors in scalar-on-quantile-function regression (SOQFR). This allows us to apply inferential tools of functional data analysis to make inference about the specific quantile levels of distributional predictors. This approach is further generalized using functional generalized additive model \citep{mclean2014functional} which can capture possible nonlinear effects of the quantile functions. Such models have strong mathematical interpretations as linear functional of quantile functions or its transformation is known to encode several important characteristics of a continuous distribution \citep{powley2013quantile}. Second, we propose to use L-moments to represent user-specific quantile functions via functional decompositions that also preserve distributional information and encode this information via moments.  L-moments introduced in \cite{hosking1990moments} are robust rank based analogs of traditional moments and they fully define the distribution. We show that SOQFR model can be reduced to a generalized linear model with L-moments (SOQFR-L). These two approaches provide two mutually consistent interpretations: in terms of quantile levels by SOQFR and in terms of L-moments by SOQFR-L.
 Third, in our motivating application, there are multiple digital biomarkers that can be grouped into five gait domains including Amplitude, Pace, Rhythm, Symmetry, and Variability. Thus, this gives rise to a design with multi-modal distributional data. We demonstrate how L-moments can be used for analyzing joint and individual sources of variation using JIVE (Joint and Individual Variation Explained) \citep{lock2013joint} method.  

The rest of this article is organized as follows. In Section 2, we present our modelling framework, introduce the mathematical background for quantile functions, and illustrate the proposed SOQFR approach. In Section 3, we introduce L-moments for distributional data and show how they can be used for SOQFR-L and JIVE. In Section 4, we demonstrate the applications of the proposed methods in the Alzheimer's disease (AD) study. Section 5 concludes with a discussion on the main contribution and some possible extensions of this work.

\section{MODELLING FRAMEWORK}
\label{mf4}

Suppose, we have 
multiple repeated observations of a variable $X$ 
per subject denoted by $X_{ij} \hspace{1 mm} (j=1,\ldots,n_i)$, for subject $i=1,\ldots,n$, where $n_i$ (the number of observations for subject $i$) is typically quite large (e.g., on an average around 100 walking bouts in our study). In some applications (e.g. activity data), $X_{ij} = X_i(t_{ij})$ can be observed across various time-points $t_{ij}$. Assume $X_{ij}$ $(j=1,\ldots,n_i) \sim F_i(x)$, a subject-specific cumulative distribution function (cdf), where $F_i(x)=P(X_{ik}\leq x)$. Then, we can define subject-specific quantile function $Q_i(p)= \inf\{x: F_i(x)\geq p\}$. The quantile function completely characterizes the distribution of the individual observations. In this article, we restrict our attention to cases, where both $F_i(x)$ s and $Q_i(p)$ are continuous, which ensures $Q_i=F_{i}^{-1}$, so $F_i(Q_i(p))=p,\hspace{2 mm} Q_i(F_i(x))=x$. This also ensures both quantile function and cdf are strictly increasing in their respective domains \citep{powley2013quantile}. Using $X_{ij}$ ($j=1,\ldots,J$) one can calculate $\hat{F}_i(x)$, the empirical cdf and then obtain the empirical quantile function $\hat{Q}_i(p)=\hat{F}_i^{-1}(p)$ ($p \in [0,1]$, the percentile resolution can be determined based on the amount of available data).
Estimation of quantile functions can be done via a linear interpolation of  order statistics \citep{parzen2004quantile} and does not require a bandwidth selection as in kernel density estimation.
In particular, we use the following estimator of quantile functions, 
$$\hat{Q}(p)=(1-w)X_{([(n+1)p])}+wX_{([(n+1)p]+1)},$$
where $X_{(1)}\leq X_{(2)}\leq \ldots,X_{(n)}$ are the corresponding order statistics from a sample $(X_1,X_2,\ldots, X_n)$ and $w$ is a weight satisfying $(n+1)p=[(n+1)p]+w$.
Quantile functions have a few nice mathematical properties  \citep{gilchrist2000statistical,powley2013quantile} that make them particularly suitable for distributional modelling. For convenience, we list some of these properties below.

\begin{itemize}
\itemsep -0.5 em 
\item A non-negative linear combination of finite number of quantile functions is a quantile function.
\item For a probability distribution defined via a quantile function $Q(\cdot)$, all integer moments can be represented as $E(X^m)=\int_{0}^{1}  Q^m(p)dp$ (assuming that all moments exist).
\item The quantile density function and the $p$-probability density function are defined as $q(p)= Q^{'}(p)$ and $f(Q(p))$, respectively. Here, $f$ is the density function corresponding to $F$.  
\item The average of subject-specific quantile functions 
$\Bar{Q}(p)=\frac{1}{n}\sum_{i=1}^{n}Q_i(p)$
 can be mapped to the Wasserstein barycenter of the measures induced by the respective distributions \citep{bigot2018upper}.
\item In $L_2\hspace{1mm}[0,1]$, a distance between any two quantile functions can be defined via the 2-Wasserstein distance $W_2$,   $d_2(F_1,F_2)=(\int_{0}^{1} (Q_1(p)-Q_2(p))^2 dp)^{1/2}$.
\end{itemize}
Expansions of quantile functions via orthogonal polynomials are directly related to the components of the Shapiro–Francia Statistic \citep{takemura1983orthogonal} and L-moments \citep{hosking1990moments} which will be discussed in greater details in Section \ref{Lmom}.

Next section illustrates the use of quantile functions as functional objects in scalar-on-function regression models. 

\subsection{Scalar-on-quantile-function regression}
\label{sofr:qf}
In this section, we assume that $Y_i, i=1,\ldots,n$, is an outcome of interest that can be continuous or discrete, coming from an exponential family. We consider the following generalized scalar-on-function regression (SOFR) with quantile functions as predictors which we will refer to as a scalar-on-quantile-function regression (SOQFR) model: 
\begin{eqnarray}
E(Y_i|X_{i1},X_{i2},\ldots,X_{in_i})=\mu_i, \hspace{2 mm}
g(\mu_i)=\alpha+\*Z_i^T\bm\gamma + \int_{0}^{1} Q_i(p)\beta(p)dp.\label{soqfr}
\end{eqnarray}
Here $g$ is a known link function, $Z_i$ are confounding scalar covariates and $Q_i(p)$'s ($i=1,2,\ldots,n$) are the subject-specific quantile functions of predictor of interest $X_{ij}$. The smooth coefficient function $\beta(p)$ represents the functional effect of the quantile function at quantile level $p$. As pointed out in \cite{reiss2017methods}, locations with largest $|\beta(p)|$ are the most influential to the response and of practical interest. In the special case of $\beta(p)=\beta$, model (\ref{soqfr}) reduces to a usual GLM on the mean ($\nu_i = \int_{0}^{1} Q_i(p)dp$) of X (the variable on which we have multiple observations)
\begin{equation}
g(\mu_i)=\alpha+\*Z_i^T\bm\gamma + \beta\int_{0}^{1} Q_i(p)dp=\alpha+\*Z_i^T\bm\gamma + \beta\nu_i. \label{meansofr}
\end{equation}
Several methods exist in the literature for estimation of the smooth coefficient function $\beta(p)$ \citep{goldsmith2011penalized,marx1999generalized} in SOFR. In this article, we follow a smoothing spline estimation method.
The penalized negative log likelihood criterion for estimation is given by
\begin{equation}
R(\alpha,\bm\gamma,\beta(\cdot))=  -2log L(\alpha,\bm\gamma,\beta(\cdot);Y_i,\*Z_i^T,Q_i(p)) + \lambda \int_{0}^{1}\{\beta^{''}(p)\}^2dp.\label{sofr:est1}
\end{equation}
The second derivative penalty on $\beta(p)$ ensures the resulting coefficient function is smooth. We model the unknown coefficient functions $\beta(p)$ using univariate basis function expansion as $\beta(p)=\sum_{k=1}^{K}\beta_{k}\theta_{k}(p)=\bm\theta(p)^{T}\bm\beta$, where $\bm\theta(p)=[\theta_{1}(p),\theta_{2}(p),\ldots ,\theta_{K }(p)]^T$ and $\bm\beta=(\beta_{1},\beta_{2},\ldots ,\beta_{K})^T$ is the vector of unknown coefficients. In this article, we use cubic B-spline basis functions, however, other basis functions can be used as well. The linear functional effect then becomes $\int_{0}^{1} Q_i(p)\beta(p)dp=\sum_{k=1}^{K}\beta_{k}\int_{0}^{1} Q_i(p)\theta_{k}(p)dp=\sum_{k=1}^{K}\beta_{k}Q_{ik}= \*Q_{i}^T\bm\theta$. The minimization criterion in (\ref{sofr:est1}) now can be reformulated as, 
\begin{equation}
R(\psi)=R(\alpha,\bm\gamma,\bm\beta)=  -2log L(\alpha,\bm\gamma,\bm\beta;Y_i,\*Z_i,\*Q_i) + \lambda\bm\beta^T\+P\bm\beta, \label{sofr:est2}
\end{equation}
where $\+P$ is the penalty matrix given by $\+P=\{\int_{0}^{1}\bm\theta^{''}(p)\bm\theta^{''}(p)^{T}dp\}$.
This minimization can be carried out using the Newton-Raphson algorithm implemented under generalized additive models (GAM) \citep{wood2016smoothing,wood2017generalized}. The smoothing parameter $\lambda$ can be chosen using REML, information criteria like AIC, BIC, or data driven methods such as Generalized CV. We use the {\tt refund} package \citep{refund} in R \citep{Rsoft} for implementation of SOFR.

\subsection{Functional Generalized Additive Regression with Quantile Functions}
The SOQFR model (\ref{soqfr}) assumes a linear association between the quantile function and the outcome. SOQFR model can be extended to functional generalized additive model (FGAM) of \cite{mclean2014functional} which can be used to capture nonlinear effects of quantile function $Q(p)$. We denote this model as FGAM-QF. Specifically, we model the link function $g(\cdot)$ as
\begin{equation}
   g(\mu_i)=\alpha+\*Z_i^T\bm\gamma + \int_{0}^{1}F(Q_i(p),p)dp.\label{FGAM-QF} 
\end{equation}
The bivariate function $F\{Q(p),p\}$ is smooth function on $\mathbb{R}\times [0,1]$, capturing effect of the subject-specific quantile function $Q(p)$ at quantile level $p$. In a special case of $F\{Q(p),p\}=Q(p)\beta(p)$, FGAM-QF reduces to model (\ref{soqfr}). The estimation procedure for FGAM-QF is discussed in Appendix 1 of Supplementary Material. It can be shown that the FGAM-QF is flexible and remains invariant under any continuous transformation of predictors \citep{mclean2014functional}.

\section{L-moments}
\label{Lmom}
L-moments were introduced and popularized by \cite{hosking1990moments}. If $X_1, X_2, \ldots, X_n$ are $n$ independent copies of $X$ and $X_{1:n}\leq X_{2:n}\leq \ldots \leq X_{n:n}$ are their corresponding order statistics, then the $r$-order L-moment is defined as follows  
\begin{equation}
    L_r=r^{-1}\sum_{k=0}^{r-1}(-1)^k {r-1 \choose k} E(X_{r-k:r})\hspace{3mm} r=1,2,\ldots 
\end{equation}
The first order L-moment, $L_1 = EX_{1:1}=EX$, so it just coincides with the traditional first moment of $X$. The second order L-moment, $L_2 = 1/2(E(X_{2:2})-E(X_{1:2}))$, equals exactly a half of mean absolute difference (Gini-coefficient) and can be seen as a robust measure of scale. The third and fourth order L-moments, $L_3 = 1/3E(X_{3:3}-2X_{2:3}+X_{1:3})$ and $L_4 = 1/4E(X_{4:4}-3X_{3:4}+3X_{2:4}-X_{1:4})$, capture higher-order distributional properties and normalized by $L_2$ can be interpreted as robust counterparts of traditional higher-order moments such as skewness and kurtosis. Sample L-moments can be calculated using corresponding U-statistics. There are three main advantages of L-moments over traditional moments. First, all L-moments exist as long as $E(X) < \infty$.  Second, L-moments are unique and fully define the distribution. Third, L-moments are defined via linear combinations of order statistics, and are typically more robust compared to the traditional moments. 

In our approach, we leverage an alternative representation of L-moments as projections of quantile functions on shifted Legendre polynomial basis. Specifically, L-moments of order $r$ can be represented as
\begin{equation}
    L_r = \int_0^1 Q(p)P_{r-1}(p)dp, 
\end{equation}
where $P_r(p)$ is the shifted Legendre polynomials (LP) of degree $r$  defined as follows
\begin{equation}
    P_r(p) = \sum_{k=0}^r s_{r,k}p^r, \quad s_{r,k} = (-1)^{r-k}{r \choose k}{r+k \choose k} = \frac{(-1)^{r-k}(r+k)!}{(k!)^2(r-k)!}.
\end{equation}
Shifted Legendre polynomials have standard orthogonality properties on $[0,1]$ as 
\begin{equation}
\int_0^1 P_s(p)P_r(p)dp = \delta_{rs}(\frac{1}{2r+1}),
\end{equation}
where $\delta_{rs}=I(r=s)$. Thus, quantile function $Q(p)$ has the following decomposition
\begin{equation}
Q_k(p) =\sum_{r=1}^k(2r-1)P_{r-1}(p)\int_0^1 Q(p)P_{r-1}(p)dp = \sum_{r=1}^k (2r-1)L_rP_{r-1}(p)  \to Q(p),  k \to \infty \label{qrep}.
\end{equation}
Note that this approximation can be poor in the tails, if the distribution is heavy tailed \citep{hosking1990moments}. 
For a non negative random variable $X$, regular moments can be expressed as $\mu_k = EX^k=\mu_k(\bar{F}) = k\int_0^{\infty} x^{k-1}\bar{F}(x) dx=\mu_k(Q)= \int_0^1 Q^{k}(p)dp$, where $\bar{F}(x)=1-F(x)$, and $F(x)$ is the cumulative distribution function for $X$. 

Representation of L-moments via Legendre polynomial basis helps to see a geometrical intuition behind the existence of all L-moments, given a finite mean. Supplementary Figure S1 shows a comparison between $\mu_k(\bar{F})$, $\mu_k(Q)$ and $L_k(Q) = \int_0^1 Q(p)P_{k-1}(p)dp$ for a log-normal and a Beta distribution. 
Note that for the integral $\mu_k(\bar{F})$, $kx^{k-1}\bar{F}(x)$ can diverge over $x$. Similarly, for the integral $\mu_k(Q)$, $Q^k(p)$ can diverge over $p$. However, functions $Q(p)P_{k-1}(p)$ always lie between $Q(p)$ and $-Q(p)$, which guarantees the existence of all L-moments as long as $\int_0^1 Q(p)dp < \infty$.

\subsection{Scalar-on-quantile-function regression (SOQFR) using L-moments}
\label{lsofr}
In this section, we develop a scalar-on-quantile-function regression method using subject-specific L-moments (SOQFR-L). The SOQFR-L approach will provide the interpretation of SOQFR results in terms of the regression coefficients for L-moments. Since the shifted Legendre polynomials form an orthogonal basis of 
$L_2[0,1]$, we can approximate $\beta(p)$ in model (\ref{soqfr}) in terms of a truncated basis expansion as $\beta(p)=\sum_{k=1}^{K}\beta_{k}P_{k-1}(p)$. Thus, the SOQFR model (\ref{soqfr}) reduces to a standard GLM on the subject-specific L-moments as
\begin{equation}
g(\mu_i)=\alpha+\*Z_i^T\bm\gamma +\sum_{k=1}^{K}\beta_{k} \int_{0}^{1} Q_i(p)P_{k-1}(p)dp=\alpha+\*Z_i^T\bm\gamma + \sum_{k=1}^{K}\beta_{k}L_{ik}.\label{lmom:sofrmod}
\end{equation}
The unknown basis coefficients $\beta_{k}$ in this SOQFR-L representation capture a linear effect of the L-moments of the subject-specific distribution. Note that the first order L-moment, $L_1$, equals mean of the subject-specific distribution, therefore, model (\ref{lmom:sofrmod}) is more general than using the subject-specific mean.

SOQFR-L approach can be seen as somewhat analogous to functional principal component regression (fPCR) method \citep{reiss2017methods}, where fPC scores are used for supervised learning. L-moments are projections on orthogonal basis functions, and hence, can be interpreted in similar additive manner while additionally providing moment-based characterization of underlying distributions. Thus, the proposed approach allows mutually consistent interpretation of the results from the quantile-level perspective from SOQFR via $\beta(p)$ and from the L-moment level perspective (robust distributional summaries) from SOQFR-L via $\beta_k$'s.  

The number of L-moments $K$ to be retained (the number of basis functions for modelling $\beta(p)$) can be treated as a tuning parameter and can be chosen in a data-driven way using cross-validation, proportion of variance explained or information criteria such as AIC or BIC. The proportion of variance explained (PVE) by the first $k$ ($k=2,3 \ldots,$) L-moments of the subject specific quantile function $Q_i(p)$ can be defined as
$$\tau_k^2= 1- \frac{\int_{0}^{1} (Q_i(p)-Q_{i}^{k}(p))^2 dp}{\int_{0}^{1} (Q_i(p)-\mu_i)^2 dp},$$
where $Q_{i}^{k}(p)$ is the approximate quantile function based on first $k$ L-moments as in equation (\ref{qrep}). Note that, $Q_{i}^{1}(p)=L_{i1}=\mu_i$, hence $\tau_1^2=0$ and $\tau_k^2 (k\geq 2)$ represents the amount of variance in the subject specific quantile function explained by the first k subject-specific L-moments. In the applications of this article, we restrict our attention to the first 4 L-moments ($K=4$) to retain the interpretability of the models in terms of the first 4 distributional summaries (similar to the first 4 traditional moments typically used in most of applications). 

Nonlinear associations can be modelled using nonlinear extensions of SOFR \citep{reiss2017methods} that can be seen as a functional analogue of the single index model\citep{stoker1986consistent}. In particular, we can use this model
\begin{eqnarray}
E(Y_i|X_{i1},X_{i2},\ldots,X_{iJ})=\mu_i, \qquad
g(\mu_i)=\alpha+\*Z_i^T\bm\gamma + h\left (\int_{0}^{1} Q_i(p)\beta(p)dp \right ), \label{lmom:sim}
\end{eqnarray}
where $h(\cdot)$ is a smooth unknown function on real line. Expanding $\beta(p)$ via Legendre polynomial basis, we get a traditional single index model where the L-moments play a role of predictors as $ g(\mu_i) =\alpha+\*Z_i^T\bm\gamma + h(\*L_i^T\bm\beta)$.
Traditional estimation methods for the single index model \citep{wang2009spline,ichimura1991semiparametric} can be applied to estimate both $h(\cdot)$ and $\bm\beta$. Another alternative way to capture nonlinear association between the outcome and quantile functional predictors is to use a generalized additive model (GAM) with L-moments as $ g(\mu_i) = \alpha+\*Z_i^T\bm\gamma + \sum_{k=1}^{K}h_k(L_{ik}).$
The GAM approach using L-moments is analogous to the ``functional additive model'' of \cite{muller2008functional}, where fPC scores were used for scalar-on-function modelling and offers additional interpretability in terms of nonlinear effects of robust distributional summaries of data.

\subsection{Modelling multi-modal distributional data via Joint and Individual Variation Explained and L-moments}
\label{jive: method}
In this section, we demonstrate how L-moments can be used to identify joint and individual sources of variation in multi-modal distributional data.
Suppose, we have repeated measures data from multiple domains $d=1,2,\ldots,D$ each consisting of $R_d$ different features on the same subjects $i=1,2,\ldots,n$. Thus, we have subject-specific quantile functions $Q_{i}^{(d,r_d)}(p)$ for $r_d$-th feature within domain $d$ ($r_d=1,2,\ldots,R_d$). In many applications, it is important to identify joint and individual sources of variation in these multi-modal distributional data. For scalar data, \cite{lock2013joint} introduced joint and individual variation explained (JIVE) for integrative analysis of data coming from multiple domains. JIVE decomposes the original block data matrix into three parts, a low rank approximation capturing the joint structure and low rank approximations capturing domain-specific individual variation and noise. For multi-modal distributional data $Q_{i}^{(d,r_d)}(p)$, we propose to use L-moments to analyze joint and individual sources of variation. Specifically, let $L_{ik}^{(d,r_d)}=\int_0^1Q_{i}^{(d,r_d)}(p)P_{k-1}(p)$ be the $k$-th L-moment ($k=1,2,\ldots,K$) for $Q_{i}^{(d,r_d)}$. For each feature, we form the vector of L-moments and denoting it as $\*L_{i}^{(d,r_d)}$. Then, for each domain $d$, we get the following vector of L-moments
$\*L_{i}^{(d)}=[ \*L_{i}^{(d,1)^T},  \*L_{i}^{(d,2)^T},\ldots,\*L_{i}^{(d,R_d)^T}]^T$,
where $\*L_{i}^{(d)}$ is a $v_d=KR_d-$dimensional vector consisting of all L-moments for all features in domain $d$. Next, we apply JIVE decomposition of these L-moments vectors as $
  \*L_{i}^{(d)}=\*J_i^d +\*A_i^d+\bm\varepsilon_i^d=\^\Phi_J^d\bm\xi_{J,i}+\^\Phi_A^d\bm\xi_{A,i}^d+\bm\varepsilon_i^d$. 
 Here, $\*J_i^d$ and $\*A_i^d$ represent the low rank joint and individual structures with rank $s$ and $s_d$, respectively, and $\bm\varepsilon_i^d$ is the residual noise. Matrices of loadings for joint and individual structures are given by $\^\Phi_J^d\in\+R^{\nu_d\times S}$, $\^\Phi_A^d\in\+R^{\nu_d\times S_d}$, and $\bm\xi_{J,i}$, $\bm\xi_{A,i}^d$ are corresponding joint and individual scores for subject $i$. The summary of the proposed JIVE approach is presented as an Algorithm in Appendix 2 of Supplementary Material. The ranks $s$, $s_d$ can be chosen using BIC or permutation tests \citep{lock2013joint}. The number of L-moments $K$ can be chosen beforehand in a data-driven way. The joint and individual scores $\bm\xi_{J,i}$, $\bm\xi_{A,i}^d$ can further be used for supervised learning purposes. We use the {\tt r.jive} package \citep{rjive} in R \citep{Rsoft} for implementation of JIVE.

\section{DIGITAL GAIT BIOMARKERS IN ALZHEIMERS' DISEASE}

Accelerometry data for this study \citep{gait2020vr} were collected using a GT3x+ tri-axial  accelerometer in a sample of $86$ older participants including $38$ mild-AD and $48$ age-matched cognitively normal controls (CNC). Descriptive statistics on several baseline variables including age, sex, BMI, years of education, VO$_2$ max (maximum rate of oxygen consumption during a treadmill test) for the whole sample, and mild-AD and CNC groups are reported in Table S1 of Supplementary Material. Briefly, the sample had  $50\%$ female and average age of 73.2 years. There were no statistical differences between mild-AD and CNC group were found across age, BMI, and  VO$_2$ max. Compared to CNC group, mild-AD group had a significantly smaller percentage of females (26.3 vs 68.7) and lower education (15.6 years vs 17.4 years). 

The accelerometer was placed on the dominant hip of the participants via elastic belt. Activity was monitored continuously for seven days and, subsequently, gait parameters were obtained using a processing pipeline developed and validated in the Parkinson's Disease (PD) field \citep{weiss2014objective}. The pipeline outputs 52 gait metrics coming from 5 gait domains of Amplitude (8 metrics), Pace (3 metrics), Rhythm (13 metrics), Symmetry (9 metrics), and Variability (19 metrics). The complete list of gait metrics, along with their description and associated domains have been described in \cite{gait2020vr} and is given in Table S4 of Supplementary Material. Each gait metric is calculated every time a subject completes a sustained bout of walking of at least 60 sec; this provides multiple observations per subject across each of the seven wear days. Figure \ref{fig:fig1} reveals distributional nature of this data for a particular gait metric ``step velocity" for AD and CNC group.


\subsection{Discrimination of AD using SOQFR}
\label{sofr:res}
One of the primary objectives of our analysis is to explore how well the distributional representation of digital gait biomarkers can discriminate between mild-AD and CNC. To do that, we perform SOQFR and FGAM-QF with logit-link to model mild-AD vs CNC. We fit multiple models using each gait metrics separately and adjust for age and sex. For evaluation of the models, we use the ``deviance explained'' criterion in GAM  \citep{wood2017generalized}, which represents the proportion of null deviance explained by the respective models. We also report the average cross-validated area under the curve (AUC) of the receiver operating characteristic as an estimate for the out-of-sample prediction performance of the considered models. In particular, we perform a repeated 10-fold cross-validation (B=100 times) and report the average cross-validated AUC (cvAUC).

We use the \texttt{refund} \citep{refund} package within R \citep{Rsoft} for implementation of SOQFR and FGAM-QF. Table \ref{tab:my-table2} displays the top ten gait metrics ranked by the proportion of deviance explained in SQOFR and FGAM-QF. The variables in FGAM-QF consistently explain higher deviance which is expected. Particularly interesting are  ``Mean\_Stride\_Time\_\_s\_'' (mean stride time) and ``Mean\_Step\_Time\_s\_'' (mean step time). Used within FGAM-QF, they explain a much higher proportion of deviance ($0.63$) compared to SOQFR ($0.37$). This might be due to possible nonlinear effects of the quantile functions for these variables. The metrics ``Step\_Velocity\_\_cm\_sec\_" (step velocity), ``Distance\_\_m\_" (distance), ``Cadence\_V\_time\_domain\_" (cadence) perform more or less similarly using either SOQFR or FGAM-QF, indicating a linear effect. The average cvAUCs from SOQFR and FGAM-QF models illustrate an improved predictive performance compared to generalized linear models with the mean of the gait metrics (adjusted for age and sex), specifically for the measures of step velocity, cadence, distance, and mean stride time.  

\begin{figure}[H]
\begin{center}
\begin{tabular}{l}
 \includegraphics[width=.9\linewidth , height=.5\linewidth]{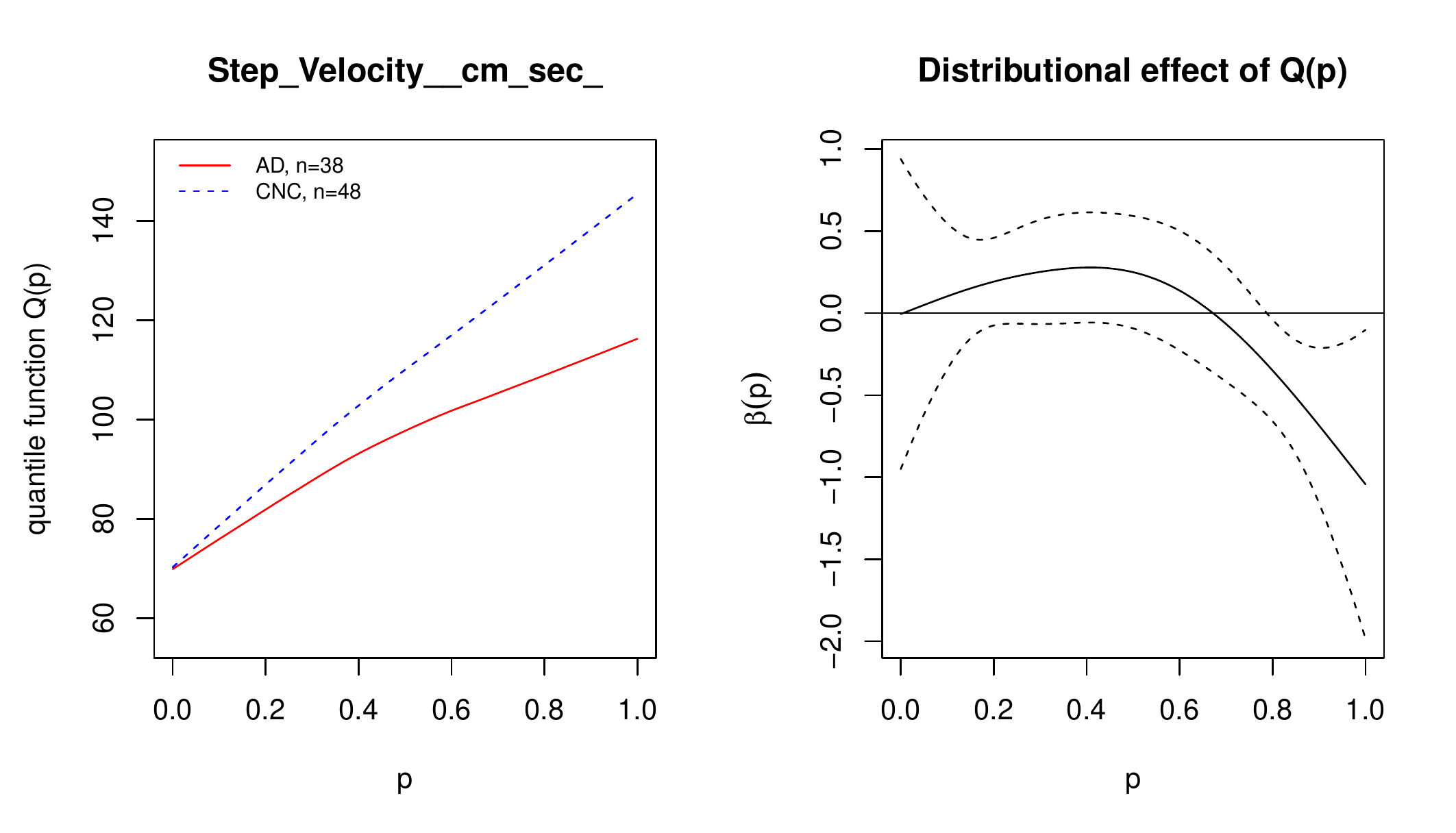} \\
 \includegraphics[width=.9\linewidth , height=.5\linewidth]{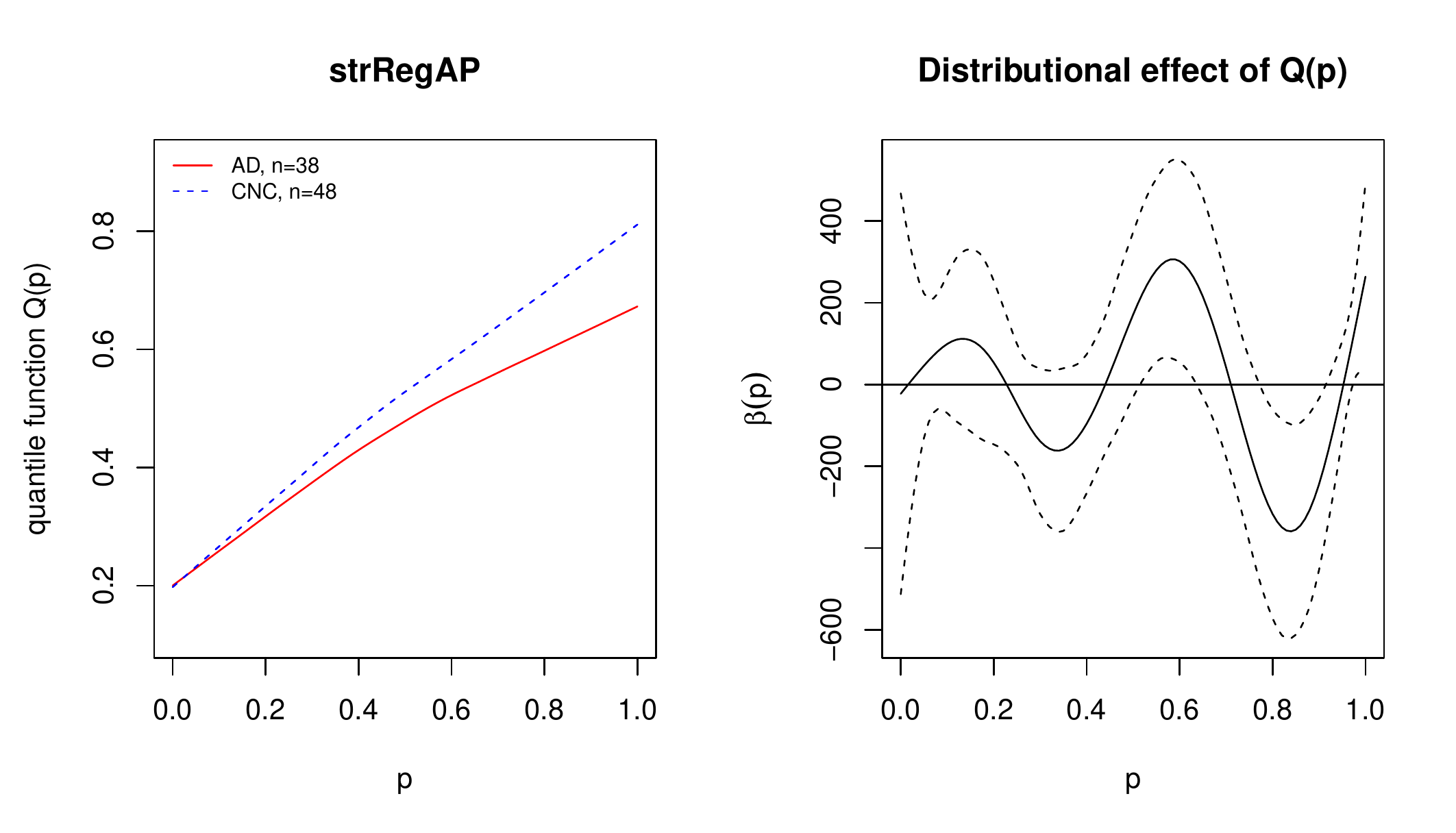}\\
\end{tabular}
\end{center}
\caption{Estimated linear functional effects $\beta(p)$ of quantile functions of step velocity (top right) and stride regularity (bottom right). The average quantile functions of the gait features for the AD (solid lines) and CNC group (dashed lines) are shown in the left column.}
\label{fig:fig2}
\end{figure}

Figure \ref{fig:fig2} displays the estimated functional effects for the top 2 metrics for SOQFR, ``strRegAP" (stride regularity) and step velocity, along with their average quantile functions. The 95$\%$ pointwise confidence intervals \citep{goldsmith2011penalized} for the estimated functional effects are also shown. We see a clear negative effect in the upper quantiles for both step velocity and stride regularity, indicating higher the maximal performance for these measures lower the odds of AD, which is very interesting from a clinical perspective. The additive quantile functional effects of the metrics mean stride time and step velocity obtained using FGAM-QF are displayed in Figure S2 of Supplementary Material along with the average quantile functions of AD and CNC groups. The sliced effect of the corresponding surfaces are shown in Supplementary Figure S3. For mean stride time, both tails can be informative. Interestingly, a higher maximal or minimal performance of this metric seems to be associated with higher odds of AD. There is a visible non-linearity in the upper tail of the estimated bivariate surface and the sliced effect $\hat{F}(q,p)$. FGAM-QF captures this non-linearity and therefore produces a superior performance for this metric in terms of deviance explained ($63\%$) compared to the SOQFR model ($37\%$).
The estimated bivariate surface of the quantile effect is more or less linear for step velocity and is highly negative in the upper tail (evident from the sliced effect). Hence high maximal performance in this metric is associated with lower odds of AD. Since the effect of the quantile function for this metric is linear, the performance and inference from FGAM-QF are similar to what we obtain from SOQFR.
\vspace*{- 10 mm}
\subsection{Comparison of SOQFR with histogram based modelling}

\begin{figure}[H]
\begin{center}
\begin{tabular}{l}
 \includegraphics[width=.75\linewidth , height=.4\linewidth]{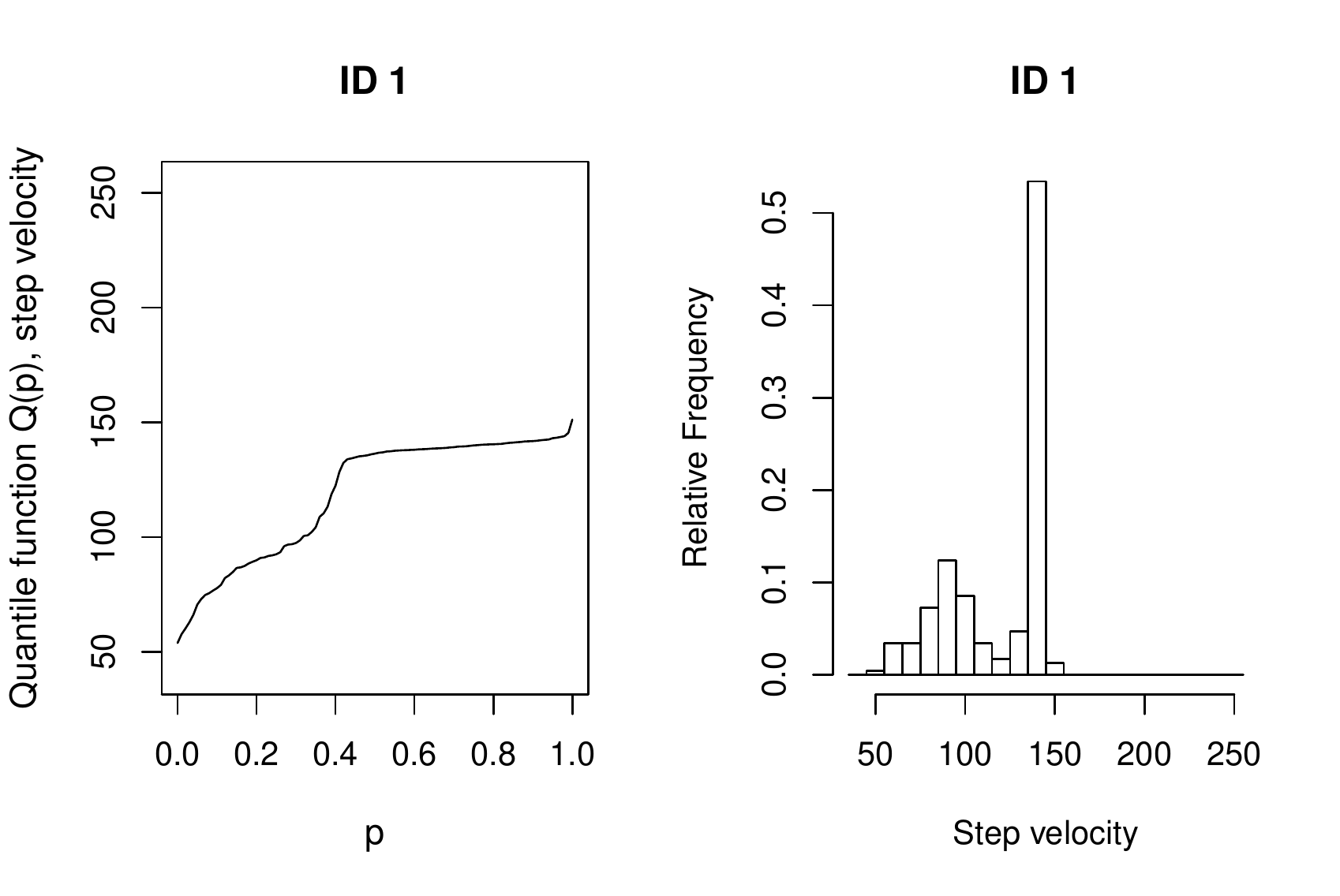} \\
 \includegraphics[width=.9\linewidth , height=.4\linewidth]{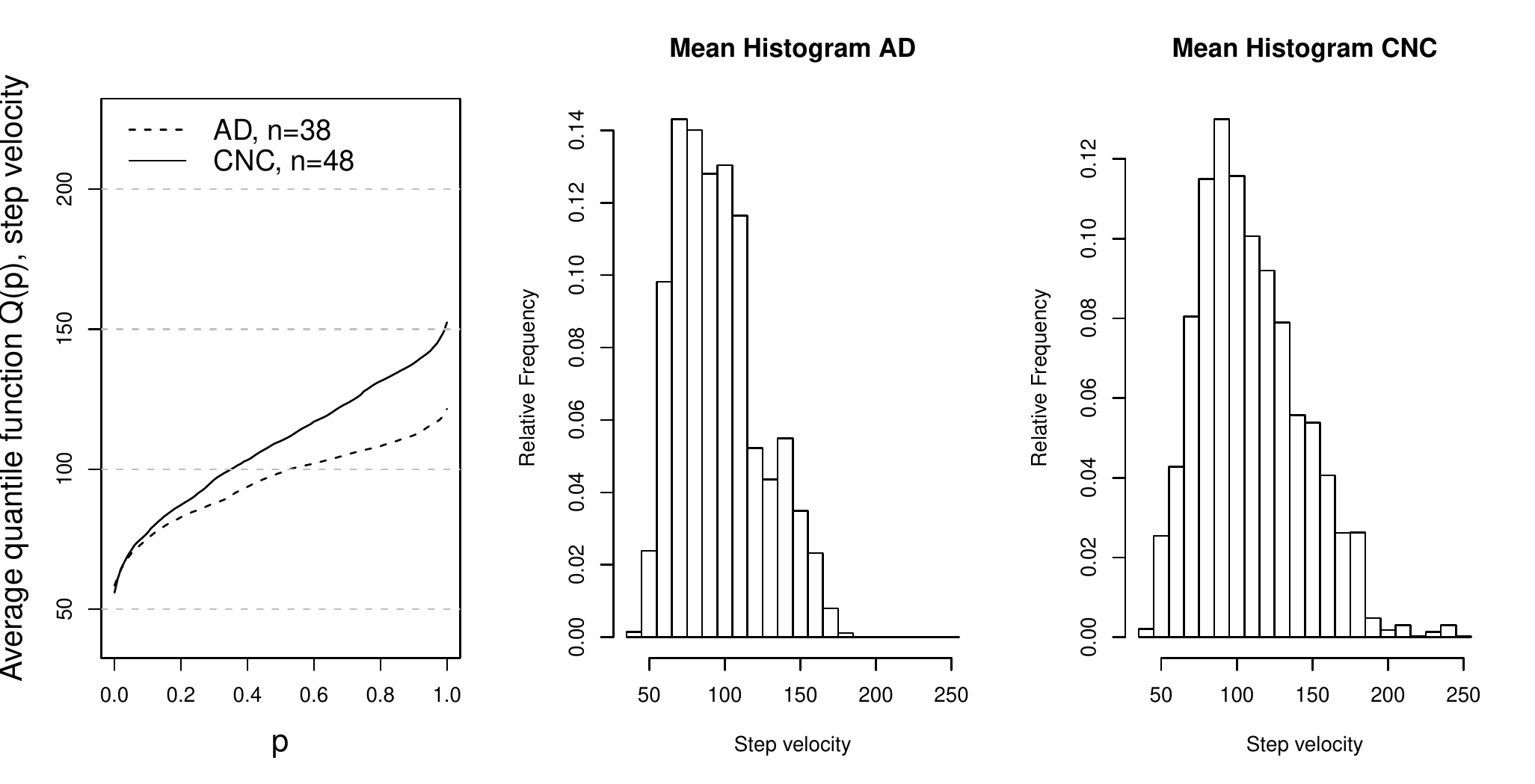}\\
 \includegraphics[width=0.8\linewidth , height=.45\linewidth]{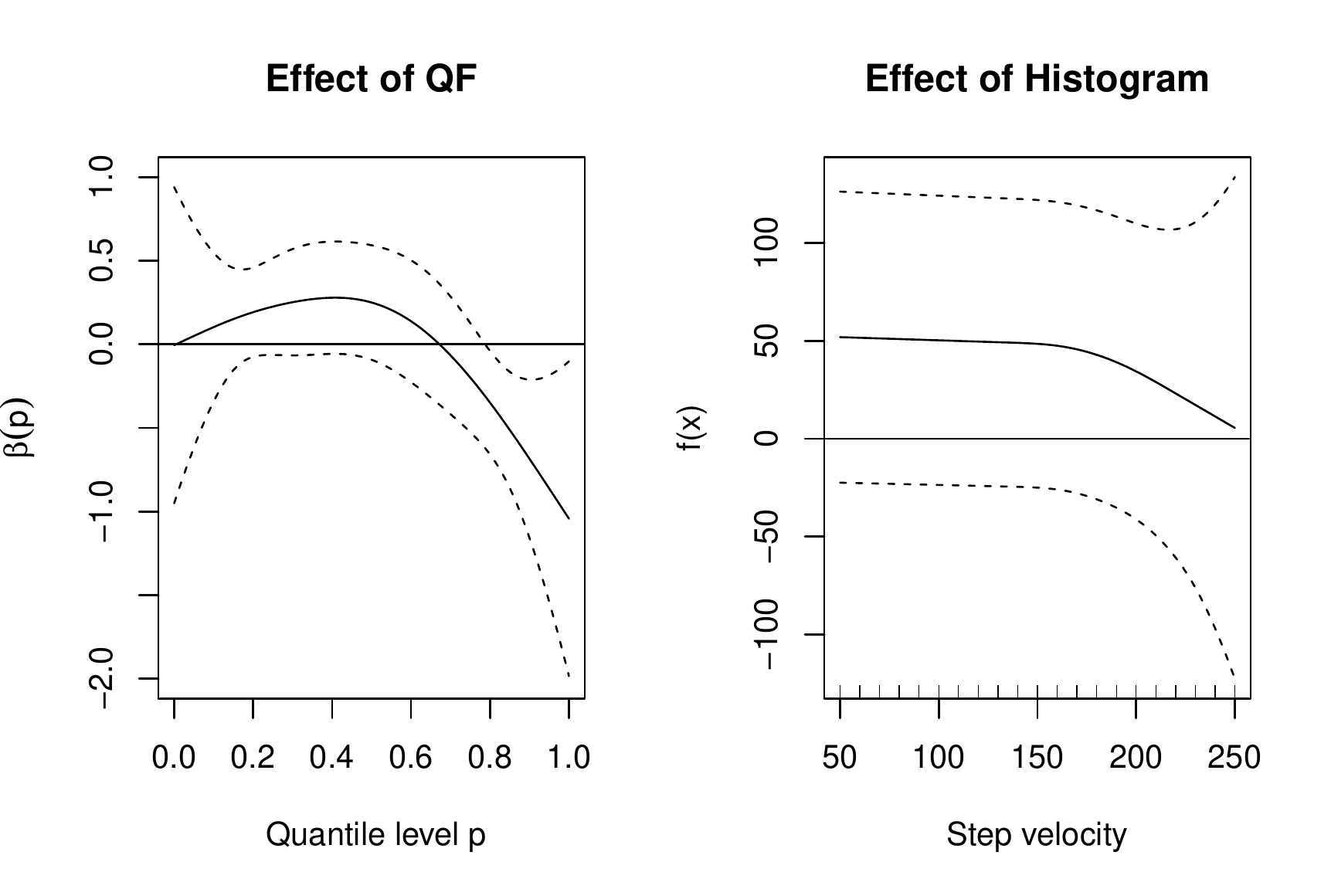}\\
\end{tabular}
\end{center}
\caption{Top panel: the subject-specific quantile function and histogram (relative frequency) of step velocity for subject ID=1. Middle panel: the group means of quantile functions (left) and histograms (right) of step velocity for AD and CNC groups. Bottom panel: estimated functional effect $\beta(p)$ for quantile functions of step velocity (left) and estimated effect $f(x)$ of subject-specific histogram (right).}
\label{fig:fig4hist}
\end{figure}

We compare the proposed SOQFR method with the histogram-based approach by \cite{augustin2017modelling}. In particular, we focus on step velocity and obtain subject-specific histograms (relative frequency) in 22 bins of equal width (10) between step velocity values of 35 and 255 (cm/second). Top panel of Figure \ref{fig:fig4hist} shows both  the quantile function and histogram of step velocity for a random subject from our study. The group average of quantile functions and the group average of histogram (relative frequency) for AD and CNC groups are shown in the middle panel. It is worth noting that the average of quantile functions is a barycenter and is well-defined in terms of 2-Wassterstein distince \citep{panaretos2020invitation}, while the average of histograms is based on Euclidean distance, not well-defined and included here for illustrative purposes. Visually, the mean histograms are not directly interpretable compared to the averages of quantile functions, which nicely captures the divergence between the two groups in terms of maximal levels of step velocity. Following \cite{augustin2017modelling}, we fit a GLM model for the binary outcome of cognitive status (mild-AD vs CNC) with  subject-specific histograms $H_i(x_j)$ of step velocity as  functional predictors and adjust the model for age and sex. Specifically, the model is as follows: $g(\mu_i)=\alpha+\*Z_i^T\bm\gamma + \sum_j H_i(x_j)f_x(x_j).$
Here $H_i(x_j)$ is the subject-specific histogram (relative frequency) of step-velocity of subject $i$ with some given number at mid-point $x_j$ and $f_x(x_j)$ captures the smooth effect of the subject-specific relative frequency at $x_j$. We compare the estimates and the performance of this model to the SOQFR model for step velocity. The proportion of deviance explained of the above histogram-based model is  $0.30$ compared to $0.50$ for SOQFR. This illustrates higher explanatory power of the subject-specific quantile function $Q_i(p)$ compared to the histogram-based representation of step velocity. The estimated smooth effect $f_x(x)$ and linear functional effect $\beta(p)$ of step velocity from SOQFR is shown in the bottom panel of Figure \ref{fig:fig4hist}. The $95\%$ credible intervals
from GAM for $f_x(x)$ \citep{wood2017generalized} include zero at all values of step velocity and hence subject specific representation of step-velocity in terms of histogram (relative frequency) is not statistically significant. On the contrary, the functional regression coefficient  $\beta(p)$ for SOQFR is statistically significant for quantile levels of $p>0.8$. This illustrates that a higher maximal level of step velocity is associated with a reduced odds of AD.
Further, we compare the predictive performance of the two approaches in terms of cross-validated area under the curve (AUC) of the receiver operating characteristic. Specifically, we perform a repeated 10-fold cross-validation (B=100 times) and report the average cross-validated AUC (cvAUC). cvAUC of the proposed SOQFR method is calculated to be $0.89$, which is much higher than cvAUC of $0.79$ for the histogram-based approach. This illustrates a higher predictive (discriminatory) power of the quantile function distributional representation of step-velocity in our application.

\vspace*{- 10 mm}
\subsection{Discrimination of AD using SOQFR-L and comparison with SOQFR}
\label{lmom:res}
In this section, we apply SOQFR-L to stride regularity, step velocity, and cadence, top-performing metrics in SOQFR. Supplementary Figure S4 shows PVE in subject-specific quantile functions of stride regularity, step velocity, and cadence. We observe that the first four L-moments, on average, explain around $92-95\%$ of the variance in those three gait measures. Compared to the traditional or central moments, L-moments are orthogonal projections on the shifted Legendre polynomials, hence they tend to encode less correlated distributional properties.  Supplementary Figure S5 shows the heatmap of correlations among the first four L-moments, traditional moments and central moments for cadence. As can be noticed, the traditional and central moments are highly positively correlated among themselves, which is not the case for L-moments. 

We fit three separate logit-link SOQFR-L models for each of the three gait metrics. The results are reported in Table \ref{tab:my-table3}. Importantly, it is not the mean (equal to $L_1$), but the second-order ($L_2$) or third-order ($L_3$) L-moments which have significant effects on odds of AD across all three gait metrics. Considering predictive performance, model A2 with step velocity is found to be the best among the three models considered, in terms of the highest proportion deviance explained ($48.75\%$). Stride regularity explains a much lower proportion of deviance ($21.58\%$) in SOQFR-L compared to SOQFR ($58\%$), which indicates there might be higher order L-moments for stride regularity which are important to consider. Upon further exploration of the first eight L-moments of stride regularity, we do find the L-moments of orders 5, 6 and 8 to be significant (results are not included in the Table) and this also improves the model performance (deviance explained=$55.81\%$ using $L_2,L_5,L_6,L_8$). In terms of out-of-sample prediction, the reported cvAUC clearly demonstrates an improved performance of SOQFR-L compared to GLM on mean of the gait measures. 

Supplementary Figure S6 compares functional regression coefficients  $\beta(p)$ estimated using SOQRFR (right column) and using SOQFR-L (left column). The coefficient functions $\beta(p)$ obtained with traditional SOQFR and SOQFR-L are very similar for step velocity and cadence (using $K=4$ L-moments). For stride regularity, we see that $\beta(p)$ is more ``complex'' and we used $K=8$ L-moments to get a better approximation. Thus, in addition to providing quantile interpretation via $\beta(p)$, SOQFR-L also provides interpretability in terms of significance of the specific L-moments. This provides more flexibility and interpretability in many applications. 

For comparison, we provide results from logit-link GLM analyses using the first four traditional and first four central moments in Supplementary Table S2 and S3, respectively. The effect of the gait measures are much weaker and not significant (at nominal level $\alpha=0.05$) for traditional moments, possibly due to high correlation among themselves. Central moments perform similarly to L-moments. However, they do not provide a consistent way to additionally interpret the effect of quantile levels as it can be done in SOQFR and SOQFR-L.
\vspace*{- 6 mm}
\subsection*{Modelling cognitive scores with SOQFR-L}
In addition to the cognitive status, this study used confirmatory factor analysis (CFA) to derive cognitive scores for attention (ATTN), verbal memory (VM) and executive function (EF), which represent a continuous scale of cognitive functioning. We study association between the L-moments of stride regularity, step velocity and cadence with the cognitive scores of ATTN, VM and EF using SOQFR-L (with identity link function), while adjusting for age, sex and education. The results 
are displayed in Table \ref{tab:my-table44}. We observe the cognitive scores of ATTN, VM and EF to be associated with sex, education and primarily the second ($L_2$) order L-moments. For all the measures considered, this association is found to be positive, indicating higher $L_2$ moments are associated with higher cognitive scores and lower odds of dementia, which matches with our earlier analysis. The signal for the cognitive scores of VM and EF are found be much stronger (adjusted R-squared $35\%-40\%$) compared to ATTN (adjusted R-squared $19\%-24\%$). All three gait measures provide more or less similar predictive performance and significant gains compared to a benchmark model on age, sex and education and competing models using only the mean of the gait metrics (around $30\%-40\%$ improvement) in terms of the adjusted R-squared criterion. We also report cross-validated R-squared of the models from repeated (B=100) 10-fold cross-validation for each of the model in Table \ref{tab:my-table44}. An improved performance of the models using SOQFR-L can be observed compared to the competing models. Additional results from GAM using the L-moments of the gait measures discussed in Section \ref{lmom:res} and are reported in Supplementary Table S5. The effect of the second order L-moment is found to be most significant for the gait measures in all the models considered.


\subsection{JIVE with L-moments}
\begin{figure}[ht]
\begin{center}
\begin{tabular}{ll}
\includegraphics[width=.4\linewidth , height=.4\linewidth]{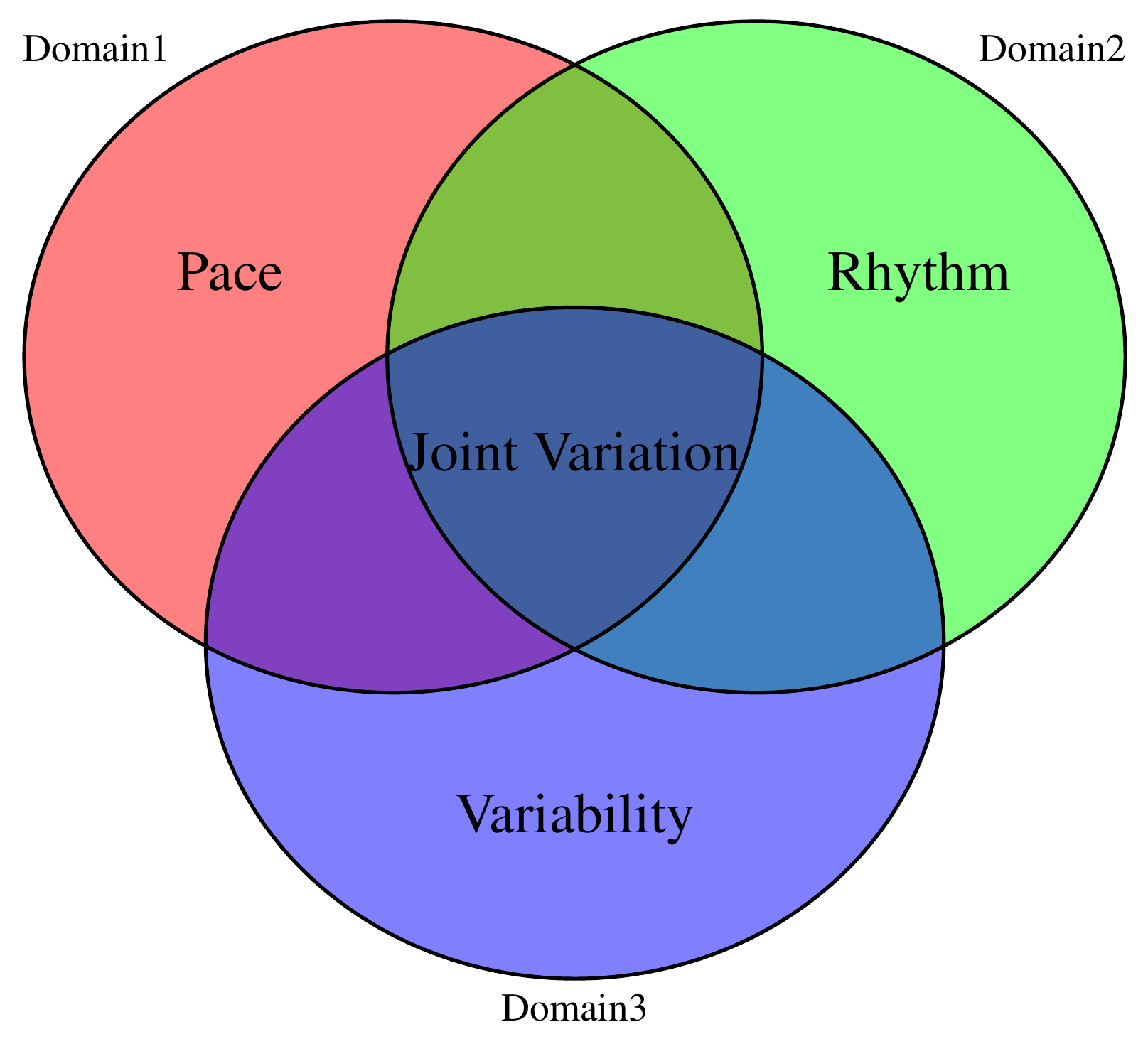} &
\includegraphics[width=.5\linewidth , height=.4\linewidth]{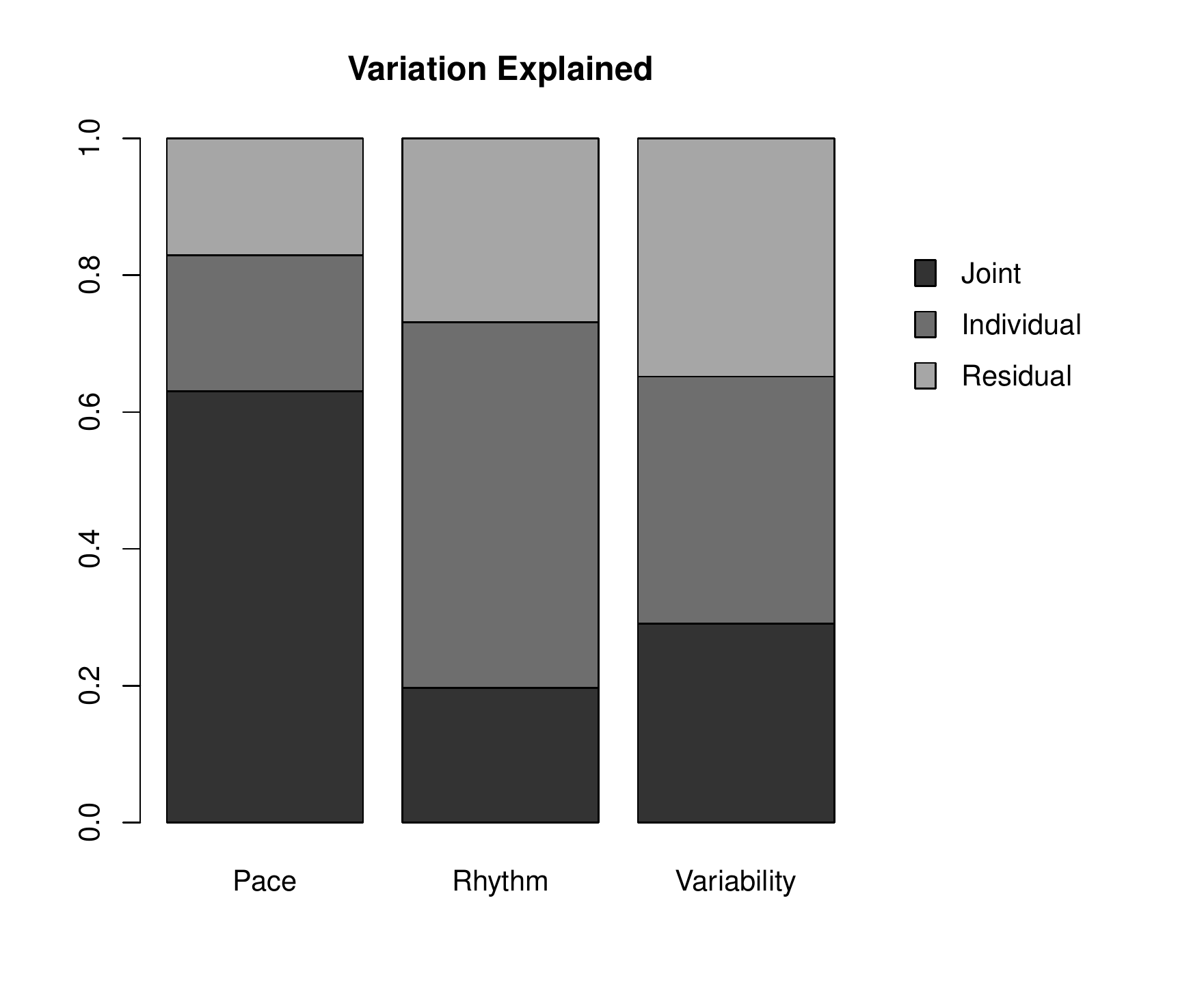}\\
\end{tabular}
\end{center}
\caption{Left Panel: Venn diagram highlighting two main sources of variation: joint (the shaded intersection region) and domain-specific (individual part of each circle). Right Panel: Joint and individual variation explained by each domain from JIVE.}
\label{fig:fig44}
\end{figure}

So far in our analysis, we have considered each gait measure separately while performing SOQFR or SOQFR-L to study their association with cognitive performance. However, gait measures are correlated among themselves and conceptually can be placed within one of the five unique gait domains mentioned earlier. The inter-dependence of the metrics within and between the domains can be beneficial in statistical modelling and can be analyzed using JIVE approach with L-moments illustrated in Section \ref{jive: method}. We focus on domains of Pace (3 features), Rhythm (13 features) and Variability (19 features) as the top performing gait metrics in the SOQFR analysis (see Table \ref{tab:my-table2}) belong to either of these three domains.
Figure \ref{fig:fig44} (left panel) displays a Venn diagram illustrating a conceptual overlap of joint and individual variation across the three domains.

First, we pre-normalize all the variables (subject-specific L-moments in our case)  via z-score transformation $z_i = \Phi^{-1}(\hat{F}_n(x_i))$, where $x_i$s are the original features and $\hat{F}_n$ is the empirical c.d.f of $\{x_i\}_{i=1}^{n}$. Further, all the data blocks are again normalized (centered and scaled) so that data blocks from different domains are comparable as suggested by \cite{lock2013joint}. Applying JIVE \citep{rjive} and determining the optimal ranks via the permutation test, we get the following ranks: (Joint, Pace, Rhythm, Variability) = $(2,2,7,9)$.
Figure \ref{fig:fig44} (right panel) displays the amount of variation explained by joint and individual components in each of the three domains. Note that JIVE results in significant dimension reduction -  reducing the dimension from $140$ (4 L-moments from each of 35 gait metrics) to just $20$. JIVE estimates of the joint and individual structures are shown in Supplementary Figure S7.

We use the JIVE scores, orthogonal by construction, to study the association with cognitive status (mild-AD). Since, we only have a relatively small sample ($n=86$), we further perform variable selection and identify the important JIVE scores using LASSO \citep{tibshirani1996regression}. The selected JIVE scores (joint-PC1, joint-PC2, Pace-PC1, Pace-PC2, Rhythm-PC1, Rhythm-PC2, Rhythm-PC5, Rhythm-PC7, Variability-PC4, Variability-PC7 and Variability-PC9) are used in modelling cognitive status via logistic regression, while adjusting for age and sex. The results are reported in Supplementary Table S6. 

\begin{figure}[H]
\hspace*{- 7 mm}
\begin{tabular}{ll}
 \scalebox{0.45}{\includegraphics{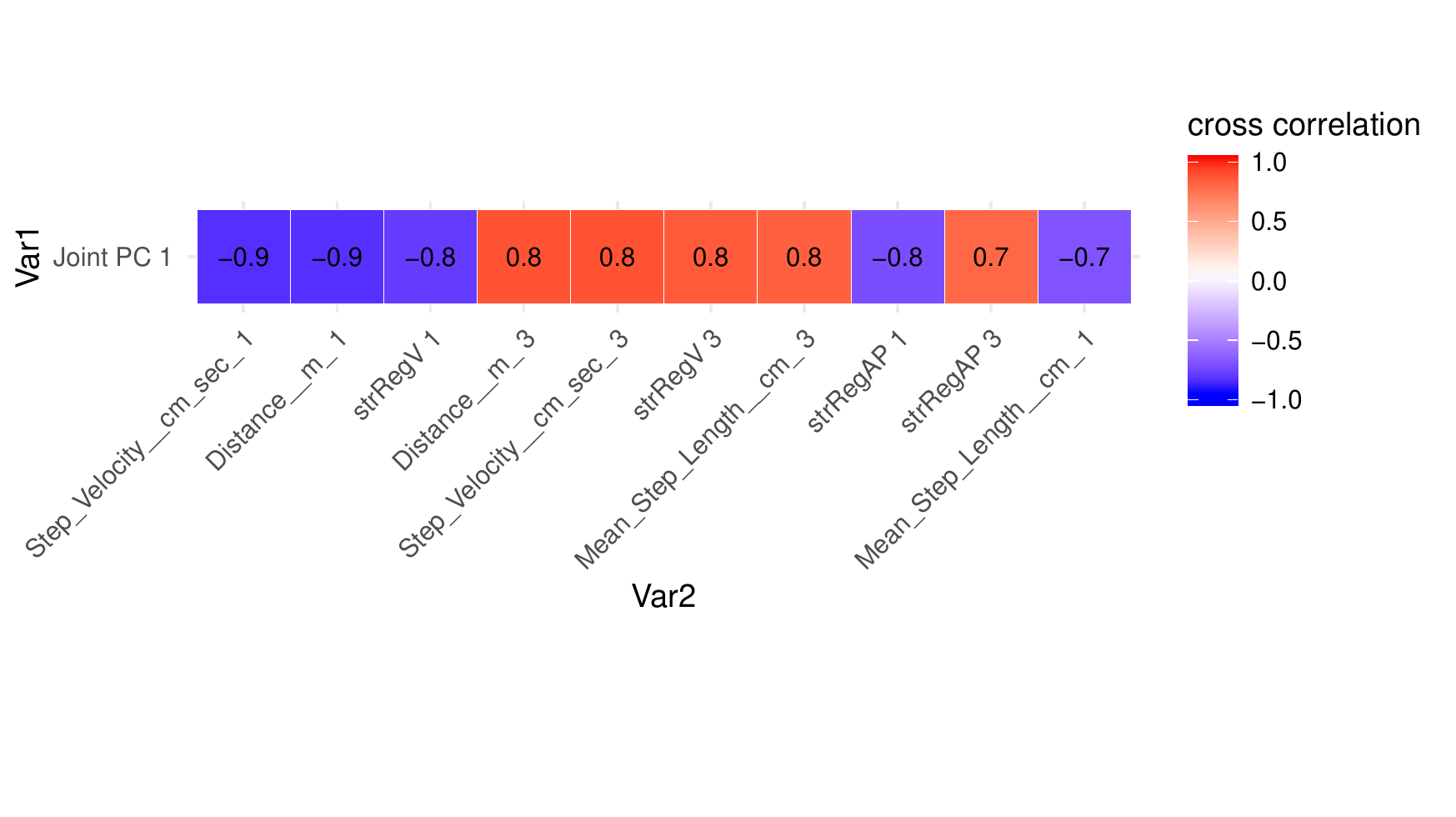}} &
 \scalebox{0.45}{\includegraphics{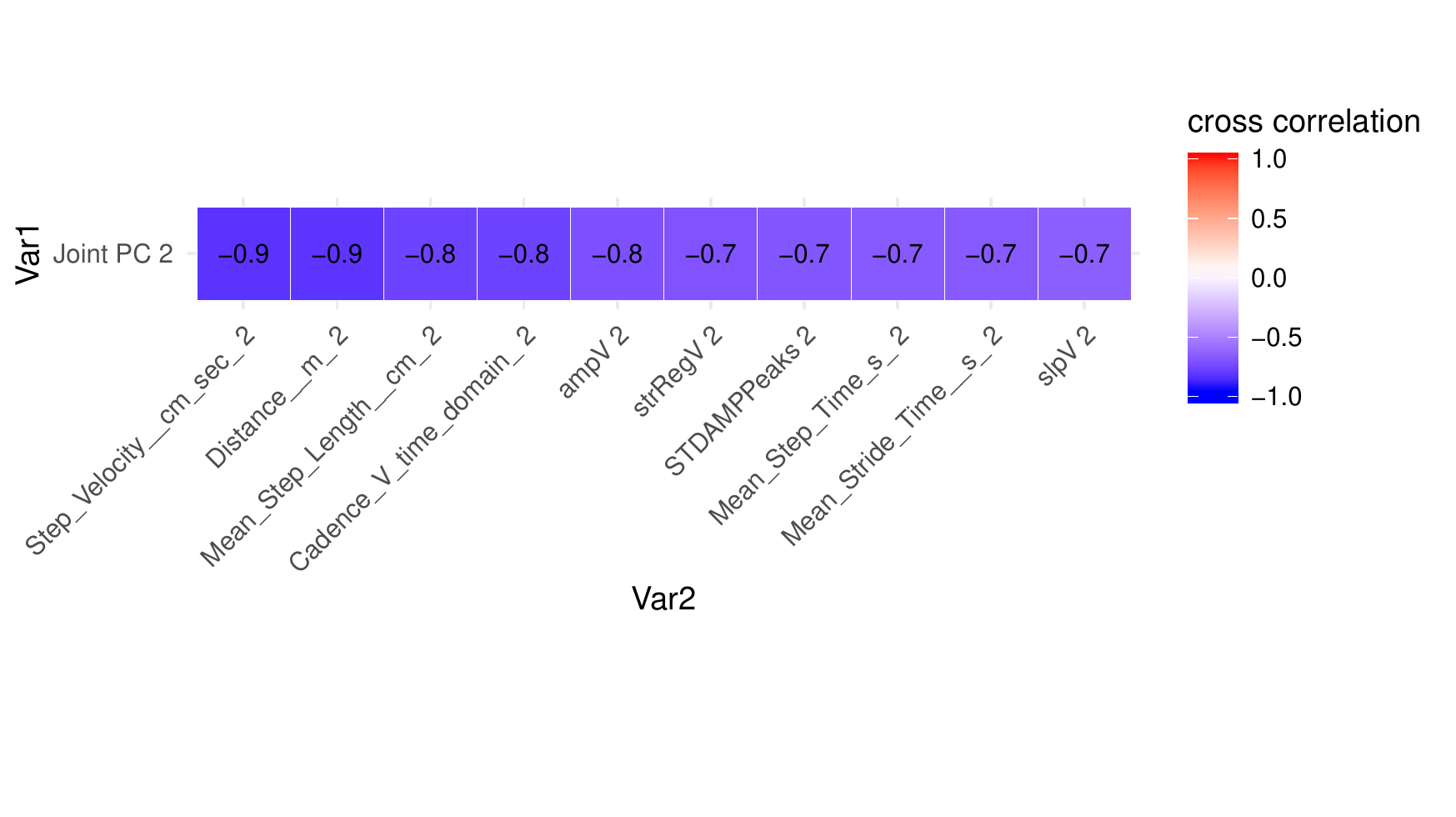}}\\
 \scalebox{0.45}{\includegraphics{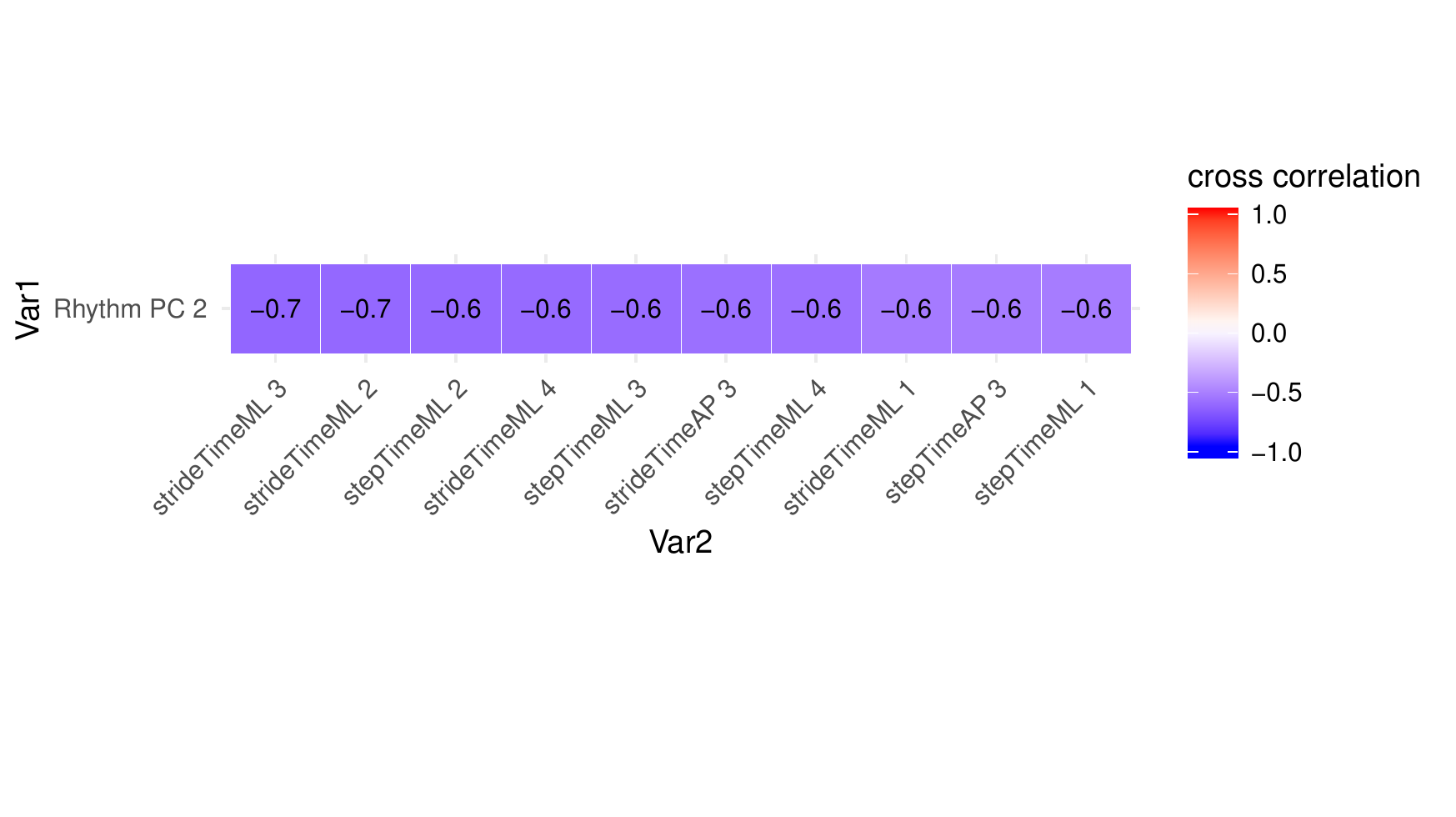}} &
 \scalebox{0.45}{\includegraphics{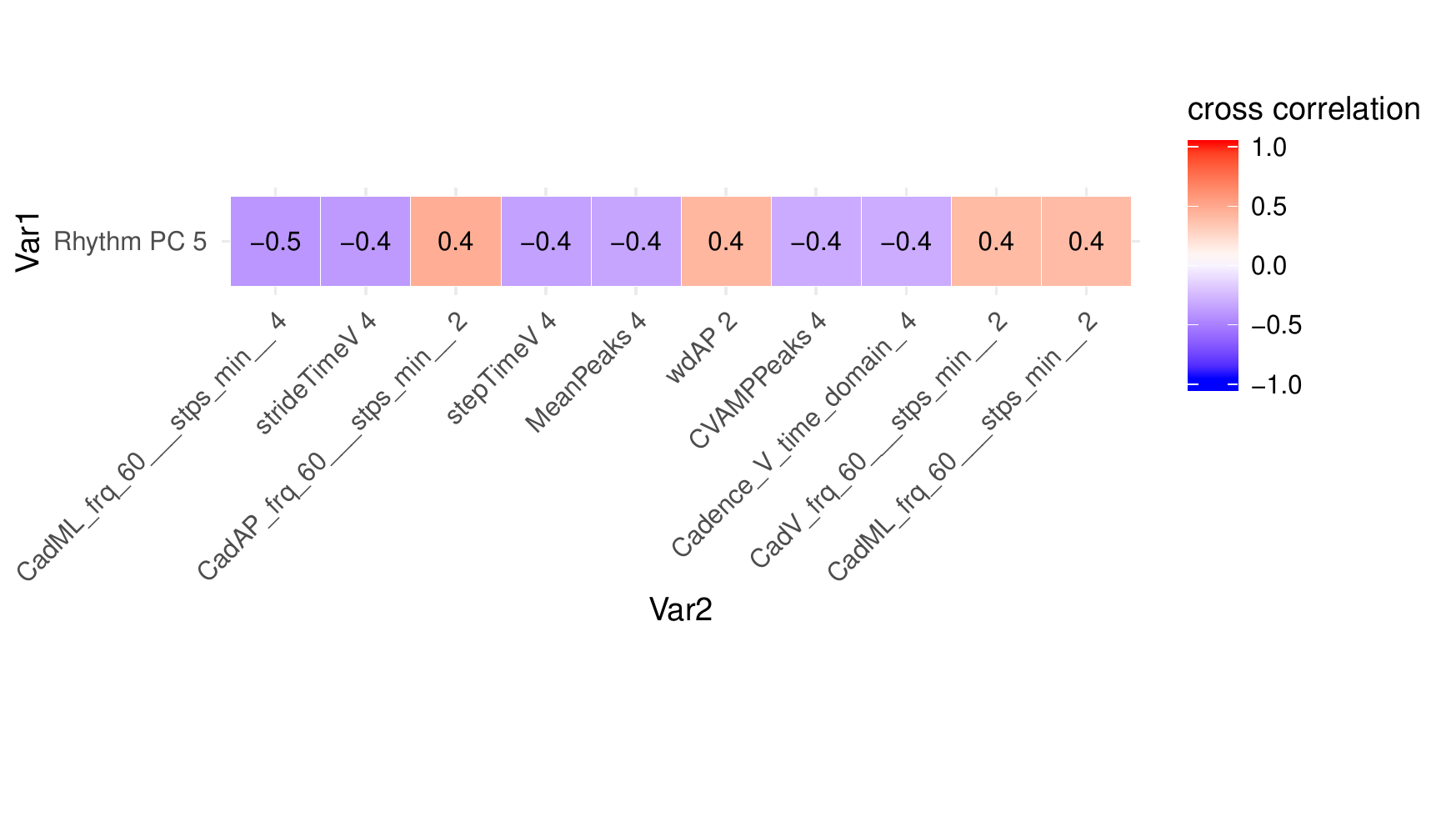}}\\
 \scalebox{0.45}{\includegraphics{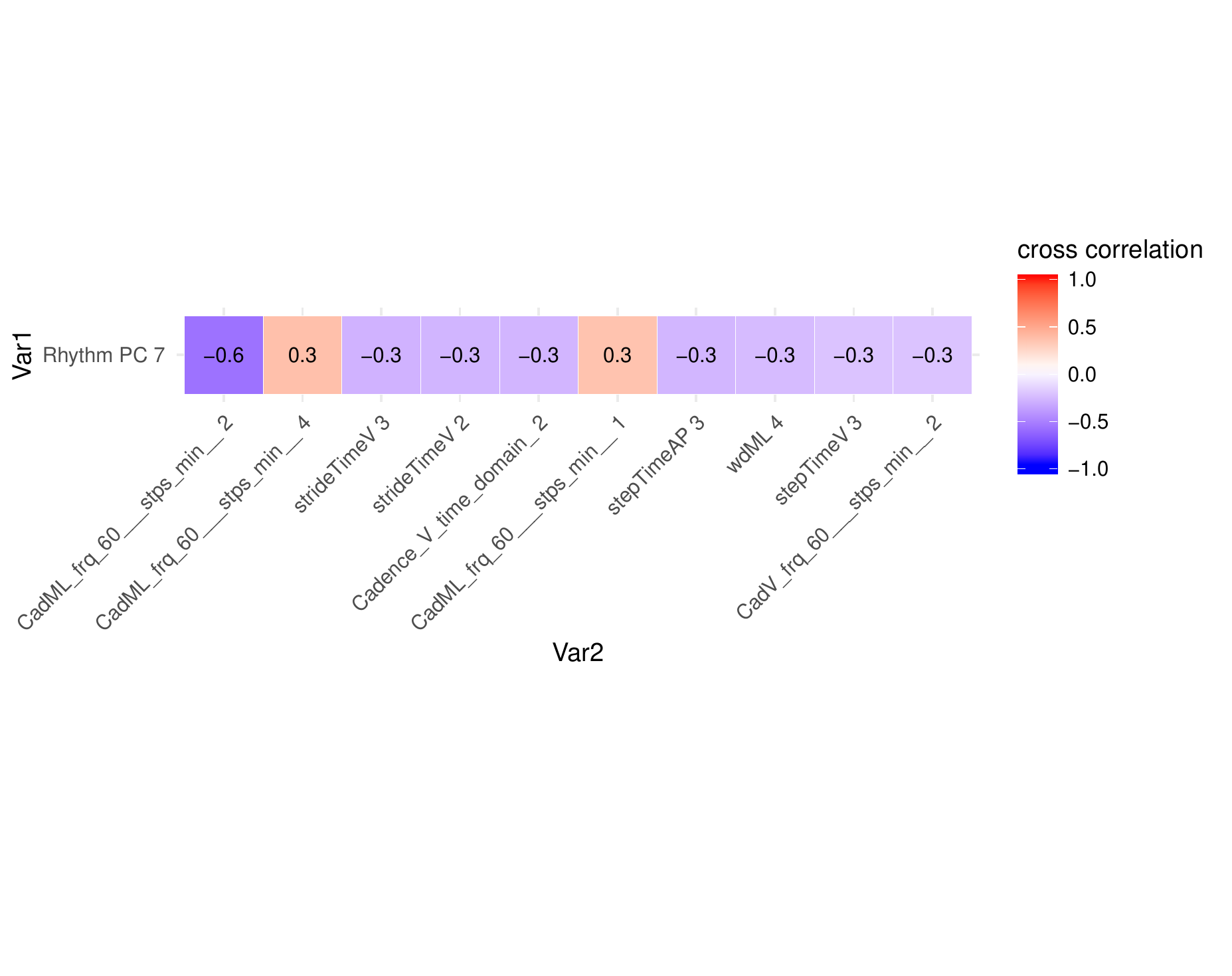}} &
                                           \\
\end{tabular}
\caption{Cross correlation between JIVE PC scores and L-moments. Shown are the top 10 L-moments ranked according to absolute value of correlation with each PC score. ``Gait measure $r$" represents $(r)$ th L-moment of the particular gait metric.}
\label{fig:fig7}
\end{figure}

To further understand the associations of JIVE PC scores with cognitive function relate them back to the original gait metrics, we use the cross-correlation between the original L-moments and the significant JIVE scores. The top 10 gait metrics (L-moments) ranked according to their correlation with each score are displayed in Figure \ref{fig:fig7}. Joint-PC1, joint-PC2 are found to be positively associated with higher odds of AD. For joint PC-1, the first order L-moments ($L_1$) from Pace (step velocity, distance) and Rhythm (stride regularity) are negatively loaded, indicating higher mean value for these variables e.g., step velocity lowers the odds of AD. Similarly, for joint PC-2, the second order L-moments ($L_2$) from Pace (step velocity, distance, mean step length) and Rhythm (cadence) are negatively loaded, indicating higher second order L-moments (representing scale) for these variables are associated with a lower risk of AD which matches with our analysis in Section \ref{lmom:res}. The association between cognitive status and Rhythm-PC2, Rhythm-PC5 are found to be negative, whereas Rhythm-PC7 is found to be positively associated. Primarily, higher order L-moments ($L_2,L_3,L_4$) gait metrics from Rhythm and Variability are loaded on these individual PCs outlining the importance of these domains. 
\vspace* {-6 mm}

\section{Discussion}
 Distributional data analysis is an emerging area of research in digital medicine that has a large number of diverse applications.
  There are many ways to represent distributional information including cumulative distribution function, density function, quantile function, and others. Although, all these distributional representations can be re-expressed through each other via differential, integral, inverse, or other more involved transformation, a specific choice for statistical modelling may depend on desirable interpretation and require different analytical machinery. We have proposed to capture distributional nature of wearable data via subject-specific quantile functions and use them or their L-moment representations in SOFR, FGAM, and JIVE methods. As we argue, our approach provide many advantages including intuitive interpretation of results in terms of both quantile levels and L-moments as well as uniform support on $[0,1]$.
 Additional motivation for the proposed quantile-function approach stems from multiple research efforts aimed at discovering digital biomarkers based on distributional properties of wearable data. Stride-velocity-q95, a 95-percentile of accelerometry-estimated stride velocity, is one of the very first digital biomarkers approved as a secondary endpoint by the European Medicines Agency (EMA), a European counterpart of the Federal Drug Administration in the US \citep{haberkamp2019european}. This measure has been demonstrated to be much more sensitive for tracking longitudinal decline in children with Duchenne muscular distrophy compared to the mean stride-velocity. Another example comes from accelerometry-estimated gait, which is often quantified via user-specific averages, measures of variability (such as standard deviation), and asymmetry (such as skewness) of gait parameters \citep{hausdorff2018everyday,shema2020wearable} estimated over a wear-period. Thus, wearable applications actively employ various distributional or quantile based user-specific summaries.
 
 We have demonstrated how proposed approaches provide deeper understanding of the associations between digital gait biomarkers and cognitive functioning in Alzheimer’s disease. Specifically, we showed that quantile functions of gait metrics including step velocity, distance, cadence, stride regularity, mean stride time and mean step time provide higher discrimination between mild-AD and non-AD (CNC) disease status. We also found that second order L-moments capturing subject-specific variability of a few gait parameters are significantly associated with cognitive domains of attention, verbal memory, and executive function. With continuous monitoring of patients with accelerometers and gyroscopes, the method proposed in this manuscript can be adapted to monitor the functional status in Alzheimer’s disease. 

These are many more areas that remain to be explored based on the current work. We chose the number of L-moments using PVE criteria. However, similarly to FPCR, it may impose constraints on the flexibility/complexity of estimated $\beta(p)$ and strategies similar to functional penalized regression \citep{goldsmith2011penalized} could be explored. In the application of this paper, we have considered one distributional predictor at a time to study associations with the particular outcome of interest. It is plausible to consider several distributional predictors together or perform variable selection within SOQFR \citep{gertheiss2013variable} for a better prediction/classification achievement. It would be also of interest to conceptualize and accommodate distribution-level interactions. One way to do this would be via the inner product of quantile functions expressed in terms of the interactions of corresponding L-moments. Another possible direction of work is to extend the SOQFR model to capture local temporal effects (e.g. for physical activity data) which depends on the time of the day. Beyond distributional aspect of wearable data, there are many other aspects that may be informative. For example, the time of day, the time since the last walking bout, the walking bout duration, and others could be possibly informative for modelling outcomes of interest. Finally, normalization of scales across multiple distributional predictors will need to be deeply studied. For example, quantile function $Q_i(p)$ could be normalized within each subject using the transformation $(Q_i(p)-Median_i)/IQR_i$ to make scales more comparable across subjects and variables.
\section{Software}
\label{sec5}
Illustration of the proposed framework via R \citep{Rsoft}, along with the dataset analyzed, is available on Github at \url{https://github.com/rahulfrodo/DDA}.

\section{Supplementary Material}
\label{sec6}
Supplementary material is available online at
\url{http://biostatistics.oxfordjournals.org}.
\newpage



\singlespacing
\bibliographystyle{biorefs}
\bibliography{refs}

\begin{thebibliography}{99}

\bibitem[Aitchison(1982)Aitchison]{aitchison1982statistical}
\textsc{Aitchison, John}. (1982).
\newblock The statistical analysis of compositional data.
\newblock {\em Journal of the Royal Statistical Society: Series B
  (Methodological)\/}~\textbf{44}(2), 139--160.

\bibitem[{Alzheimer's Association}(2020){Alzheimer's
  Association}]{addoi:10.1002/alz.12068}
\textsc{{Alzheimer's Association}}. (2020).
\newblock 2020 alzheimer's disease facts and figures.
\newblock {\em Alzheimer's \& Dementia\/}~\textbf{16}(3), 391--460.

\bibitem[Augustin \emph{and others}(2017)Augustin, Mattocks, Faraway, Greven
  and Ness]{augustin2017modelling}
\textsc{Augustin, Nicole~H, Mattocks, Calum, Faraway, Julian~J, Greven, Sonja
  and Ness, Andy~R}. (2017).
\newblock Modelling a response as a function of high-frequency count data: The
  association between physical activity and fat mass.
\newblock {\em Statistical methods in medical research\/}~\textbf{26}(5),
  2210--2226.

\bibitem[Bakrania \emph{and others}(2017)Bakrania, Yates, Edwardson, Bodicoat,
  Esliger, Gill, Kazi, Velayudhan, Sinclair and
  Sattar]{bakrania2017associations}
\textsc{Bakrania, Kishan, Yates, Thomas, Edwardson, Charlotte~L, Bodicoat,
  Danielle~H, Esliger, Dale~W, Gill, Jason~MR, Kazi, Aadil, Velayudhan, Latha,
  Sinclair, Alan~J and Sattar, Naveed}. (2017).
\newblock Associations of moderate-to-vigorous-intensity physical activity and
  body mass index with glycated haemoglobin within the general population: a
  cross-sectional analysis of the 2008 health survey for england.
\newblock {\em BMJ Open\/}~\textbf{7}(4), e014456.

\bibitem[Bigot \emph{and others}(2018)Bigot, Gouet, Klein and
  Lopez]{bigot2018upper}
\textsc{Bigot, J{\'e}r{\'e}mie, Gouet, Ra{\'u}l, Klein, Thierry and Lopez,
  Alfredo}. (2018).
\newblock Upper and lower risk bounds for estimating the wasserstein barycenter
  of random measures on the real line.
\newblock {\em Electronic Journal of Statistics\/}~\textbf{12}(2), 2253--2289.

\bibitem[Chen \emph{and others}(2021)Chen, Lin and
  M{\"u}ller]{chen2021wasserstein}
\textsc{Chen, Yaqing, Lin, Zhenhua and M{\"u}ller, Hans-Georg}. (2021).
\newblock Wasserstein regression.
\newblock {\em Journal of the American Statistical
  Association\/}~(just-accepted), 1--40.

\bibitem[Dryden and Mardia(2016)Dryden and Mardia]{dryden2016statistical}
\textsc{Dryden, Ian~L and Mardia, Kanti~V}. (2016).
\newblock {\em Statistical shape analysis: with applications in R\/}, Volume
  995. John Wiley \& Sons.

\bibitem[Dumuid \emph{and others}(2020)Dumuid, Pedi{\v{s}}i{\'c},
  Palarea-Albaladejo, Mart{\'\i}n-Fern{\'a}ndez, Hron and
  Olds]{dumuid2020compositional}
\textsc{Dumuid, Dorothea, Pedi{\v{s}}i{\'c}, {\v{Z}}eljko, Palarea-Albaladejo,
  Javier, Mart{\'\i}n-Fern{\'a}ndez, Josep~Antoni, Hron, Karel and Olds,
  Timothy}. (2020).
\newblock Compositional data analysis in time-use epidemiology: what, why, how.
\newblock {\em International Journal of Environmental Research and Public
  Health\/}~\textbf{17}(7), 2220.

\bibitem[Dumuid \emph{and others}(2019)Dumuid, Pedi{\v{s}}i{\'c}, Stanford,
  Mart{\'\i}n-Fern{\'a}ndez, Hron, Maher, Lewis and
  Olds]{dumuid2019compositional}
\textsc{Dumuid, Dorothea, Pedi{\v{s}}i{\'c}, {\v{Z}}eljko, Stanford,
  Tyman~Everleigh, Mart{\'\i}n-Fern{\'a}ndez, Josep-Antoni, Hron, Karel, Maher,
  Carol~A, Lewis, Lucy~K and Olds, Timothy}. (2019).
\newblock The compositional isotemporal substitution model: a method for
  estimating changes in a health outcome for reallocation of time between
  sleep, physical activity and sedentary behaviour.
\newblock {\em Statistical Methods in Medical Research\/}~\textbf{28}(3),
  846--857.

\bibitem[Gertheiss \emph{and others}(2013)Gertheiss, Maity and
  Staicu]{gertheiss2013variable}
\textsc{Gertheiss, Jan, Maity, Arnab and Staicu, Ana-Maria}. (2013).
\newblock Variable selection in generalized functional linear models.
\newblock {\em Stat\/}~\textbf{2}(1), 86--101.

\bibitem[Ghodrati and Panaretos(2021)Ghodrati and
  Panaretos]{ghodrati2021distribution}
\textsc{Ghodrati, Laya and Panaretos, Victor~M}. (2021).
\newblock Distribution-on-distribution regression via optimal transport maps.
\newblock {\em arXiv preprint arXiv:2104.09418\/}.

\bibitem[Gilchrist(2000)Gilchrist]{gilchrist2000statistical}
\textsc{Gilchrist, Warren}. (2000).
\newblock {\em Statistical modelling with quantile functions\/}. CRC Press.

\bibitem[Goldsmith \emph{and others}(2011)Goldsmith, Bobb, Crainiceanu, Caffo
  and Reich]{goldsmith2011penalized}
\textsc{Goldsmith, Jeff, Bobb, Jennifer, Crainiceanu, Ciprian~M, Caffo, Brian
  and Reich, Daniel}. (2011).
\newblock Penalized functional regression.
\newblock {\em Journal of Computational and Graphical
  Statistics\/}~\textbf{20}(4), 830--851.

\bibitem[Goldsmith \emph{and others}(2016)Goldsmith, Liu, Jacobson and
  Rundle]{goldsmith2016new}
\textsc{Goldsmith, Jeff, Liu, Xinyue, Jacobson, Judith and Rundle, Andrew}.
  (2016).
\newblock New insights into activity patterns in children, found using
  functional data analyses.
\newblock {\em Medicine and Science in Sports and Exercise\/}~\textbf{48}(9),
  1723.

\bibitem[Goldsmith \emph{and others}(2018)Goldsmith, Scheipl, Huang, Wrobel,
  Gellar, Harezlak, McLean, Swihart, Xiao, Crainiceanu and Reiss]{refund}
\textsc{Goldsmith, Jeff, Scheipl, Fabian, Huang, Lei, Wrobel, Julia, Gellar,
  Jonathan, Harezlak, Jaroslaw, McLean, Mathew~W., Swihart, Bruce, Xiao, Luo,
  Crainiceanu, Ciprian} \emph{and others}. (2018).
\newblock {\em refund: Regression with Functional Data\/}.
\newblock R package version 0.1-17.

\bibitem[Goldsmith \emph{and others}(2015)Goldsmith, Zipunnikov and
  Schrack]{goldsmith2015generalized}
\textsc{Goldsmith, Jeff, Zipunnikov, Vadim and Schrack, Jennifer}. (2015).
\newblock Generalized multilevel function-on-scalar regression and principal
  component analysis.
\newblock {\em Biometrics\/}~\textbf{71}(2), 344--353.

\bibitem[Haberkamp \emph{and others}(2019)Haberkamp, Moseley, Athanasiou,
  de~Andres-Trelles, Elferink, Rosa and Magrelli]{haberkamp2019european}
\textsc{Haberkamp, Marion, Moseley, Jane, Athanasiou, Dimitrios,
  de~Andres-Trelles, Fernando, Elferink, Andr{\'e}, Rosa, M{\'a}rio~Miguel and
  Magrelli, Armando}. (2019).
\newblock European regulators’ views on a wearable-derived performance
  measurement of ambulation for duchenne muscular dystrophy regulatory trials.
\newblock {\em Neuromuscular Disorders\/}~\textbf{29}(7), 514--516.

\bibitem[Hausdorff \emph{and others}(2018)Hausdorff, Hillel, Shustak, Del~Din,
  Bekkers, Pelosin, Nieuwhof, Rochester and Mirelman]{hausdorff2018everyday}
\textsc{Hausdorff, Jeffrey~M, Hillel, Inbar, Shustak, Shiran, Del~Din, Silvia,
  Bekkers, Esther~MJ, Pelosin, Elisa, Nieuwhof, Freek, Rochester, Lynn and
  Mirelman, Anat}. (2018).
\newblock Everyday stepping quantity and quality among older adult fallers with
  and without mild cognitive impairment: initial evidence for new motor markers
  of cognitive deficits?
\newblock {\em The Journals of Gerontology: Series A\/}~\textbf{73}(8),
  1078--1082.

\bibitem[Hebert \emph{and others}(2013)Hebert, Weuve, Scherr and
  Evans]{hebert2013alzheimer}
\textsc{Hebert, Liesi~E, Weuve, Jennifer, Scherr, Paul~A and Evans, Denis~A}.
  (2013).
\newblock Alzheimer disease in the united states (2010--2050) estimated using
  the 2010 census.
\newblock {\em Neurology\/}~\textbf{80}(19), 1778--1783.

\bibitem[Hosking(1990)Hosking]{hosking1990moments}
\textsc{Hosking, Jonathan~RM}. (1990).
\newblock L-moments: Analysis and estimation of distributions using linear
  combinations of order statistics.
\newblock {\em Journal of the Royal Statistical Society: Series B
  (Methodological)\/}~\textbf{52}(1), 105--124.

\bibitem[Hron \emph{and others}(2016)Hron, Menafoglio, Templ, Hruzova and
  Filzmoser]{hron2016simplicial}
\textsc{Hron, Karel, Menafoglio, Alessandra, Templ, Matthias, Hruzova, K and
  Filzmoser, Peter}. (2016).
\newblock Simplicial principal component analysis for density functions in
  bayes spaces.
\newblock {\em Computational Statistics \& Data Analysis\/}~\textbf{94},
  330--350.

\bibitem[Huang \emph{and others}(2019)Huang, Bai, Ivanescu, Harris, Maurer,
  Green and Zipunnikov]{huang2019multilevel}
\textsc{Huang, Lei, Bai, Jiawei, Ivanescu, Andrada, Harris, Tamara, Maurer,
  Mathew, Green, Philip and Zipunnikov, Vadim}. (2019).
\newblock Multilevel matrix-variate analysis and its application to
  accelerometry-measured physical activity in clinical populations.
\newblock {\em Journal of the American Statistical
  Association\/}~\textbf{114}(526), 553--564.

\bibitem[Ichimura(1991)Ichimura]{ichimura1991semiparametric}
\textsc{Ichimura, Hidehiko}. (1991).
\newblock Semiparametric least squares (sls) and weighted sls estimation of
  single-index models.

\bibitem[Irpino and Verde(2013)Irpino and Verde]{irpino2013metric}
\textsc{Irpino, Antonio and Verde, Rosanna}. (2013).
\newblock A metric based approach for the least square regression of
  multivariate modal symbolic data.
\newblock In:  {\em Statistical Models for Data Analysis\/}. Springer, pp.\
  161--169.

\bibitem[Kourtis \emph{and others}(2019)Kourtis, Regele, Wright and
  Jones]{kourtis2019digital}
\textsc{Kourtis, Lampros~C, Regele, Oliver~B, Wright, Justin~M and Jones,
  Graham~B}. (2019).
\newblock Digital biomarkers for alzheimer’s disease: the mobile/wearable
  devices opportunity.
\newblock {\em NPJ Digital Medicine\/}~\textbf{2}(1), 1--9.

\bibitem[Lock \emph{and others}(2013)Lock, Hoadley, Marron and
  Nobel]{lock2013joint}
\textsc{Lock, Eric~F, Hoadley, Katherine~A, Marron, James~Stephen and Nobel,
  Andrew~B}. (2013).
\newblock Joint and individual variation explained (jive) for integrated
  analysis of multiple data types.
\newblock {\em The Annals of Applied Statistics\/}~\textbf{7}(1), 523.

\bibitem[Marx and Eilers(1999)Marx and Eilers]{marx1999generalized}
\textsc{Marx, Brian~D and Eilers, Paul~HC}. (1999).
\newblock Generalized linear regression on sampled signals and curves: a
  p-spline approach.
\newblock {\em Technometrics\/}~\textbf{41}(1), 1--13.

\bibitem[Matabuena and Petersen(2021)Matabuena and
  Petersen]{matabuena2021distributional}
\textsc{Matabuena, Marcos and Petersen, Alex}. (2021).
\newblock Distributional data analysis with accelerometer data in a nhanes
  database with nonparametric survey regression models.
\newblock {\em arXiv\/}.

\bibitem[Matabuena \emph{and others}(2021)Matabuena, Petersen, Vidal and
  Gude]{matabuena2020glucodensities}
\textsc{Matabuena, Marcos, Petersen, Alexander, Vidal, Juan~C and Gude,
  Francisco}. (2021).
\newblock Glucodensities: a new representation of glucose profiles using
  distributional data analysis.
\newblock {\em Statistical Methods in Medical Research\/}~\textbf{30}(6),
  1445--1464.

\bibitem[Mc~Ardle \emph{and others}(2020)Mc~Ardle, Del~Din, Galna, Thomas and
  Rochester]{mc2020differentiating}
\textsc{Mc~Ardle, R{\'\i}ona, Del~Din, Silvia, Galna, Brook, Thomas, Alan and
  Rochester, Lynn}. (2020).
\newblock Differentiating dementia disease subtypes with gait analysis:
  feasibility of wearable sensors?
\newblock {\em Gait \& Posture\/}~\textbf{76}, 372--376.

\bibitem[Mc~Ardle \emph{and others}(2019)Mc~Ardle, Galna, Donaghy, Thomas and
  Rochester]{mc2019alzheimer}
\textsc{Mc~Ardle, R{\'\i}ona, Galna, Brook, Donaghy, Paul, Thomas, Alan and
  Rochester, Lynn}. (2019).
\newblock Do alzheimer's and lewy body disease have discrete pathological
  signatures of gait?
\newblock {\em Alzheimer's \& Dementia\/}~\textbf{15}(10), 1367--1377.

\bibitem[McKeague and Chang(2019)McKeague and Chang]{mck}
\textsc{McKeague, Ian and Chang, Hsin-wen}. (2019).
\newblock Functional data analysis for activity profiles from wearable devices.
\newblock
  \url{https://www.ima.umn.edu/materials/2019-2020/DW9.16-17.19/28237/talk_Minneapolis.pdf}.

\bibitem[McLean \emph{and others}(2014)McLean, Hooker, Staicu, Scheipl and
  Ruppert]{mclean2014functional}
\textsc{McLean, Mathew~W, Hooker, Giles, Staicu, Ana-Maria, Scheipl, Fabian and
  Ruppert, David}. (2014).
\newblock Functional generalized additive models.
\newblock {\em Journal of Computational and Graphical
  Statistics\/}~\textbf{23}(1), 249--269.

\bibitem[Morris(2015)Morris]{morris2015functional}
\textsc{Morris, Jeffrey~S}. (2015).
\newblock Functional regression.
\newblock {\em Annual Review of Statistics and Its Application\/}~\textbf{2},
  321--359.

\bibitem[Morris \emph{and others}(2006)Morris, Arroyo, Coull, Ryan, Herrick and
  Gortmaker]{morris2006using}
\textsc{Morris, Jeffrey~S, Arroyo, Cassandra, Coull, Brent~A, Ryan, Louise~M,
  Herrick, Richard and Gortmaker, Steven~L}. (2006).
\newblock Using wavelet-based functional mixed models to characterize
  population heterogeneity in accelerometer profiles: a case study.
\newblock {\em Journal of the American Statistical
  Association\/}~\textbf{101}(476), 1352--1364.

\bibitem[M{\"u}ller and Yao(2008)M{\"u}ller and Yao]{muller2008functional}
\textsc{M{\"u}ller, Hans-Georg and Yao, Fang}. (2008).
\newblock Functional additive models.
\newblock {\em Journal of the American Statistical
  Association\/}~\textbf{103}(484), 1534--1544.

\bibitem[O'Connell and Lock(2017)O'Connell and Lock]{rjive}
\textsc{O'Connell, Michael~J. and Lock, Eric~F.} (2017).
\newblock {\em r.jive: Perform JIVE Decomposition for Multi-Source Data\/}.
\newblock R package version 2.1.

\bibitem[Panaretos and Zemel(2020)Panaretos and Zemel]{panaretos2020invitation}
\textsc{Panaretos, Victor~M and Zemel, Yoav}. (2020).
\newblock {\em An invitation to statistics in Wasserstein space\/}. Springer
  Nature.

\bibitem[Parzen(2004)Parzen]{parzen2004quantile}
\textsc{Parzen, Emanuel}. (2004).
\newblock Quantile probability and statistical data modeling.
\newblock {\em Statistical Science\/}~\textbf{19}(4), 652--662.

\bibitem[Petersen and M{\"u}ller(2016)Petersen and
  M{\"u}ller]{petersen2016functional}
\textsc{Petersen, Alexander and M{\"u}ller, Hans-Georg}. (2016).
\newblock Functional data analysis for density functions by transformation to a
  hilbert space.
\newblock {\em The Annals of Statistics\/}~\textbf{44}(1), 183--218.

\bibitem[Petersen \emph{and others}(2021)Petersen, Zhang and
  Kokoszka]{petersen2021modeling}
\textsc{Petersen, Alexander, Zhang, Chao and Kokoszka, Piotr}. (2021).
\newblock Modeling probability density functions as data objects.
\newblock {\em Econometrics and Statistics\/}.

\bibitem[Powley(2013)Powley]{powley2013quantile}
\textsc{Powley, Bradford~W}. (2013).
\newblock Quantile function methods for decision analysis [Ph.D. Thesis].
  Stanford University.

\bibitem[{R Core Team}(2018){R Core Team}]{Rsoft}
\textsc{{R Core Team}}. (2018).
\newblock {\em R: A Language and Environment for Statistical Computing\/}.
\newblock R Foundation for Statistical Computing, Vienna, Austria.

\bibitem[Reider \emph{and others}(2020)Reider, Bai, Scharfstein and
  Zipunnikov]{reider2020methods}
\textsc{Reider, Lisa, Bai, Jiawei, Scharfstein, Daniel~O and Zipunnikov,
  Vadim}. (2020).
\newblock Methods for step count data: Determining “valid” days and
  quantifying fragmentation of walking bouts.
\newblock {\em Gait \& Posture\/}~\textbf{81}, 205--212.

\bibitem[Reiss \emph{and others}(2017)Reiss, Goldsmith, Shang and
  Ogden]{reiss2017methods}
\textsc{Reiss, Philip~T, Goldsmith, Jeff, Shang, Han~Lin and Ogden, R~Todd}.
  (2017).
\newblock Methods for scalar-on-function regression.
\newblock {\em International Statistical Review\/}~\textbf{85}(2), 228--249.

\bibitem[Shema-Shiratzky \emph{and others}(2020)Shema-Shiratzky, Hillel,
  Mirelman, Regev, Hsieh, Karni, Devos, Sosnoff and
  Hausdorff]{shema2020wearable}
\textsc{Shema-Shiratzky, Shirley, Hillel, Inbar, Mirelman, Anat, Regev, Keren,
  Hsieh, Katherine~L, Karni, Arnon, Devos, Hannes, Sosnoff, Jacob~J and
  Hausdorff, Jeffrey~M}. (2020).
\newblock A wearable sensor identifies alterations in community ambulation in
  multiple sclerosis: contributors to real-world gait quality and physical
  activity.
\newblock {\em Journal of Neurology\/}~\textbf{267}(7), 1912--1921.

\bibitem[Stoker(1986)Stoker]{stoker1986consistent}
\textsc{Stoker, Thomas~M}. (1986).
\newblock Consistent estimation of scaled coefficients.
\newblock {\em Econometrica: Journal of the Econometric Society\/}, 1461--1481.

\bibitem[Takemura(1983)Takemura]{takemura1983orthogonal}
\textsc{Takemura, Akimichi}. (1983).
\newblock Orthogonal expansion of quantile function and components of the
  shapiro-francia statistic.
\newblock {\em Technical Report}, Stanford University CA Department of
  Statistics.

\bibitem[Talsk{\'a} \emph{and others}(2021)Talsk{\'a}, Hron and
  Grygar]{talska2021compositional}
\textsc{Talsk{\'a}, Ren{\'a}ta, Hron, Karel and Grygar, Tom{\'a}{\v{s}}~Matys}.
  (2021).
\newblock Compositional scalar-on-function regression with application to
  sediment particle size distributions.
\newblock {\em Mathematical Geosciences\/}, 1--29.

\bibitem[Tibshirani(1996)Tibshirani]{tibshirani1996regression}
\textsc{Tibshirani, Robert}. (1996).
\newblock Regression shrinkage and selection via the lasso.
\newblock {\em Journal of the Royal Statistical Society: Series B
  (Methodological)\/}~\textbf{58}(1), 267--288.

\bibitem[Van~den Boogaart \emph{and others}(2014)Van~den Boogaart, Egozcue and
  Pawlowsky-Glahn]{van2014bayes}
\textsc{Van~den Boogaart, Karl~Gerald, Egozcue, Juan~Jos{\'e} and
  Pawlowsky-Glahn, Vera}. (2014).
\newblock Bayes hilbert spaces.
\newblock {\em Australian \& New Zealand Journal of
  Statistics\/}~\textbf{56}(2), 171--194.

\bibitem[Varma \emph{and others}(2017)Varma, Dey, Leroux, Di, Urbanek, Xiao and
  Zipunnikov]{varma2017re}
\textsc{Varma, Vijay~R, Dey, Debangan, Leroux, Andrew, Di, Junrui, Urbanek,
  Jacek, Xiao, Luo and Zipunnikov, Vadim}. (2017).
\newblock Re-evaluating the effect of age on physical activity over the
  lifespan.
\newblock {\em Preventive Medicine\/}~\textbf{101}, 102--108.

\bibitem[Varma \emph{and others}(2021)Varma, Ghosal, Hillel, Volfson, Weiss,
  Urbanek, Hausdorff, Zipunnikov and Watts]{gait2020vr}
\textsc{Varma, Vijay~R, Ghosal, Rahul, Hillel, Inbar, Volfson, Dmitri, Weiss,
  Jordan, Urbanek, Jacek, Hausdorff, Jeffrey~M., Zipunnikov, Vadim and Watts,
  Amber}. (2021).
\newblock Continuous gait monitoring discriminates community dwelling mild ad
  from cognitively normal controls.
\newblock {\em Alzheimer's \& Dementia: Translational Research \& Clinical
  Interventions\/}~\textbf{7}(1), e12131.

\bibitem[Varma and Watts(2017)Varma and Watts]{varma2017daily}
\textsc{Varma, Vijay~R and Watts, Amber}. (2017).
\newblock Daily physical activity patterns during the early stage of
  alzheimer’s disease.
\newblock {\em Journal of Alzheimer's Disease\/}~\textbf{55}(2), 659--667.

\bibitem[Verde and Irpino(2010)Verde and Irpino]{verde2010ordinary}
\textsc{Verde, Rosanna and Irpino, Antonio}. (2010).
\newblock Ordinary least squares for histogram data based on wasserstein
  distance.
\newblock In:  {\em Proceedings of COMPSTAT'2010\/}. Springer, pp.\  581--588.

\bibitem[Wang and Yang(2009)Wang and Yang]{wang2009spline}
\textsc{Wang, Li and Yang, Lijian}. (2009).
\newblock Spline estimation of single-index models.
\newblock {\em Statistica Sinica\/}~\textbf{19}(2), 765--783.

\bibitem[Weiss \emph{and others}(2014)Weiss, Herman, Giladi and
  Hausdorff]{weiss2014objective}
\textsc{Weiss, Aner, Herman, Talia, Giladi, Nir and Hausdorff, Jeffrey~M}.
  (2014).
\newblock Objective assessment of fall risk in parkinson's disease using a
  body-fixed sensor worn for 3 days.
\newblock {\em PloS one\/}~\textbf{9}(5), e96675.

\bibitem[Wood(2017)Wood]{wood2017generalized}
\textsc{Wood, Simon~N}. (2017).
\newblock {\em Generalized additive models: an introduction with R\/}. CRC
  press.

\bibitem[Wood \emph{and others}(2016)Wood, Pya and
  S{\"a}fken]{wood2016smoothing}
\textsc{Wood, Simon~N, Pya, Natalya and S{\"a}fken, Benjamin}. (2016).
\newblock Smoothing parameter and model selection for general smooth models.
\newblock {\em Journal of the American Statistical
  Association\/}~\textbf{111}(516), 1548--1563.

\bibitem[Wrobel \emph{and others}(2019)Wrobel, Zipunnikov, Schrack and
  Goldsmith]{wrobel2019registration}
\textsc{Wrobel, Julia, Zipunnikov, Vadim, Schrack, Jennifer and Goldsmith,
  Jeff}. (2019).
\newblock Registration for exponential family functional data.
\newblock {\em Biometrics\/}~\textbf{75}(1), 48--57.

\bibitem[Xiao \emph{and others}(2015)Xiao, Huang, Schrack, Ferrucci, Zipunnikov
  and Crainiceanu]{xiao2015quantifying}
\textsc{Xiao, Luo, Huang, Lei, Schrack, Jennifer~A, Ferrucci, Luigi,
  Zipunnikov, Vadim and Crainiceanu, Ciprian~M}. (2015).
\newblock Quantifying the lifetime circadian rhythm of physical activity: a
  covariate-dependent functional approach.
\newblock {\em Biostatistics\/}~\textbf{16}(2), 352--367.

\bibitem[Yang(2020)Yang]{yang2020random}
\textsc{Yang, Hojin}. (2020).
\newblock Random distributional response model based on spline method.
\newblock {\em Journal of Statistical Planning and Inference\/}~\textbf{207},
  27--44.

\bibitem[Yang \emph{and others}(2020)Yang, Baladandayuthapani, Rao and
  Morris]{yang2020quantile}
\textsc{Yang, Hojin, Baladandayuthapani, Veerabhadran, Rao, Arvind~UK and
  Morris, Jeffrey~S}. (2020).
\newblock Quantile function on scalar regression analysis for distributional
  data.
\newblock {\em Journal of the American Statistical
  Association\/}~\textbf{115}(529), 90--106.

\bibitem[Yogev-Seligmann \emph{and others}(2008)Yogev-Seligmann, Hausdorff and
  Giladi]{yogev2008role}
\textsc{Yogev-Seligmann, Galit, Hausdorff, Jeffrey~M and Giladi, Nir}. (2008).
\newblock The role of executive function and attention in gait.
\newblock {\em Movement Disorders\/}~\textbf{23}(3), 329--342.

\bibitem[Zhang and M{\"u}ller(2011)Zhang and M{\"u}ller]{zhang2011functional}
\textsc{Zhang, Zhen and M{\"u}ller, Hans-Georg}. (2011).
\newblock Functional density synchronization.
\newblock {\em Computational Statistics \& Data Analysis\/}~\textbf{55}(7),
  2234--2249.

\end{thebibliography}


\begin{thebibliography}{99}

\bibitem[Goldsmith \emph{and others}(2018)Goldsmith, Scheipl, Huang, Wrobel,
  Gellar, Harezlak, McLean, Swihart, Xiao, Crainiceanu and Reiss]{refund}
\textsc{Goldsmith, Jeff, Scheipl, Fabian, Huang, Lei, Wrobel, Julia, Gellar,
  Jonathan, Harezlak, Jaroslaw, McLean, Mathew~W., Swihart, Bruce, Xiao, Luo,
  Crainiceanu, Ciprian} \emph{and others}. (2018).
\newblock {\em refund: Regression with Functional Data\/}.
\newblock R package version 0.1-17.

\bibitem[Marx and Eilers(1998)Marx and Eilers]{marx1998direct}
\textsc{Marx, Brian~D and Eilers, Paul~HC}. (1998).
\newblock Direct generalized additive modeling with penalized likelihood.
\newblock {\em Computational Statistics \& Data Analysis\/}~\textbf{28}(2),
  193--209.

\bibitem[McLean \emph{and others}(2014)McLean, Hooker, Staicu, Scheipl and
  Ruppert]{mclean2014functional}
\textsc{McLean, Mathew~W, Hooker, Giles, Staicu, Ana-Maria, Scheipl, Fabian and
  Ruppert, David}. (2014).
\newblock Functional generalized additive models.
\newblock {\em Journal of Computational and Graphical
  Statistics\/}~\textbf{23}(1), 249--269.

\bibitem[Wood(2017)Wood]{wood2017generalized}
\textsc{Wood, Simon~N}. (2017).
\newblock {\em Generalized additive models: an introduction with R\/}. CRC
  press.

\end{thebibliography}
\newpage

\begin{table}[H]
\small
\centering
\caption{Displayed are the proportion of deviance explained (D.E) by SOQFR and FGAM-QF modelling cognitive status using quantile functions of gait metrics in presence of age and sex. The top 10 metrics (ranked by proportion of deviance explained) have been reported for each of the method. We also report cross-validated (10-fold) AUC (cvAUC) of the models from repeated (B=100) cross-validation. cvAUC from generalized linear models (logit link) using only mean of the gait measures as predictor (along with age and sex) are provided for comparison in the parenthesis.}
\label{tab:my-table2}
\begin{tabular}{cccccc}
\hline
\multicolumn{1}{p{3cm}}{\centering Variable\\ (ranked for SOQFR)} & \multicolumn{1}{p{1cm}}{\centering D.E\\         SOQFR} &  \multicolumn{1}{p{1cm}}{\centering cvAUC\\         SOQFR} & \multicolumn{1}{p{3cm}}{\centering Variable\\(ranked for FGAM-QF)} & \multicolumn{1}{p{1cm}}{\centering D.E\\         FGAM-QF} &  \multicolumn{1}{p{1cm}}{\centering cvAUC\\         FGAM-QF}\\ \hline
strRegAP                                                    & 0.58 &0.79  (0.74)                                                                    & Mean\_Stride\_Time\_\_s\_                                   & 0.63      &0.93 (0.73)                                                                \\ \hline
Step\_Velocity\_\_cm\_sec\_                                 & 0.50 &0.89  (0.80)                                                                    & Mean\_Step\_Time\_s\_                                       & 0.63    &0.93 (0.73)                                                                  \\ \hline
Distance\_\_m\_                                             & 0.49 &0.89 (0.80)                                                                    & strRegAP                                                    & 0.59      &0.83 (0.74)                                                                \\ \hline
Cadence\_V\_time\_domain\_                                  & 0.48 &0.90 (0.76)                                                                      & Step\_Velocity\_\_cm\_sec\_                                 & 0.50        &0.89 (0.80)                                                              \\ \hline
strideTimeV                                                 & 0.40 &0.80 (0.70)                                                                     & Mean\_Step\_Length\_\_cm\_                                  & 0.50 &0.83 (0.79)                                                       \\ \hline
frqML                                                       & 0.38 &0.84   (0.71)                                                                   & frqV                                                        & 0.50          & 0.86 (0.73)                                                            \\ \hline
CV\_Step\_time                                              & 0.38  & 0.85 (0.68)                                                                    & stepTimeV                                                   & 0.50             & 0.85 (0.70)                                                            \\ \hline
frqV                                                        & 0.37 &0.82 (0.73)                                                                    & strideTimeV                                                 & 0.50                & 0.85 (0.70)                                                      \\ \hline
Mean\_Stride\_Time\_\_s\_                                   & 0.37 &0.81 (0.73)                                                                     & Distance\_\_m\_                                             & 0.49             &0.89 (0.80)                                                          \\ \hline
Mean\_Step\_Time\_s\_                                       & 0.37 &0.81 (0.73)                                                                     & Cadence\_V\_time\_domain\_                                  & 0.48                & 0.90 (0.76)                                                      \\ \hline
\end{tabular}
\end{table}

\begin{table}[H]
\centering
\caption{The results of SOQFR-L, modelling mild-AD status (adjusted for age and sex) on first four L-moments of stride regularity (Model A1) , step velocity  (Model A2) and cadence (Model A3) respectively. Deviance explained using only mean of gait measures (adjusted for age and sex) are provided for comparison in the parenthesis. We also report cross-validated (10-fold) AUC (cvAUC) of the models from repeated (B=100) cross-validation. cvAUC from models using only mean of the gait measures as predictor (along with age and sex) are provided for comparison in the parenthesis.}
\label{tab:my-table3}
\begin{tabular}{lllllllll}
\hline
\multicolumn{3}{c}{Model A1 (stride regularity)} & \multicolumn{3}{c}{Model A2 (step velocity)} & \multicolumn{3}{c}{Model A3 (cadence)} \\ \hline
Coef                   &  est                   & p-value                  & Coef              &  est               & p-value                            & Coef                &  est                 & p-value                    \\ \hline
$\beta_0$                     & 4.31                       & 0.17327                  & $\beta_0$                  & 18.203                 & 0.00267                            & $\beta_0$                    & 6.338                    & 0.25626                    \\ \hline
age                           & -0.044                     & 0.22957                  & age                       & -0.151                 & 0.00836                            & age                         & -0.048                   & 0.29464                    \\ \hline
sex (M)                    & 2.113                      & 0.00015                  & sex (M)                & 3.706                  & $5.11\times 10^{-5}$              & sex (M)                  & 2.740                    & 0.00155                    \\ \hline
$L_1$                         & -0.604                     & 0.85739                  & $L_1$                     & -0.044                 & 0.16441                            & $L_1$                       & 0.009                    & 0.84854                    \\ \hline
$L_2$                         & -21.801                    & 0.03457                  & $L_2$                     & -0.480                 & 0.00127                            & $L_2$                       & -1.055                   & $4.2\times 10^{-5}$                 \\ \hline
$L_3$                         & 4.621                      & 0.84022                  & $L_3$                     & -0.601                 & 0.03620                            & $L_3$                       & 0.002                    & 0.99676                    \\ \hline
$L_4$                         & 12.761                     & 0.63239                  & $L_4$                     & -0.212                 & 0.61004                            & $L_4$                       & -0.121                   & 0.90079                    \\ \hline
\multicolumn{1}{c}{\begin{tabular}[c]{@{}c@{}}Deviance \\  explained\end{tabular}} & \multicolumn{2}{c}{$21.58 (15.59)\%$} & \multicolumn{1}{c}{\begin{tabular}[c]{@{}c@{}}Deviance \\  explained\end{tabular}} & \multicolumn{2}{c}{$48.75 (24.20)\%$}   & \multicolumn{1}{c}{\begin{tabular}[c]{@{}c@{}}Deviance \\  explained\end{tabular}} & \multicolumn{2}{c}{$45.11 (17.73)\%$} \\\hline 
\multicolumn{1}{c}{\begin{tabular}[c]{@{}c@{}}cvAUC\end{tabular}} & \multicolumn{2}{c}{$0.74 (0.74)$} & \multicolumn{1}{c}{\begin{tabular}[c]{@{}c@{}}cvAUC\end{tabular}} & \multicolumn{2}{c}{$0.89 (0.80)$}   & \multicolumn{1}{c}{\begin{tabular}[c]{@{}c@{}}cvAUC\end{tabular}} & \multicolumn{2}{c}{$0.87 (0.76)$} \\
\end{tabular}
\end{table}

\newpage

\begin{table}[H]
\small
\caption{The results from linear regression models of cognitive scores (ATTN, VM and EF) on first four L-moments of stride regularity, step velocity and cadence, adjusting for age, sex and education. Benchmark models using just age, sex, and education produce adjusted-$R^2_{ATTN}=0.163$, $R^2_{VM}=  0.2489$, $R^2_{EF}=  0.2754$. We also report cross-validated (10-fold) R-squared of the models from repeated (B=100) cross-validation. Both metrics (adj-Rsq and cv-Rsq) from models using only mean of the gait measures (along with age, sex and education) are provided for comparison in the parenthesis.}
\label{tab:my-table44}
\hspace*{- 8 mm}
\begin{tabular}{cccccccccc}
\hline
Y                                                              & \multicolumn{3}{c}{Model B1 (stride regularity)}  & \multicolumn{3}{c}{Model B2 (step velocity)}           & \multicolumn{3}{c}{Model B3 (cadence)}   \\ \hline
\multirow{10}{*}{ATTN}                                                   & Coef      & est    & p-value              & Coef      & est                  & p-value              & Coef      & est    & p-value              \\ \cline{2-10} 
                                                                         & $\beta_0$ & -2.471 & 0.017               & $\beta_0$ & -1.964               & 0.089               & $\beta_0$ & 0.738  & 0.591               \\ \cline{2-10} 
                                                                         & age       & 0.005  & 0.640               & age       & 0.007                & 0.507               & age       & 0.001  & 0.924               \\ \cline{2-10} 
                                                                         & sex (M)   & -0.34  & 0.031               & sex (M)   & -0.454               & 0.006               & sex (M)   & -0.549 & 0.002               \\ \cline{2-10} 
                                                                         & edu & 0.082  & 0.001               & edu & 0.069                & 0.009               & edu & 0.074  & 0.004               \\ \cline{2-10} 
                                                                         & $L_1$     & 0.879  & 0.379               & $L_1$     & 0.001                & 0.897               & $L_1$     & -0.025 & 0.035               \\ \cline{2-10} 
                                                                         & $L_2$     & 5.879  & 0.051               & $L_2$     & 0.039                & 0.138               & $L_2$     & 0.124  & 0.004               \\ \cline{2-10} 
                                                                         & $L_3$     & 5.616  & 0.424               & $L_3$     & 0.015                & 0.756               & $L_3$     & -0.137 & 0.122               \\ \cline{2-10} 
                                                                         & $L_4$     & -6.456 & 0.415               & $L_4$     & -0.004               & 0.961               & $L_4$     & 0.025  & 0.882               \\ \cline{2-10} 
                                                                         & adj-Rsq   & \multicolumn{2}{c}{0.240 (0.172)}   & adj-Rsq   & \multicolumn{2}{c}{0.192 (0.173)}                 & adj-Rsq   & \multicolumn{2}{c}{0.226 (0.154)} 
                                                            \\ \cline{2-10} 
                                                                         & cv-Rsq   & \multicolumn{2}{c}{0.273 (0.211)}   & cv-Rsq   & \multicolumn{2}{c}{0.214 (0.209)}                 & cv-Rsq   & \multicolumn{2}{c}{0.264 (0.209)}               \\ \hline
\multirow{9}{*}{VM}                                                      & $\beta_0$ & -4.469 & 0.036               & $\beta_0$ & -3.283               & 0.159               & $\beta_0$ & 0.655  & 0.819               \\ \cline{2-10} 
                                                                         & age       & 0.001  & 0.973               & age       & 0.009                & 0.707               & age       & -0.009 & 0.666               \\ \cline{2-10} 
                                                                         & sex (M)   & -1.296 &  $1\times 10^{-4}$                & sex (M)   & -1.591               & $5.8\times 10^{-6}$ & sex (M)   & -1.583 & $3.4\times 10^{-5}$ \\ \cline{2-10} 
                                                                         & edu & 0.171  & 0.001               & edu & 0.119                & 0.025               & edu & 0.123  & 0.022               \\ \cline{2-10} 
                                                                         & $L_1$     & 1.173  & 0.571               & $L_1$     & $7.4\times 10^{-6}$ & 0.999               & $L_1$     & -0.030 & 0.213               \\ \cline{2-10} 
                                                                         & $L_2$     & 17.118 & 0.007               & $L_2$     & 0.131                & 0.014               & $L_2$     & 0.315  & 0.001               \\ \cline{2-10} 
                                                                         & $L_3$     & 11.616 & 0.424               & $L_3$     & 0.094                & 0.350               & $L_3$     & -0.158 & 0.389               \\ \cline{2-10} 
                                                                         & $L_4$     & 4.214  & 0.797               & $L_4$     & 0.119                & 0.531               & $L_4$     & 0.215  & 0.545               \\ \cline{2-10} 
                                                                         & adj-Rsq   & \multicolumn{2}{c}{0.365 (0.251)}   & adj-Rsq   & \multicolumn{2}{c}{0.356 (0.269)}                  & adj-Rsq   & \multicolumn{2}{c}{0.346 (0.246)}
                                       \\ \cline{2-10} 
                                                                         & cv-Rsq   & \multicolumn{2}{c}{0.390 (0.311)}   & cv-Rsq   & \multicolumn{2}{c}{0.390 (0.332)}                  & cv-Rsq   & \multicolumn{2}{c}{0.383 (0.317)}                                    \\ \hline
\multirow{9}{*}{EF} & $\beta_0$ & -3.307 & 0.048               & $\beta_0$ & -3.623               & 0.048               & $\beta_0$ & 0.844  & 0.698               \\ \cline{2-10} 
                                                                         & age       & 0.004  & 0.814               & age       & 0.016                & 0.389               & age       & -0.002 & 0.912               \\ \cline{2-10} 
                                                                         & sex (M)   & -1.140 & $2.4\times 10^{-5}$ & sex (M)   & -1.329               & $1.5\times 10^{-6}$ & sex (M)   & -1.368 & $3.4\times 10^{-6}$ \\ \cline{2-10} 
                                                                         & edu & 0.134  & 0.001               & edu & 0.108                & 0.009               & edu & 0.103  & 0.012               \\ \cline{2-10} 
                                                                         & $L_1$     & -0.402 & 0.804               & $L_1$     & 0.005                & 0.697               & $L_1$     & -0.036 & 0.053               \\ \cline{2-10} 
                                                                         & $L_2$     & 15.992 & 0.001               & $L_2$     & 0.087                & 0.034               & $L_2$     & 0.294  & $2.8\times 10^{-5}$ \\ \cline{2-10} 
                                                                         & $L_3$     & -3.396 & 0.765               & $L_3$     & 0.051                & 0.512               & $L_3$     & -0.264 & 0.061               \\ \cline{2-10} 
                                                                         & $L_4$     & -1.960 & 0.879               & $L_4$     & -0.064               & 0.664               & $L_4$     & 0.442  & 0.104               \\ \cline{2-10} 
                                                                         & adj-Rsq   & \multicolumn{2}{c}{0.377 (0.280)}   & adj-Rsq   & \multicolumn{2}{c}{0.376 (0.302)}                 & adj-Rsq   & \multicolumn{2}{c}{0.397 (0.271)} 
                                      \\ \cline{2-10} 
                                                                         & cv-Rsq   & \multicolumn{2}{c}{0.414 (0.347)}   & cv-Rsq   & \multicolumn{2}{c}{0.425 (0.358)}                 & cv-Rsq   & \multicolumn{2}{c}{0.441 (0.345)}                                    
                                                                         \\ \hline
\end{tabular}
\end{table}

\end{document}


\title{\textbf{Supplementary Material}\\
Distributional data analysis via quantile functions and its application to modelling digital biomarkers of gait in Alzheimer’s Disease}








\date{}

\maketitle
\section{Web Appendix 1, Estimation of FGAM-QF}
For identifiability of the FGAM-QF, the following constraint is imposed: $\sum_{i=1}^{n}\int_{0}^{1}F(Q_i(p),p)=0$ \citep{mclean2014functional,wood2017generalized}.
Note that the proposed approach of using $F\{Q(p),p\}$ capturing the smooth effect of the subject-specific quantile function at each quantile level is different from using the transformation approach in \cite{mclean2014functional}, which is focused on studying the smooth diurnal effect $F\{G_t(X(t)),t\}$ using population level distribution function $G_t(x)$. Nevertheless, the same model fitting procedure can be used. In particular, we model the bivariate function $F(\cdot,\cdot)$ using a tensor product of univariate B-spline basis functions. Suppose, $\{B_{Q,k}(q)\}^{K}_{k=1}$ and $\{B_{P,\ell}(p)\}^{L}_{\ell=1}$  be a set of known basis functions over $q$ (where $Q(p)=q$) and $p$, respectively. Then, $F(\cdot,\cdot)$ is modelled using a tensor product of two basis functions as $F\{Q_{i}(p),p\}=\sum_{k=1}^{K}\sum_{\ell=1}^{L}\theta_{k,\ell}B_{Q,k}\{Q_{i}(p)\}B_{p,\ell}(p)$. Using this expansion model (5) (in the paper) can be reformulated as
\begin{eqnarray}
g(\mu_i)&=&\alpha+\*Z_i^T\bm\gamma + \sum_{k=1}^{K}\sum_{\ell=1}^{L}\theta_{k,\ell}\int_{0}^{1}B_{Q,k}\{Q_{i}(p)\}B_{p,\ell}(p)dp\notag\\
&=& \alpha+\*Z_i^T\bm\gamma +\*W_{i}^T\bm\theta,\label{FGAM2}
\end{eqnarray}
where we denote the $KL-$dimensional vector of $B_{Q,k}\{Q_{i}(p)\}B_{P,\ell}(p)$'s as $\*W_i$, and $\bm\theta$ is the vector of unknown basis coefficients $\theta_{k,\ell}$'s. Then, the penalized negative log likelihood criterion for estimation is given by
\begin{equation}
S(\psi)=R(\alpha,\bm\gamma,\bm\theta)=  -2log L(\alpha,\bm\gamma,\bm\theta;Y_i,\*Z_i,\*W_i) + \bm\theta^T\+P\bm\theta.\label{FGAM:cri}
\end{equation}
Here, $\+P=\lambda_qD_{q}^TD_{q}\otimes I_{K}+\lambda_pD_{p}^TD_{p}\otimes I_{L}$ is a penalty matrix consisting of second order row and column penalties imposing smoothness on $F(\cdot,\cdot)$ in both directions \citep{marx1998direct}. The parameters are estimated using penalized iteratively re-weighted least squares (P-IRLS) as in \cite{mclean2014functional}. We use the {\tt refund} package \citep{refund} for implementation of FGAM-QF.

\section{Web Appendix 2, JIVE Algorithm using L-moments}


\textbf{Algorithm 1: JIVE using L-Moments of Quantile functions}
\hrule
\begin{itemize}
\item[] 1. \textbf{Goal}: To estimate joint and individual structure in multi-modal distributional data
\item[] 2. \textbf{Input}: $\+L$, the block data matrix containing L-moments $\*L_{i}^{(d)}$
\item[] 3. for $iter=1$ to $A$ do
\item[] 4. Determine ranks $s$ and $s_d$ s.
\item[] 5. Estimate $\+J$ by rank $s$ SVD of $\+L$, set $\+J=\+U\+S\+V^{T}$
\item[] 6. for $d=1$ to $D$ do
\item[] 7.  $\tilde{\+L}^d=\+L^d- \+J^d$
\item[] 8. Estimate $\+A^d$ by rank $s_d$ SVD of $\tilde{\+L}^d(\+I-\+V\+V^{T})$, set ${\+L}^d=\+L^d- \+A^d$
\item[] 9. Set $\+L=\begin{bmatrix}
    \+L^{1}\\
    \+L^{2}\\
    \vdots\\
     \+L^{D}
    \end{bmatrix}$
\item[] 10. \textbf{Return} $\+J$, $\+A^d$.
\hrule
\end{itemize}

\section{Supplementary Tables}
Supplementary Tables S1-S6 referenced in the paper are given below. 

\begin{table}[H]
\large
\centering
\caption{Summary statistics for the complete, AD and CNC samples. No statistical difference between the AD and CNC groups are observed across age, BMI, or V$0_2$ max. However, AD group had a smaller percentage of females ($26.3$ vs $68.8$ for CNC) and lower education ($15.6$ years vs $17.4$ years for CNC).}
\label{tab:my-table1}
\begin{tabular}{cccccccc}
\hline
Characteristic     & \multicolumn{2}{c|}{Complete sample} & \multicolumn{2}{c|}{AD} & \multicolumn{2}{c|}{CNC} & P value         \\ \hline
                   & Mean/Freq           & SD             & Mean/Freq     & SD      & Mean/Freq       & SD        &                 \\ \hline
Age                & 73.21               & 7.13           & 73.24         & 7.71    & 73.19           & 6.73      & 0.975           \\ \hline
\% Female          & 50                  & N/A            & 26.32         & N/A     & 68.75           & N/A       & \textless 0.001 \\ \hline
Years of edu & 16.63               & 3.21           & 15.59         & 2.78    & 17.44           & 3.31      & 0.0066          \\ \hline
BMI                & 26.39               & 6.03           & 26.38         & 7.87    & 26.4            & 4.13      & 0.9854          \\ \hline
VO2 max            & 22.06               & 5.36           & 21.61         & 5.24    & 22.39           & 5.48      & 0.5175          \\ \hline
\end{tabular}
\end{table}

\vspace{15 mm}

\begin{table}[H]
\small
\caption{Displayed are the results from logistic regression models of cognitive status (adjusted for age and sex) on first four regular moments ($\mu^{'}_r$  ) of stride regularity (Model A11) , step velocity (Model A21) and cadence (Model A31) respectively.}
\label{tab:my-tableS2}
\hspace*{- 7 mm}
\begin{tabular}{lllllllll}
\hline
\multicolumn{3}{c}{Model A11 (stride regularity)} & \multicolumn{3}{c}{Model A21 (step velocity)} & \multicolumn{3}{c}{Model A31 (cadence)} \\  \hline
Coef                   &  est                   & p-value                  & Coef              &  est               & p-value                            & Coef                &  est                 & p-value                    \\ \hline
$\beta_0$                     & -4.098                       & 0.439                  & $\beta_0$                  & -95.18                 & 0.094                            & $\beta_0$                    & -713.7                   & 0.202                    \\ \hline
age                           & -0.039                     & 0.281                  & age                       & -0.058                 & 0.167                            & age                         & -0.032                  & 0.373                    \\ \hline
sex (M)                    & 2.2                      &$7.88\times 10^{-5}$                  & sex (M)                & 2.61                  & $1.91\times 10^{-5}$              & sex (M)                  & 1.727                    & 0.002                    \\ \hline
$\mu^{'}_1$                         & 45.79                     & 0.231                  & $\mu^{'}_1$                   & 4.062                 & 0.075                            & $\mu^{'}_1$                         & 27.18                    & 0.201                    \\ \hline
$\mu^{'}_2$                          & -97.62                    & 0.272                  & $\mu^{'}_2$                       & -0.059                 & 0.0712                            & $\mu^{'}_2$                         & -0.381                   & 0.203                \\ \hline
$\mu^{'}_3$                           & 66.44                     & 0.352                  & $\mu^{'}_3$                      & $3.61\times 10^{-4}$                 & 0.069                            & $\mu^{'}_3$                         & 0.002                    & 0.206                    \\ \hline
$\mu^{'}_4$                           & -10.35                     & 0.450                  & $\mu^{'}_4$                       &$-7.9\times 10^{-7}$                 & 0.067                            & $\mu^{'}_4$                         & $-5.28\times 10^{-6}$                   & 0.210                    \\\hline
\multicolumn{1}{c}{\begin{tabular}[c]{@{}c@{}}Deviance \\  explained\end{tabular}} & \multicolumn{2}{c}{$21.01\%$} & \multicolumn{1}{c}{\begin{tabular}[c]{@{}c@{}}Deviance \\  explained\end{tabular}} & \multicolumn{2}{c}{$30.29\%$}   & \multicolumn{1}{c}{\begin{tabular}[c]{@{}c@{}}Deviance \\  explained\end{tabular}} & \multicolumn{2}{c}{$20.27\%$} \\ 
\end{tabular}
\end{table}

\newpage

\begin{table}[H]
\small
\caption{Displayed are the results from logistic regression models of cognitive status (adjusted for age and sex) on mean and three central moments ($\mu_2$, $\mu_3$, $\mu_4$  ) of stride regularity (Model A12) , step velocity  (Model A22) and cadence (Model A32) respectively.}
\label{tab:my-tableS3}
\begin{tabular}{lllllllll}
\hline
\multicolumn{3}{c}{Model A12 (stride regularity)} & \multicolumn{3}{c}{Model A22 (step velocity)} & \multicolumn{3}{c}{Model A32 (cadence)} \\ \hline
Coef                   &  est                   & p-value                  & Coef              &  est               & p-value                            & Coef                &  est                 & p-value                    \\ \hline
$\beta_0$                     & 4.735                       & 0.158                  & $\beta_0$                  & 15.58                 & 0.011                            & $\beta_0$                    & 3.302                   & 0.610                    \\ \hline
age                           & -0.056                     & 0.155                  & age                       & -0.015                 & 0.012                            & age                         & -0.051                  & 0.295                    \\ \hline
sex (M)                    & 2.405                      &$6.03\times 10^{-5}$                  & sex (M)                & 4.01                  & $4.79\times 10^{-5}$              & sex (M)                  & 2.898                    & 0.001                    \\ \hline
$\mu^{'}_1$                         & -1.097                     & 0.692                  & $\mu^{'}_1$                   & -0.036                 & 0.168                            & $\mu^{'}_1$                         & 0.020                    & 0.696                    \\ \hline
$\mu_2$                          & -65.056                    & 0.003                  & $\mu_2$                       & -0.009                 & 0.024                            & $\mu_2$                         & -0.045                   & 0.003                \\ \hline
$\mu_3$                           & -45.54                     & 0.515                  & $\mu_3$                      & $-1.52\times 10^{-4}$                 & 0.141                            & $\mu_3$                         &$-3\times 10^{-4}$                   & 0.595                    \\ \hline
$\mu_4$                           & 46.60                     & 0.2438                  & $\mu_4$                       &$1.1\times 10^{-7}$                 & 0.949                            & $\mu_4$                         & $1.66\times 10^{-5}$                   & 0.484                    \\ \hline
\multicolumn{1}{c}{\begin{tabular}[c]{@{}c@{}}Deviance \\  explained\end{tabular}} & \multicolumn{2}{c}{$28.1\%$} & \multicolumn{1}{c}{\begin{tabular}[c]{@{}c@{}}Deviance \\  explained\end{tabular}} & \multicolumn{2}{c}{$48.14\%$}   & \multicolumn{1}{c}{\begin{tabular}[c]{@{}c@{}}Deviance \\  explained\end{tabular}} & \multicolumn{2}{c}{$47.15\%$} \\ 
\end{tabular}
\end{table}

{ 
\renewcommand{\arraystretch}{3}
\begin{landscape}
\begin{table}[ht]
\caption{Complete list of the gait measures, their descriptions and associated domains.}
\vspace{1 cm}
\label{tab:my-table}
\begin{adjustbox}{width=1.38\textwidth}
\fontsize{55}{22}\selectfont
\begin{tabular}{llllll}
\hline
Gait Measure                                                                             & Description                                                                                                          & Domain      & Gait Measure                                                                           & Description                                                                                                             & Domain      \\ \hline
Activity Level [g]                                                                   & mean signal vector magnitude                                                                                         & Amplitude   & Cadence V (time domain) [steps/min] & number of steps per minute & Rhythm      \\ \hline
rng V,ML,AP [g]                                                                      & acceleration range                                                                                                   & Amplitude   & rms V,ML,AP [g]                                                                    & acceleration root mean square                                                                                           & Amplitude   \\ \hline
frq V,ML,AP [Hz]            & dominant frequency of power spectrum  & Rhythm      & amp V,ML,AP [g2/Hz]                                              & amplitude of the dominant frequency  & Variability \\ \hline
wd V,ML,AP [Hz]                                                                      & width of the dominant frequency & Variability & slp V,ML,AP [g2/Hz2]                           & amp/wd                                                                                                                  & Variability \\ \hline
stpReg V,ML,AP [unitless]                                                            & step regularity                                                                                                      & Symmetry    & strReg V,ML,AP [unitless]                                                          & stride regularity                                                                                                       & Variability \\ \hline
stepSym V,ML,AP [unitless]                                                           & step symmetry (=stpReg/strReg)                                                                                       & Symmetry    & stepTime V,ML,AP [s]                                                               & step time (calculated from autocorrelation function) & Rhythm      \\ \hline
strideTime V,ML,AP [s]                                                               & stride time (calculated from autocorrelation function)   & Rhythm      & HR v,ml,ap [unitless]                                                              & harmonic ratio                                                                                                          & Symmetry    \\ \hline
Cad V,ML,AP(frequency domain) [steps/min] & frq *60                                                                                                              & Rhythm      & Mean Stride Time [s]                                                               & mean stride time                                                                                                        & Rhythm      \\ \hline
CV Stride Time [$\%$]                                                                  & 100*(standard Deviation/mean)                                                                                        & Variability & Mean Step Time[s]                                                                  & mean step time                                                                                                          & Rhythm      \\ \hline
CV Step time [$\%$]                                                                    & 100*(standard Deviation/mean)                                                                                        & Variability & Mean Step Length [cm]                                                              & mean step length              & Pace        \\ \hline
CV\_Step\_length                                                                         & 100*(standard Deviation/mean)                                                                                        & Variability & Step Velocity [cm/sec]                                                             & mean step length/mean step time                                                                                         & Pace        \\ \hline
Distance [m]                                                                         & sum of step length                                                                                                   & Pace        & MeanPeaks                                                                              & mean of peaks                                                                                                                      & Rhythm      \\ \hline
STDPeaks                                                                                 & sd of peaks                                                                                                                   & Variability & CVPeaks                                                                                & CV of peaks                                                                                                                      & Variability \\ \hline
MeanAMPPeaks                                                                             & mean peak amplitudes (time domain)                                                                                   & Amplitude   & STDAMPPeaks                                                                            & sd of peak amplitudes  (time domain)                                         & Variability \\ \hline
CVAMPPeaks                                                                               & 100*(standard Deviation/mean)                                                                                        & Variability &                                                                                        &                                                                                                                         &             \\ \hline
\end{tabular}
\end{adjustbox}
\end{table}
\end{landscape}
}

\begin{table}[H]
\small
\caption{Displayed are the results from generalized additive models (GAM) of cognitive scores (ATTN, VM and EF) on first four L-moments of stride regularity , step velocity  and cadence after adjusting for age, sex and education. Benchmark models using age, sex and education produce (adjusted) $R^2_{ATTN}=  0.163$, $R^2_{VM}=  0.2489$, $R^2_{EF}=  0.2754$. A significant improvement (around $17\%-38\%$ gain) is noticed in terms of adjusted R-squared compared to models B1-B3 (Table 3 in the paper) indicating potential non-linearity in the effects of the subject-specific L-moments.}
\label{tab:my-table44}
\centering
\begin{tabular}{ccccccc}
\hline
Y                                                              & \multicolumn{2}{c}{Model B1 (stride regularity)}  & \multicolumn{2}{c}{Model B2 (step velocity)}           & \multicolumn{2}{c}{Model B3 (cadence)}   \\ \hline
\multirow{10}{*}{ATTN}                                                   & Coef        & p-value              & Coef                     & p-value              & Coef         & p-value              \\ \cline{2-7} 
                                                                         & $\beta_0$  & 0.006               & $\beta_0$ & 0.067               & $\beta_0$ &  0.217               \\ \cline{2-7} 
                                                                         & age   & 0.049               & age& 0.340 & age       &  0.831               \\ \cline{2-7} 
                                                                         & sex (M)  &0.022               & sex (M)                & 0.009               & sex (M)  & 0.007               \\ \cline{2-7} 
                                                                         & edu  & 0.004               & edu               &  0.005            & edu   &0.003              \\ \cline{2-7} 
                                                                         & $h_1(L_1)$       & 0.024               & $h_1(L_1)$                    & 0.876             & $h_1(L_1)$     &0.092              \\ \cline{2-7} 
                                                                         & $h_2(L_2)$     & 0.002               & $h_2(L_2)$  &   0.0471        & $h_2(L_2)$     & 0.005            \\ \cline{2-7} 
                                                                         & $h_3(L_3)$      &  0.179               & $h_3(L_3)$      &  0.482              & $h_3(L_3)$   & 0.422              \\ \cline{2-7} 
                                                                         & $h_4(L_4)$     &  0.177              & $h_4(L_4)$                 & 0.636 & $h_4(L_4)$     & 0.874              \\ \cline{2-7} 
                                                                         & adj-Rsq   & \multicolumn{1}{c}{0.395}   & adj-Rsq   & \multicolumn{1}{c}{0.232}                 & adj-Rsq   & \multicolumn{1}{c}{0.259}   \\ \hline
\multirow{9}{*}{VM}                                                      & $\beta_0$ & 0.173               & $\beta_0$               & 0.159              & $\beta_0$ & 0.614              \\ \cline{2-7} 
                                                                         & age        &0.687               & age    & 0.435              & age  & 0.740               \\ \cline{2-7} 
                                                                         & sex (M)  &  $0.00057$                & sex (M)    & $1.6\times 10^{-5}$ & sex (M)   &  0.00026  \\ \cline{2-7} 
                                                                         & edu   &  0.00344              & edu                &  0.0126               & edu   & 0.011               \\ \cline{2-7} 
                                                                         & $h_1(L_1)$     & 0.265              & $h_1(L_1)$    & 0.444               & $h_1(L_1)$    & 0.678                 \\ \cline{2-7} 
                                                                         & $h_2(L_2)$    & $2.59\times10^{-5}$               & $h_2(L_2)$                &  0.00875           & $h_2(L_2)$     & $3.29\times10^{-5}$               \\ \cline{2-7} 
                                                                         & $h_3(L_3)$      &  0.071            & $h_3(L_3)$     &0.170               & $h_3(L_3)$    & 0.748                \\ \cline{2-7} 
                                                                         & $h_4(L_4)$ & 0.0149                & $h_4(L_4)$      & 0.567               & $h_4(L_4)$   & 0.715                \\ \cline{2-7} 
                                                                         & adj-Rsq   & \multicolumn{1}{c}{0.472}   & adj-Rsq   & \multicolumn{1}{c}{0.435}                  & adj-Rsq   & \multicolumn{1}{c}{0.444}   \\ \hline
\multirow{9}{*}{EF} & $\beta_0$ &0.064               & $\beta_0$              & 0.079              & $\beta_0$  & 0.293               \\ \cline{2-7} 
                                                                         & age      &  0.337               & age  & 0.261              & age      & 0.826            \\ \cline{2-7} 
                                                                         & sex (M)  & $0.00012$ & sex (M) & $2.8\times 10^{-6}$ & sex (M)  & $9.04\times 10^{-5}$ \\ \cline{2-7} 
                                                                         & edu  & 0.00226               & edu      & 0.007  & edu  & 0.005               \\ \cline{2-7} 
                                                                         & $h_1(L_1)$  & 0.273              & $h_1(L_1)$ & 0.252 & $h_1(L_1)$ &0.138               \\ \cline{2-7} 
                                                                         & $h_2(L_2)$&0.00051               & $h_2(L_2)$    &  0.0164               & $h_2(L_2)$    & $0.00013$ \\ \cline{2-7} 
                                                                         & $h_3(L_3)$ &0.474               & $h_3(L_3)$   &  0.866 & $h_3(L_3)$  & 0.163              \\ \cline{2-7} 
                                                                         & $h_4(L_4)$& 0.413              & $h_4(L_4)$   & 0.916  & $h_4(L_4)$   & 0.058              \\ \cline{2-7} 
                                                                         & adj-Rsq   & \multicolumn{1}{c}{0.491}   & adj-Rsq   & \multicolumn{1}{c}{0.407}                 & adj-Rsq   & \multicolumn{1}{c}{0.455}   \\ \hline
\end{tabular}
\end{table}

\begin{table}[H]
\small
\centering
\caption{Results from logistic regression of cognitive status on JIVE scores, adjusting for age and sex.}
\label{tab:my-table4}
\begin{tabular}{llllll}
\hline
Predictor  & Beta     & p-value  & Predictor       & Beta     & p-value  \\ \hline
age        & -0.06167 & 0.453812 & Rhythm-PC2      & -0.86731 & 0.045292 \\ \hline
sex        & 6.13591  & 0.000378 & Rhythm-PC5      & -1.87063 & 0.014383 \\ \hline
joint-PC1  & 1.05257  & 0.044063 & Rhythm-PC7      & 1.65990  & 0.006741 \\ \hline
joint-PC2  & 2.93640  & 0.000728 & Variability-PC4 & -0.43651 & 0.441194 \\ \hline
Pace-PC1   & 0.82707  & 0.214616 & Variability-PC7 & 0.35954  & 0.418637 \\ \hline
Pace-PC2   & 1.41231  & 0.070124 & Variability-PC9 & -0.69622 & 0.182581 \\ \hline
Rhythm-PC1 & 0.45852  & 0.347566 &                 &          &          \\ \hline
\end{tabular}
\end{table}
\section{Supporting Figures}
Web Figures S1-S7 referenced in the paper are given below. 

\begin{figure}[H]
\centering
\includegraphics[width=1\linewidth , height=.7\linewidth]{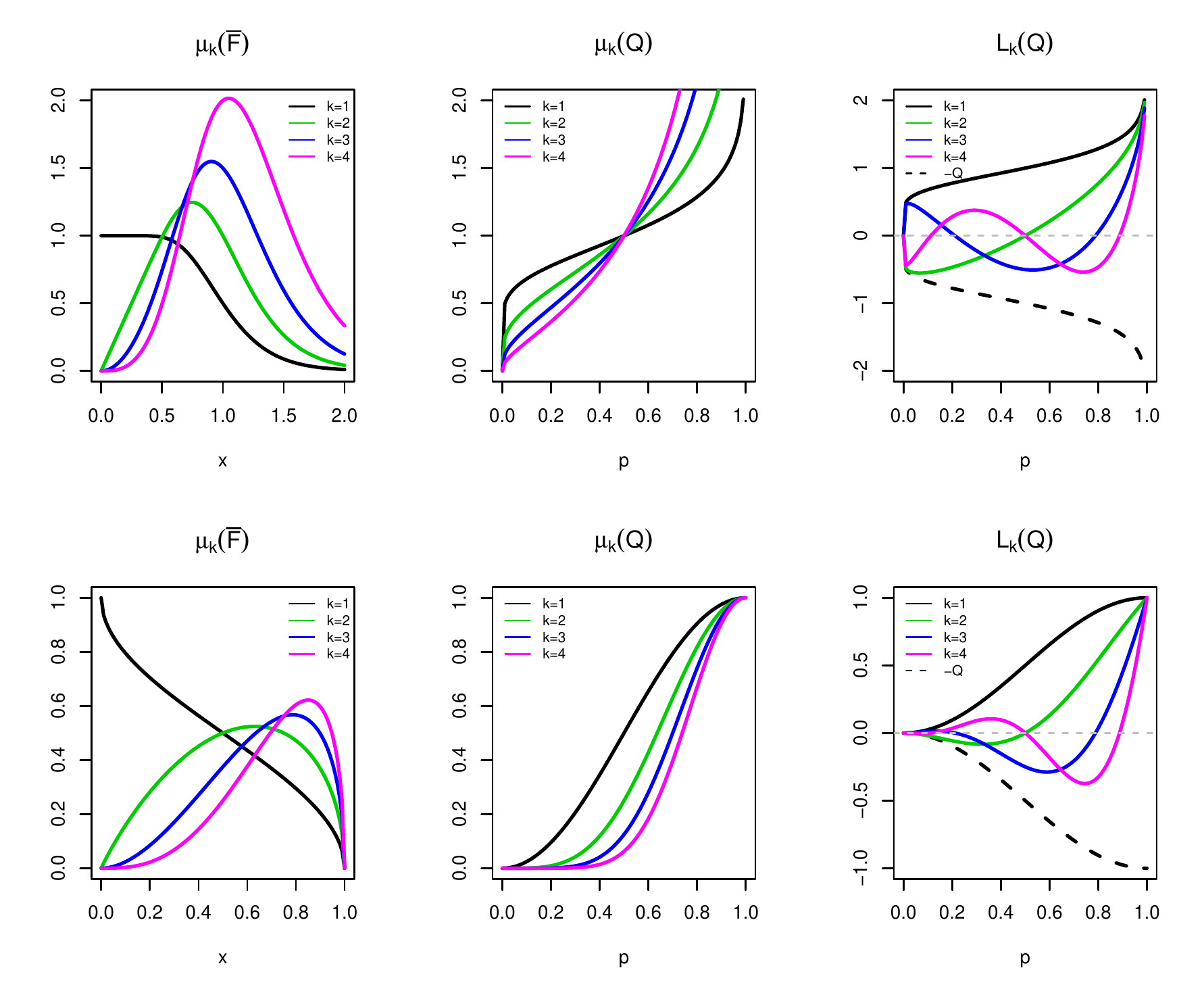}
\caption{Top: lognormal distribution with SD = 0.3. Bottom: beta distribution with $\alpha=\beta=0.5$. The y-axis values represent  $kx^{k-1}\bar{F}(x)$ for $\mu_k(\bar{F})$, $Q^k(p)$ for $\mu_k(Q)$, and $Q(p)P_{k-1}(p)$ for $L_k(Q)$.}
\label{fig:fig2l}
\end{figure}

\begin{figure}[H]
\begin{center}
\begin{tabular}{l}
 \includegraphics[width=.9\linewidth , height=.5\linewidth]{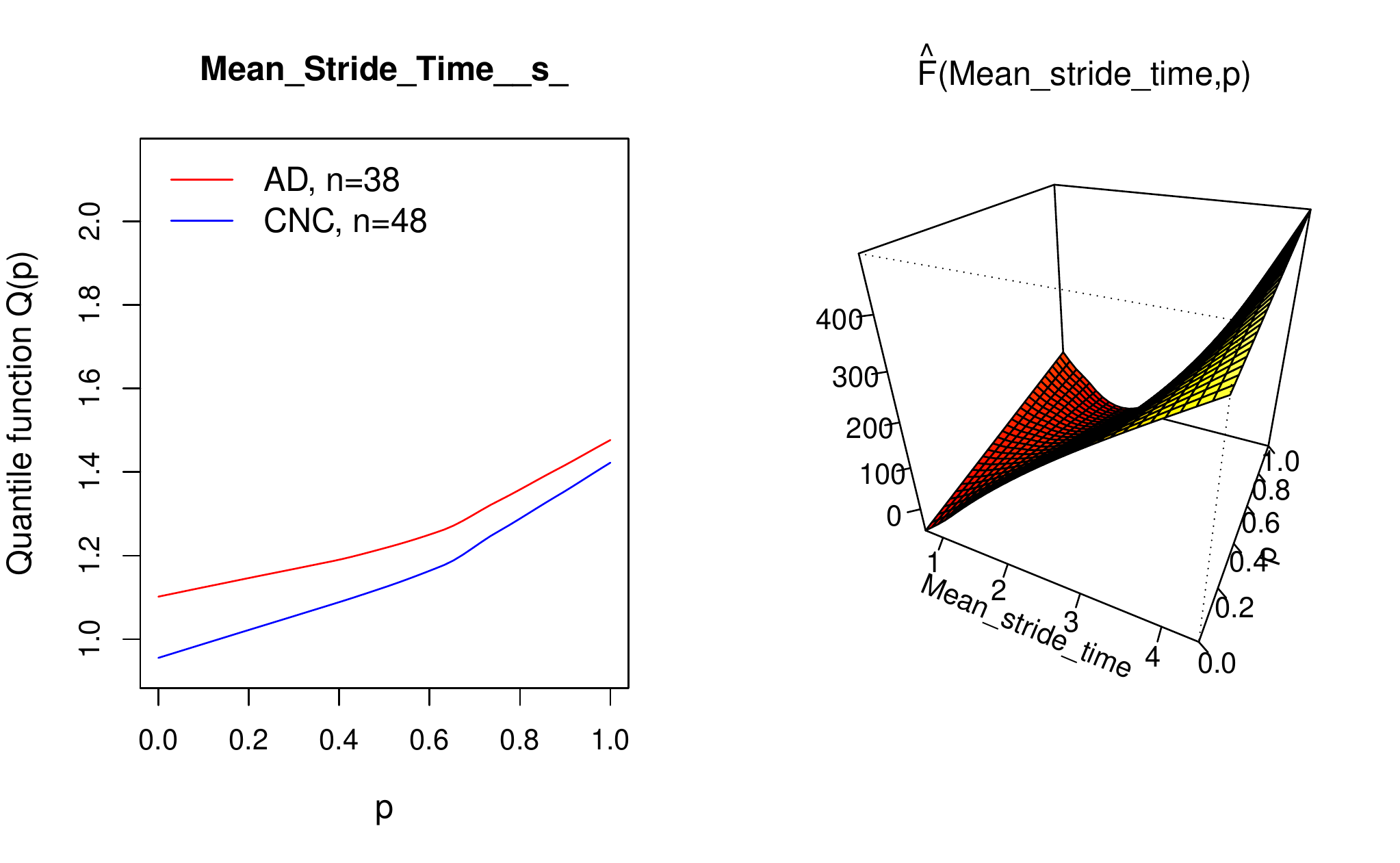}\\
 \includegraphics[width=.9\linewidth , height=.5\linewidth]{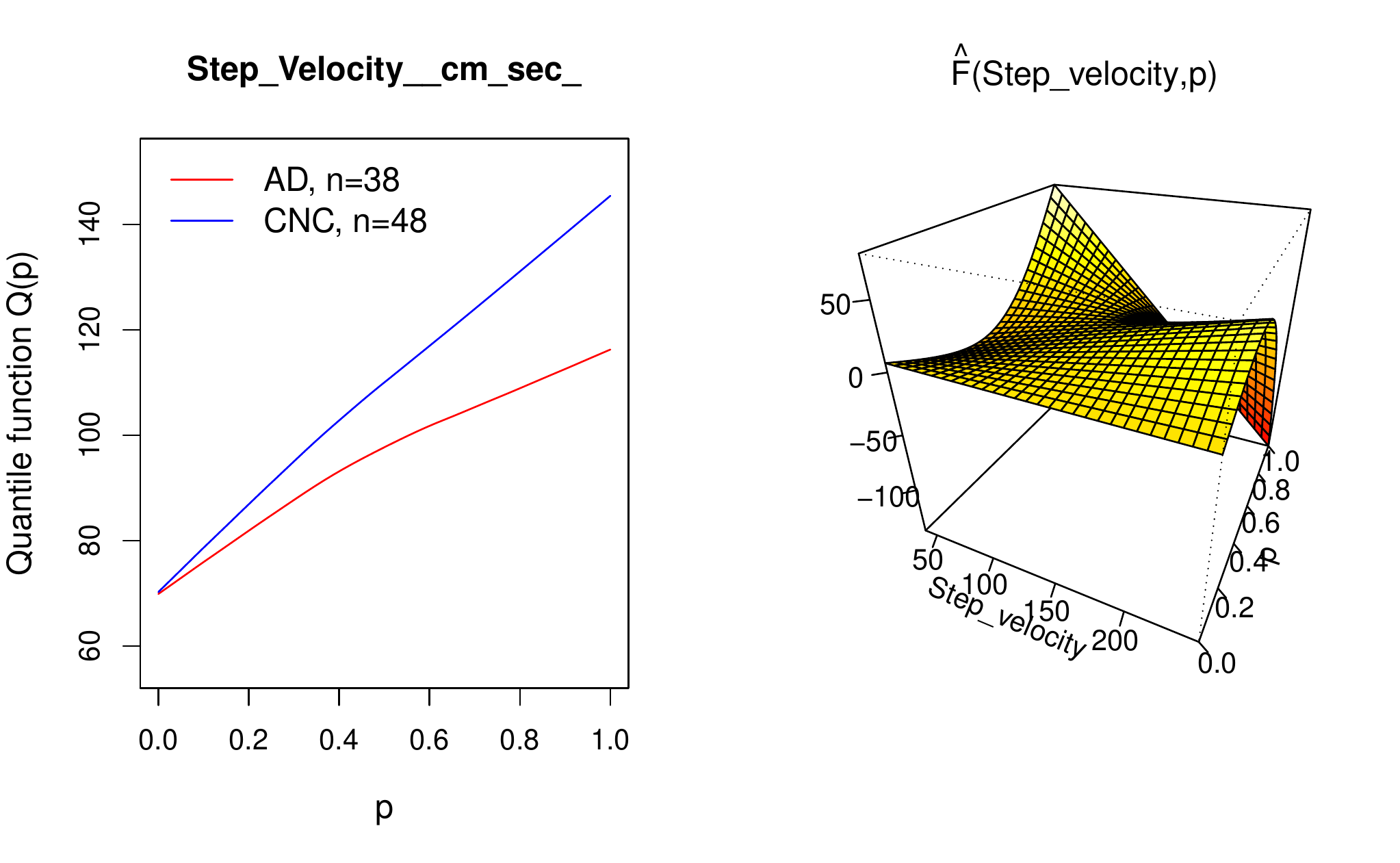}\\
\end{tabular}
\end{center}
\caption{Displayed are estimated additive effects of quantile functions of mean stride time (top) and step velocity (bottom) from FGAM-QF. The average quantile functions of the metrics for the two groups (AD and CNC) are shown in left and the estimated bivariate surfaces of quantile effect $\hat{F}(q,p)$ are shown in right.  }
\label{fig:fig4new}
\end{figure}

\begin{figure}[H]
\begin{center}
\begin{tabular}{ll}
\hspace*{- 9 mm}
 \includegraphics[width=.51\linewidth , height=.56\linewidth]{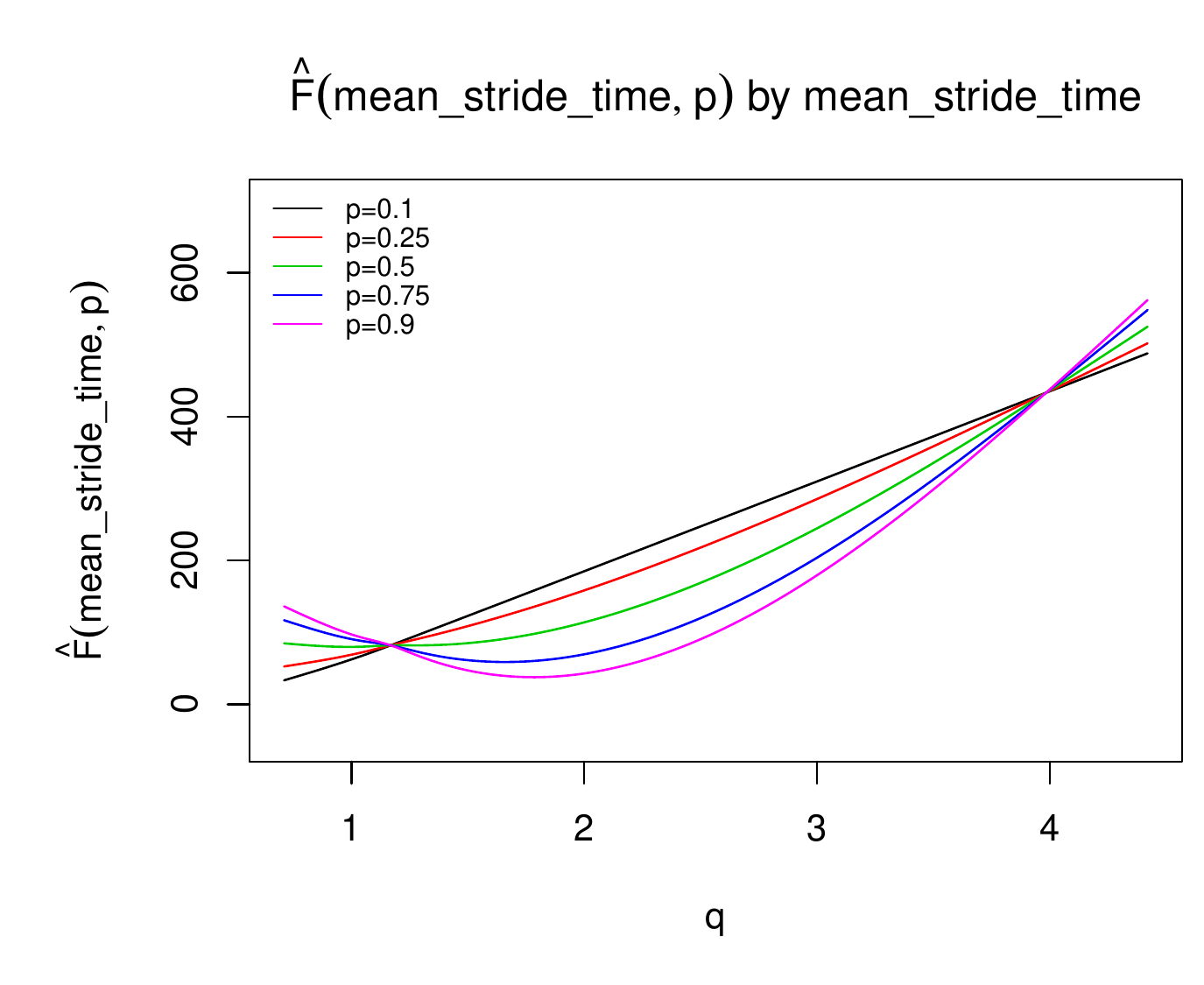} &
 \includegraphics[width=.5\linewidth , height=.55\linewidth]{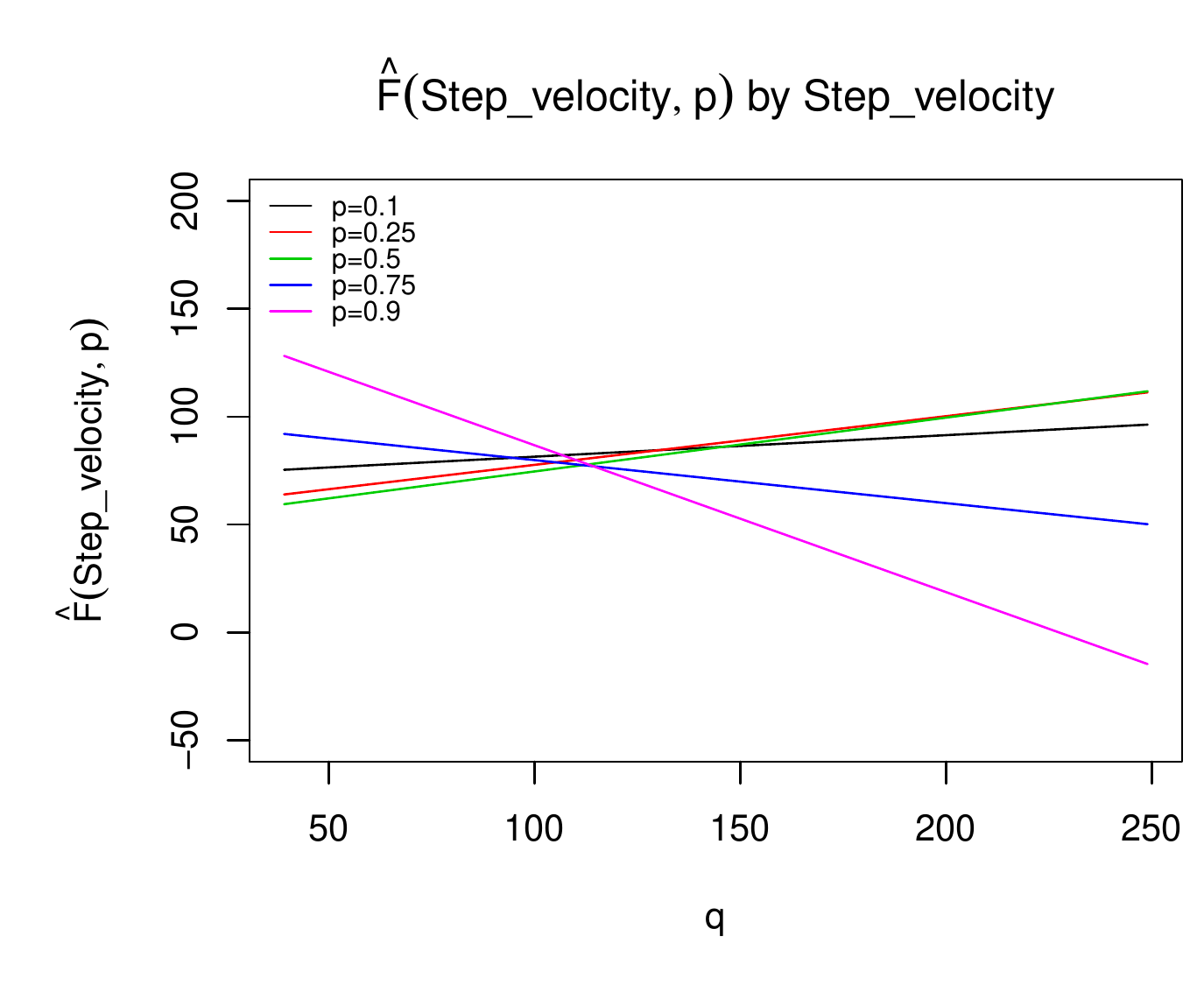}\\
\end{tabular}
\end{center}
\caption{Displayed are the sliced effect of the FGAM-QF surface $\hat{F}(q,p)$ for the two gait metrics mean stride time and step velocity, for $p=0.1,0.25,0.5,0.75,0.9$.}
\label{fig:fig4newb}
\end{figure}
\newpage

\begin{figure}[H]
\begin{center}

 \includegraphics[width=.7\linewidth , height=.7\linewidth]{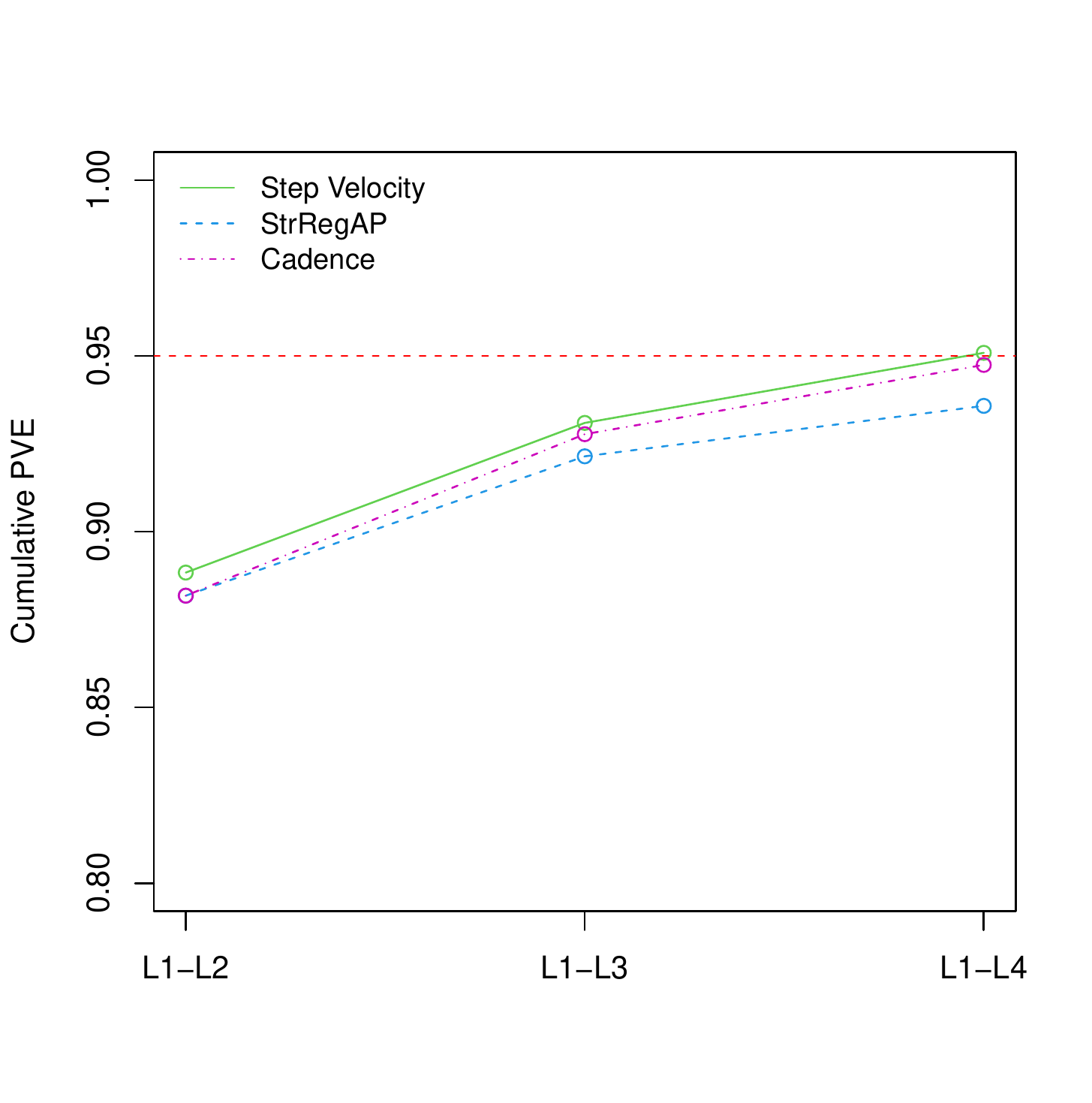} 
\end{center}
\caption{Cumulative proportion of variance explained (PVE) by the first four L-moments for step velocity, cadence, and stride regularity (StrRegAP) averaged across all samples.}
\label{fig:figs1}
\end{figure}

\begin{figure}[H]
\centering
\includegraphics[width=1\linewidth , height=.7\linewidth]{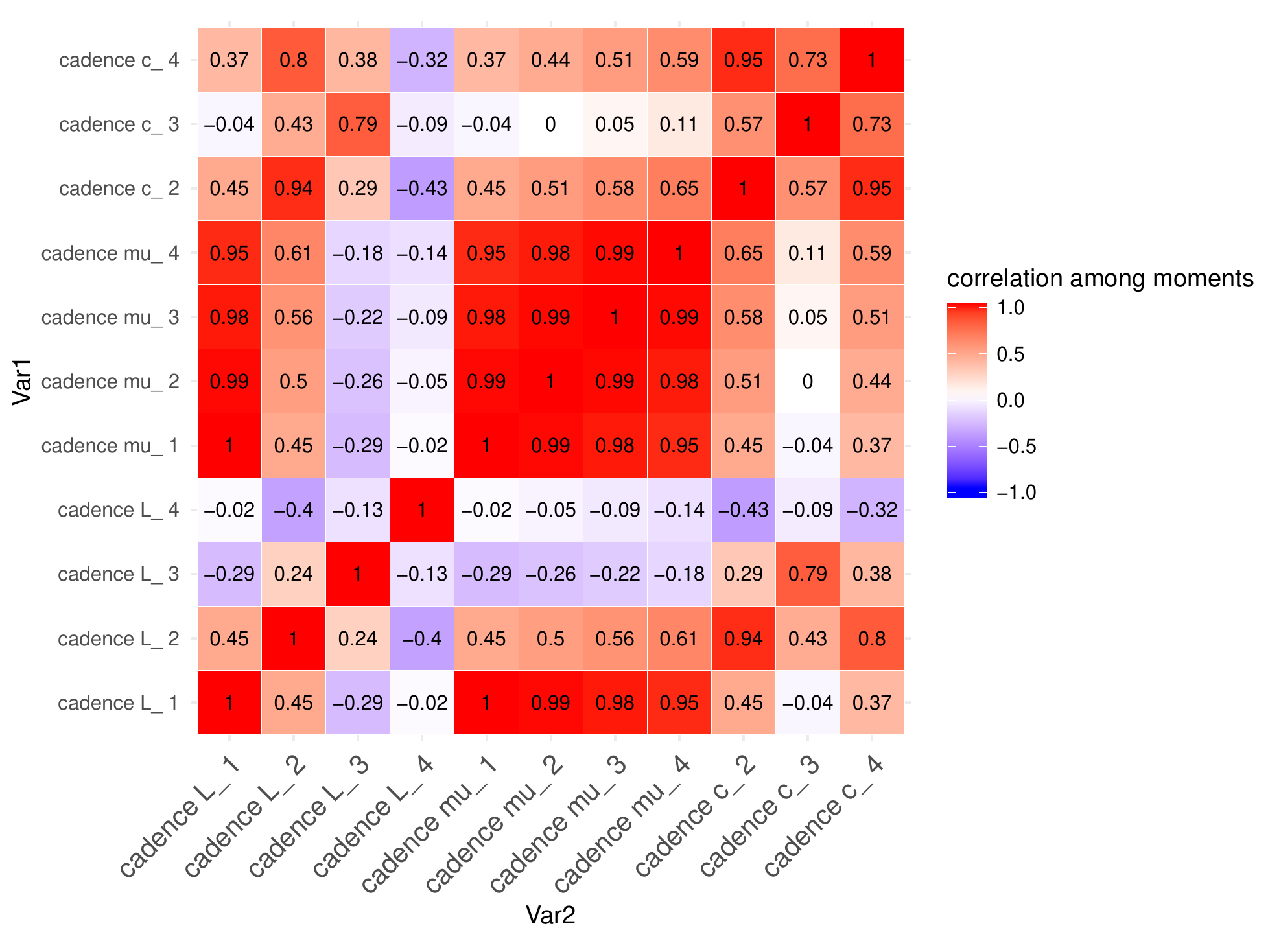}
\caption{Displayed is the heatmap of sample correlation among the first four L-moment (L), regular moments (mu) and central moments (c) of the gait metric cadence.}
\label{fig:fig2l}
\end{figure}

\begin{figure}[H]
\begin{center}
\begin{tabular}{l}
 \includegraphics[width=.66\linewidth , height=.34\linewidth]{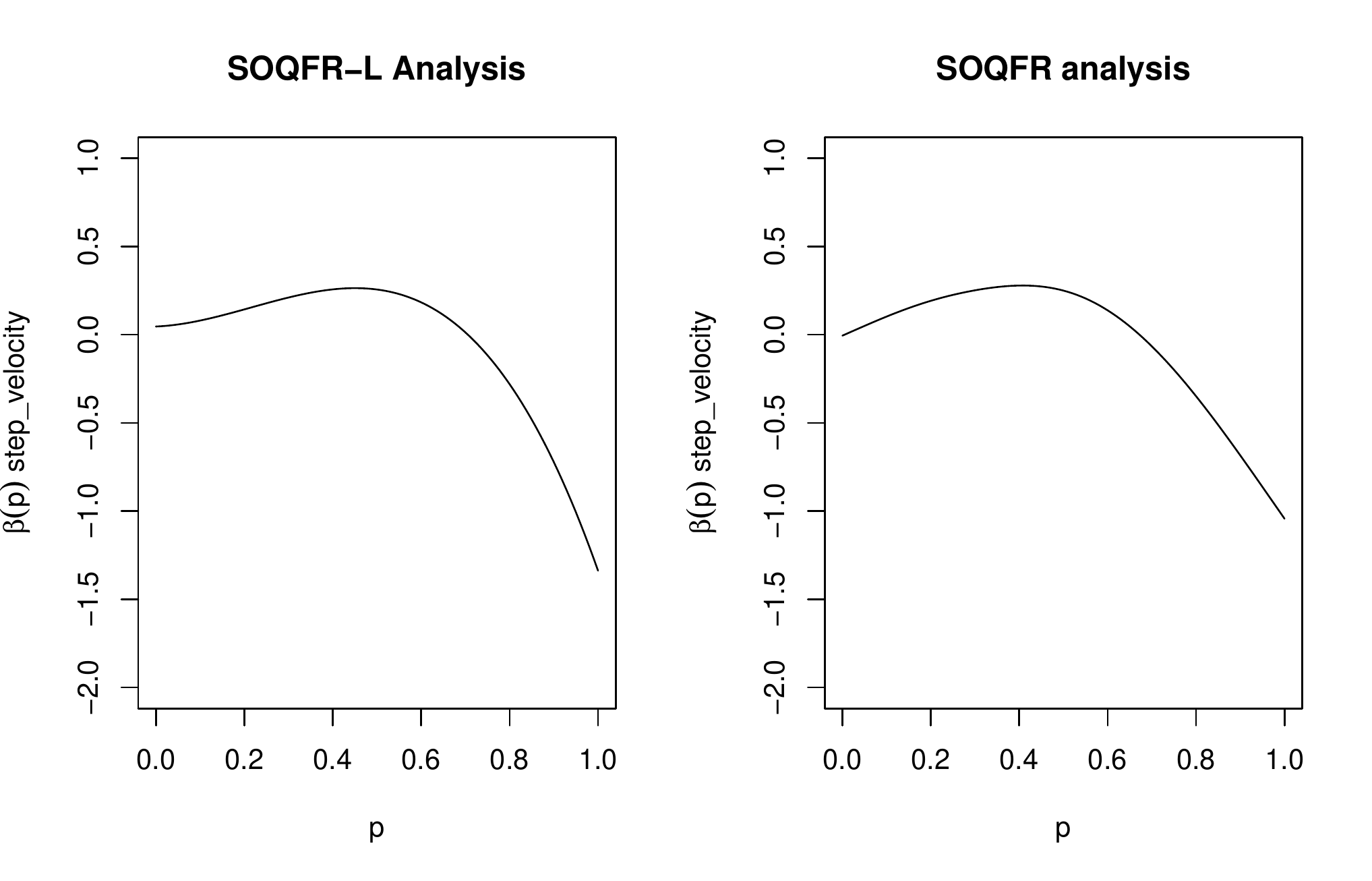} \\
 \includegraphics[width=.66\linewidth , height=.34\linewidth]{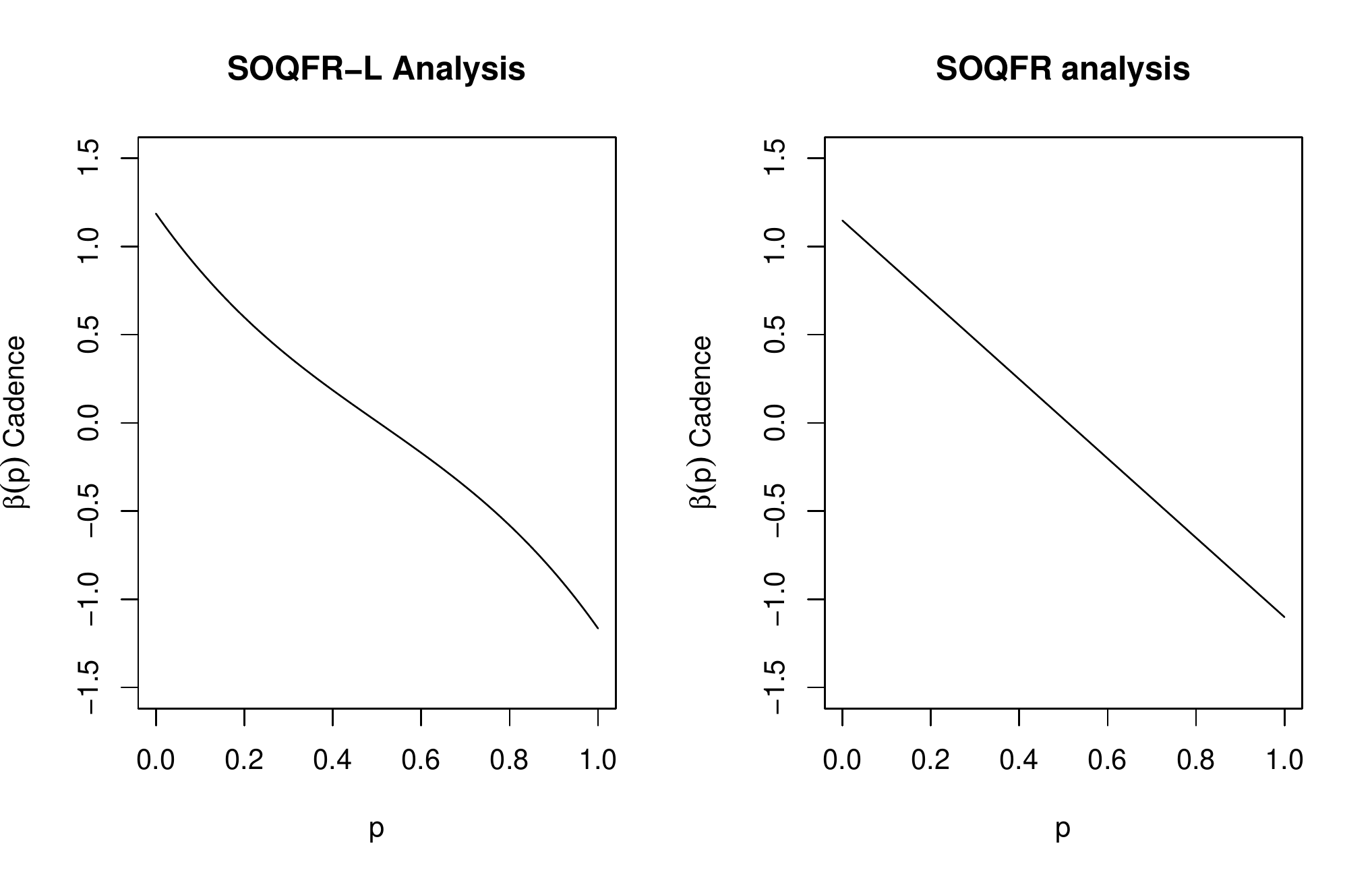}\\
  \includegraphics[width=.66\linewidth , height=.34\linewidth]{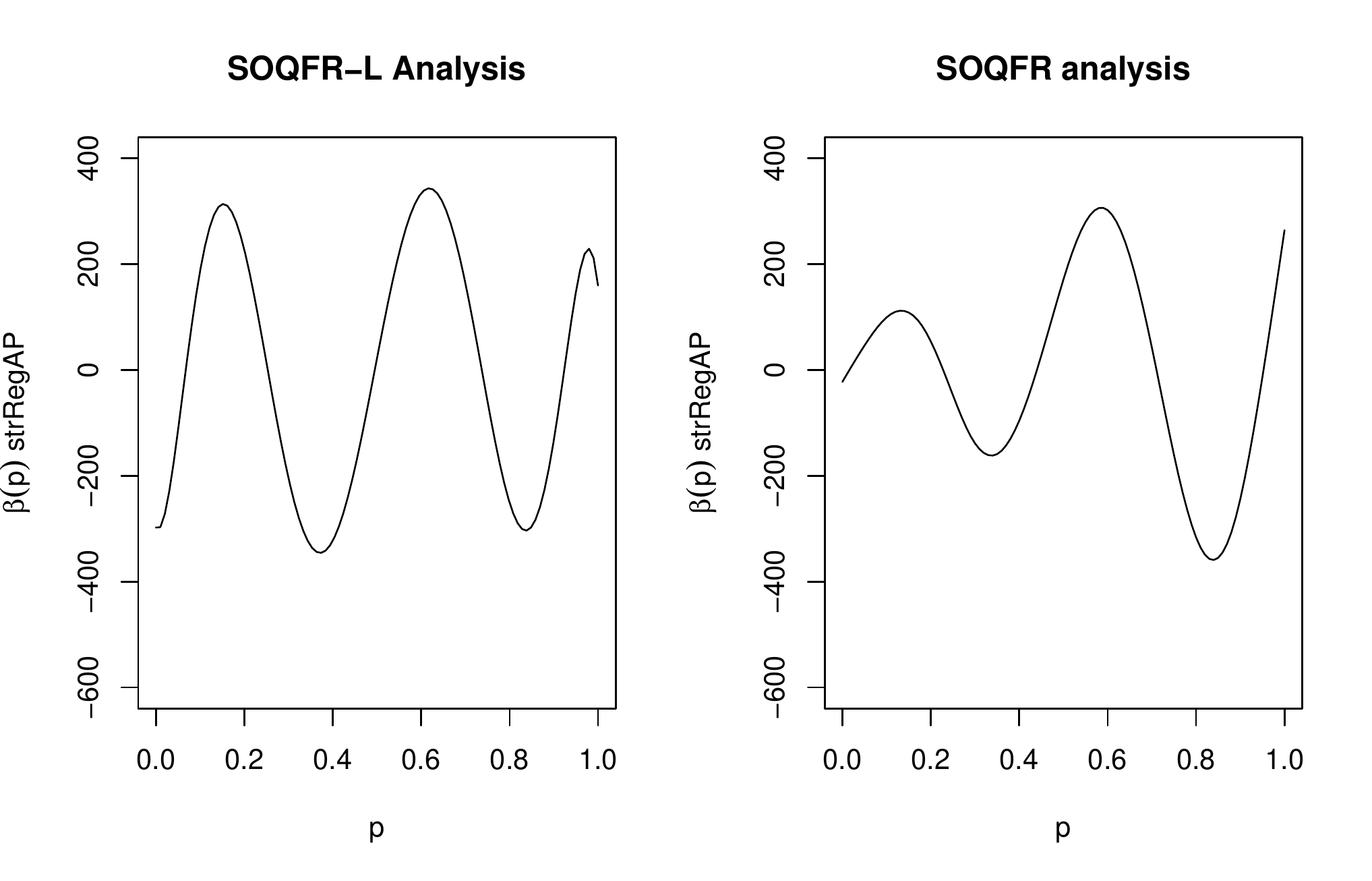}\\
\end{tabular}
\end{center}
\caption{Displayed are functional regression coefficients $\beta(p)$ estimated using SOQFR-L (left column) and SOQFR (right column) of step velocity (top), cadence (middle) and stride regularity (bottom).}

\label{fig:figs1}
\end{figure}
\begin{figure}[H]
\centering
\includegraphics[width=1\linewidth , height=.7\linewidth]{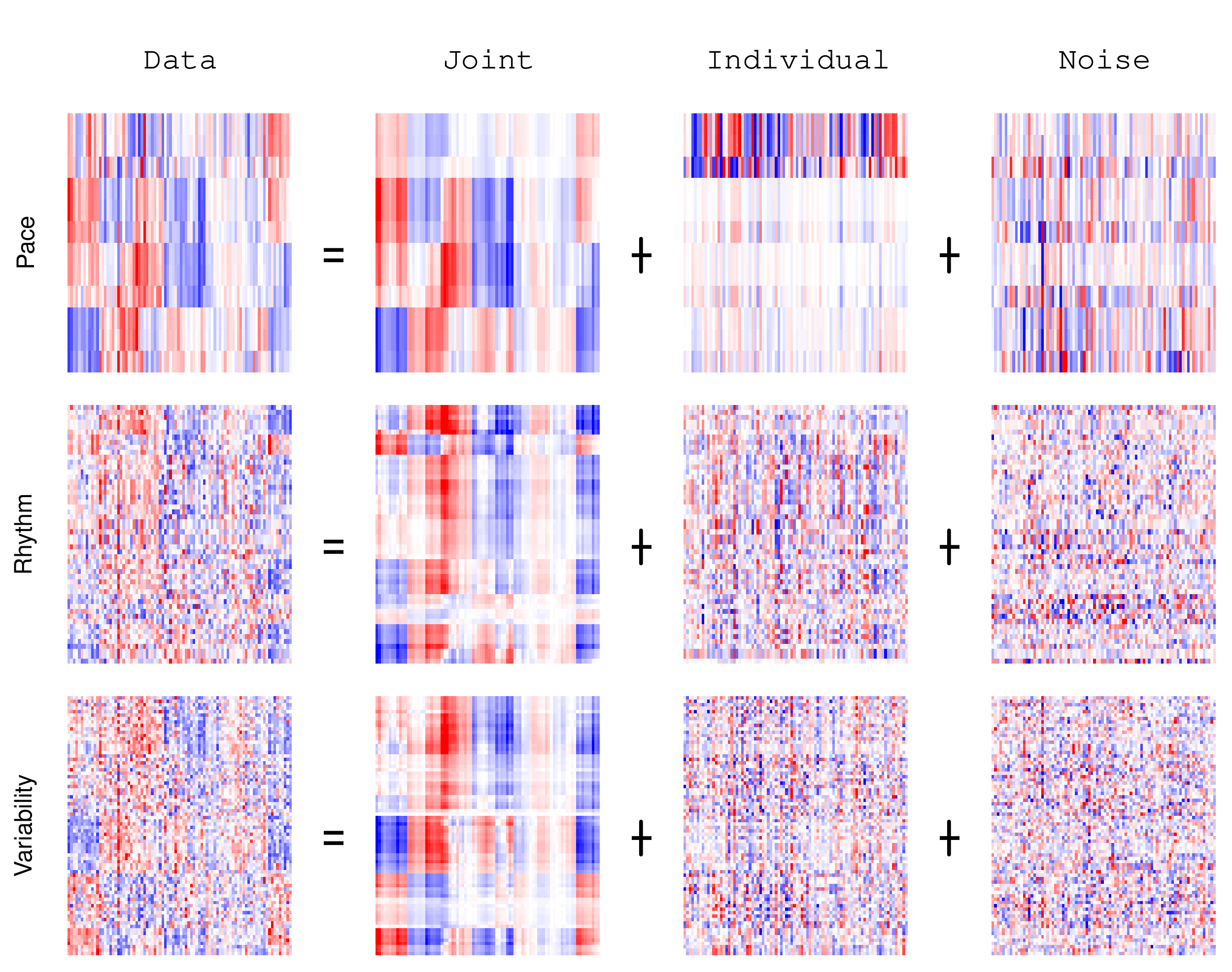}
\caption{Heatmaps showing JIVE decomposition of the L-moments of gait measures. Columns represent subjects and rows represent features. Rows and columns are
ordered by complete linkage clustering of the joint structure.}
\label{fig:fig5}
\end{figure}
\bibliographystyle{biorefs}
\bibliography{score.bib}